\documentclass[biber]{nowfnt} % creates the journal version, needs biber version
\usepackage{caption}
\usepackage{xspace}
\usepackage{colortbl}
\usepackage{multirow}
\usepackage{hyperref}
\usepackage{makecell}
\usepackage{pifont}
\usepackage{tabularx}
\usepackage{booktabs}

\usepackage{adjustbox}
\usepackage[dvipsnames]{xcolor}
\usepackage{tikz}
\usetikzlibrary{calc,shadings,patterns}
\usetikzlibrary{arrows.meta,positioning}
\usepackage{enumitem}
\usepackage{pgfplots}
\usepackage{graphicx}
\usepackage{subcaption}
\usepackage{amsmath, bm}
\usepackage{amssymb}
\usepackage{cleveref}
\usepackage{longtable}
\usepackage{tcolorbox}
\usepgfplotslibrary{statistics}
\usepackage{float}

\newcommand{\raw}{\ding{51}}              % ✓
\newcommand{\feat}{$\blacklozenge$}        % ◆
\newcommand{\link}{$\nearrow$}             % ↗
\newcommand{\enrich}{$\bigstar$}           % ★

\PassOptionsToPackage{table,dvipsnames}{xcolor}

\DeclareMathOperator*{\argmax}{arg\,max}

\definecolor{semBrightGreen}{RGB}{0,255,0}   % Q1, A*
\definecolor{semGreen}{RGB}{0,150,0}         % A
\definecolor{semYellow}{RGB}{255,220,0}      % B, Q2
\definecolor{semOrange}{RGB}{255,140,0}      % C, Q3
\definecolor{semGray}{RGB}{160,160,160}      % NR
\definecolor{RYB2}{RGB}{245,245,245}
\definecolor{RYB1}{RGB}{218,232,252}
\definecolor{RYB4}{RGB}{108,142,191}

\usetikzlibrary{positioning,fit,shapes.geometric,backgrounds,calc}
\pgfplotsset{compat=1.18}

\newcommand{\cmark}{\ding{51}}

\definecolor{forestgreen(web)}{rgb}{0.13, 0.55, 0.13}
\definecolor{lightgray}{rgb}{0.83, 0.83, 0.83}

\newcommand{\Tau}{\mathcal{T}}

\title{Data Processing for Offline Evaluation in Recommender Systems: a Survey}

\subtitle{Standard Practices and \\Multimodal Feature Extraction}

\maintitleauthorlist{
Alberto Carlo Maria Mancino\\
Politecnico di Bari\\
alberto.mancino@poliba.it
\and
Angela Di Fazio\\
Politecnico di Bari\\
angela.difazio@poliba.it
\and
Danilo Danese\\
Politecnico di Bari\\
danilo.danese@poliba.it
\and
Matteo Attimonelli\\
Politecnico di Bari; Sapienza Università di Roma\\
matteo.attimonelli@poliba.it
\and
Daniele Malitesta\\
LUISS Guido Carli University\\
dmalitesta@luiss.it
\and
Antonio Ferrara\\
Politecnico di Bari\\
antonio.ferrara@poliba.it
\and
Claudio Pomo\\
Politecnico di Bari\\
claudio.pomo@poliba.it
\and
Tommaso Di Noia\\
Politecnico di Bari\\
tommaso.dinoia@poliba.it
}

\usepackage{mwe}

\author[1]{Mancino, Alberto Carlo Maria}
\author[1]{Di Fazio, Angela}
\author[1]{Danese, Danilo}
\author[1,2]{Attimonelli, Matteo}
\author[3]{Malitesta, Daniele}
\author[1]{Ferrara, Antonio}
\author[1]{Pomo, Claudio}
\author[1]{Di Noia, Tommaso}

\affil[1]{Politecnico di Bari}
\affil[2]{Sapienza Università di Roma}
\affil[3]{{LUISS Guido Carli University}}

\articledatabox{\nowfntstandardcitation}

\begin{document}

\makeabstracttitle

\begin{abstract}
Offline evaluation is the dominant experimental paradigm in recommender systems research, enabling reproducible and cost-effective comparisons on historical interaction data. Yet, while considerable attention has been devoted to recommendation models and evaluation methodologies, the data processing decisions that precede model training have received less scrutiny. These decisions determine the information available to recommendation algorithms and can affect the comparability and reproducibility of experimental results.

This survey provides a systematic, cross-domain characterisation of data processing practices for the offline evaluation of recommender systems. We examine the data-centric pipeline, from dataset selection and interaction representation to data preparation, multimodal feature extraction, and train-validation-test splitting. Our analysis spans recommendation paradigms, including collaborative, sequential, session-based, graph-based, knowledge-aware, context-aware, multimodal, federated, cross-domain, contrastive-learning, and LLM-based recommendation. Beyond reviewing existing practices, we introduce a unified framework and taxonomy for describing data transformations and feature-extraction strategies, distinguishing data preparation from the extraction of representations from multimodal side information.

Our empirical analysis reveals a landscape dominated by a narrow set of dataset-level transformations, particularly support-driven filtering, while representation-dependent transformations remain less common. We further identify substantial heterogeneity in how auxiliary information is prepared and represented, as well as inconsistencies in the specification of data splitting protocols, where similar labels may conceal different experimental conditions.
\end{abstract}
\chapter{Introduction}

Recommender systems are commonly trained on the premise that user preferences can be inferred from historical interactions and subsequently leveraged to suggest relevant items. To assess the validity of this premise, three main evaluation paradigms are employed: offline experiments, user studies, and online experiments~\citep{DBLP:reference/sp/2022rsh}. Among them, \emph{offline evaluation} has become the dominant paradigm in academic research, mainly due to its accessibility (no user recruitment or live deployment is required) and its capacity to ensure reproducibility. By sharing protocols and datasets, researchers worldwide can replicate experimental setups and directly compare new models with established baselines. The prevalence of this paradigm is confirmed by \citet{DBLP:journals/csur/ZangerleB23}, who report that more than 70\% of published recommender systems studies rely on offline evaluation.

Despite its centrality, offline evaluation is far from being problem-free. Over time, the literature has pointed out several methodological weaknesses: dataset–task mismatch~\citep{DBLP:conf/recsys/HidasiC23a, DBLP:journals/tois/HerlockerKTR04}, dependence on heavily preprocessed data~\citep{DBLP:conf/recsys/HidasiC23a, DBLP:conf/recsys/GusakVKVF25}, information leakage~\citep{DBLP:conf/recsys/Sun23, DBLP:journals/tois/JiS0L23}, biased sampling of negative instances~\citep{DBLP:journals/tois/ChenMZWLM23}, evaluation loops~\citep{DBLP:conf/nips/SinhaGR16, 10.1145/3728372}, metric–task mismatch~\citep{DBLP:conf/recsys/TammDV21}, and lack of statistical testing~\citep{DBLP:conf/recsys/WongOZMCR21}. These factors hinder comparability across studies and may even compromise the validity of conclusions drawn from offline experiments. More fundamentally, the usefulness of offline evaluation relies on the assumption that its outcomes are predictive of real-world system performance, often referred to as \textit{evaluation in the wild}. Poorly designed protocols can further amplify the natural discrepancy between historical logs and live user behaviour, raising concerns about the external validity of offline findings.

In recent years, research efforts have primarily focused on the final stages of the recommendation pipeline, namely, \textit{model reproducibility}\citep{DBLP:conf/recsys/DacremaCJ19,DBLP:journals/tois/DacremaBCJ21, DBLP:journals/umuai/BelloginS21} and \textit{evaluation methodology}\citep{DBLP:journals/csur/ZangerleB23, DBLP:journals/umuai/PuCH12, DBLP:conf/recsys/MalitestaPAMNS24}. In contrast, the earlier stages, covering data selection, preparation, and splitting, have received comparatively little systematic attention. This lack of focus is particularly evident in scenarios involving side information. Tasks such as content-based~\citep{DBLP:journals/ijet/JavedSHIAL21}, knowledge-aware~\citep{DBLP:journals/tkde/GuoZQZXXH22}, and multimodal~\citep{DBLP:journals/tors/MalitestaCPMNS25} recommendation rely on features beyond the user–item matrix, yet the preprocessing of these additional signals is rarely standardised or transparently documented. As a result, experimental settings often differ substantially across studies, making it difficult to isolate the impact of model innovations from that of data preparation choices and hindering fair comparisons between competing approaches.

For instance, two studies may evaluate their models on different versions of the same dataset or adopt different train–test splitting strategies, making direct comparisons challenging. The problem becomes even more pronounced in multimodal recommendation, where side information can be generated through heterogeneous feature extraction pipelines (e.g., ResNet-50 versus VGG-16 for images, or Sentence-BERT versus precomputed textual embeddings). Consequently, observed performance differences may stem as much from data preparation choices as from the recommendation models themselves.

Addressing this gap, the present survey provides a comprehensive overview of datasets and preprocessing practices in the offline evaluation of recommender systems, with particular emphasis on side information. Our goal is twofold: (i) to characterise the current landscape and assess the impact of data preparation choices on experimental outcomes, and (ii) to stimulate discussion towards more consistent and transparent practices. By systematically reviewing existing approaches, we aim to provide researchers with a reference for navigating preprocessing decisions and interpreting experimental results. Ultimately, we hope this survey will contribute to more reproducible and comparable recommender systems research.

\section{Strategy for Literature Search}
\label{sec:strategy}

This survey aims to provide a comprehensive overview of data selection and preprocessing practices across the field of recommender systems research. Since these practices are often domain-dependent, being influenced by factors such as application context, research objectives, and evaluation protocols, our analysis spans both established and emerging recommendation paradigms.

To ensure broad coverage of the field, we considered both foundational and rapidly evolving recommendation domains. Specifically, we included \textbf{collaborative}, \textbf{sequential}, \textbf{session-based}, \textbf{graph-based}, \textbf{knowledge-based}, and \textbf{context-aware} recommendation, together with emerging areas such as \textbf{contrastive learning}, \textbf{federated learning}, \textbf{cross-domain}, \textbf{multimodal}, and \textbf{large language model (LLM)-based} recommendation.
Rather than performing a direct paper-level search, we leveraged recent survey articles as entry points, allowing us to systematically cover major recommendation domains through expert-curated representations of the literature.

To ensure a transparent and reproducible selection process, candidate surveys were retrieved from DBLP and Semantic Scholar using combinations of the keywords “survey”, “recommender systems”, “recommendation”, and “recommenders”, together with domain-specific terms in the title or abstract (see \Cref{tab:surveys}). Duplicate entries were merged, and citation metadata were manually verified when inconsistencies were detected.

\begin{table}[t]
\caption{List of surveys on major recommendation domains analysed in this work. Each survey is associated with the keyword(s), which also identify the corresponding recommendation domain, and with the publication venue.}
\begin{adjustbox}{width=\textwidth}
\begin{tabular}{llll}
\toprule
\textbf{Survey}                    & \textbf{Keyword(s)} & \textbf{Venue}  \\ 
\midrule
\cite{DBLP:journals/tkde/WuHWZW23}      & \textit{collaborative filtering}    &  TKDE \\
\cite{DBLP:journals/is/BokaNN24}        & \textit{sequential}                 &  Information Sciences \\
\cite{DBLP:journals/csur/WangCWSOL22}   & \textit{session-based}              &  ACM Computing Surveys \\
\cite{DBLP:journals/csur/WuSZXC23}      & \textit{graph-based}                &  ACM Computing Surveys \\
\cite{DBLP:journals/tkde/GuoZQZXXH22}   & \textit{knowledge}                  &  TKDE \\
\cite{DBLP:journals/csr/KulkarniR20}    & \textit{context-aware}              &  Computer Science Review \\
\cite{DBLP:journals/tois/JingZZW24}     & \textit{contrastive learning}       &  TOIS \\
\cite{DBLP:journals/csur/LiuHXZGWLT25}  & \textit{multimodal}                 &  ACM Computing Surveys \\
\cite{DBLP:journals/tois/ZangZLZY23}    & \textit{cross-doman/cross domain} &  TOIS \\
\cite{DBLP:journals/tnn/SunXLHKWJC25}   & \textit{federated}                  &  TNNLS \\
\cite{DBLP:journals/tkde/ZhaoFLLMWWWZTL24} & \textit{large language models}      &  TKDE \\ \bottomrule
\end{tabular}
\end{adjustbox}
\label{tab:surveys}
\end{table}

We retained only surveys published between 2020 and 2025 that had received at least 30 citations, thereby focusing on recent and community-recognised works. To further ensure quality, we excluded papers not appearing in Q1 journals (according to Scimago Journal Rank\footnote{\url{https://www.scimagojr.com/}}) or in A/A*-ranked conferences (according to the CORE ranking\footnote{\url{https://portal.core.edu.au/conf-ranks/}}). When multiple surveys satisfied all criteria within the same domain, we selected the most recent one to reflect the latest developments in that area.

The final selection comprises \textbf{11 surveys} spanning the breadth of recommender systems research and covering both foundational and emerging domains. The complete list of selected surveys and their associated domains is reported in \Cref{tab:surveys}. For full transparency and reproducibility, we publicly release the code used for survey retrieval and filtering\footnote{\url{https://github.com/albertomancino/RecSysDataProcessingSurvey}}.

\section{Selected Papers}
\label{sec:selected_papers}
\begin{figure*}[t]
    \centering
    \includegraphics[width=\textwidth]{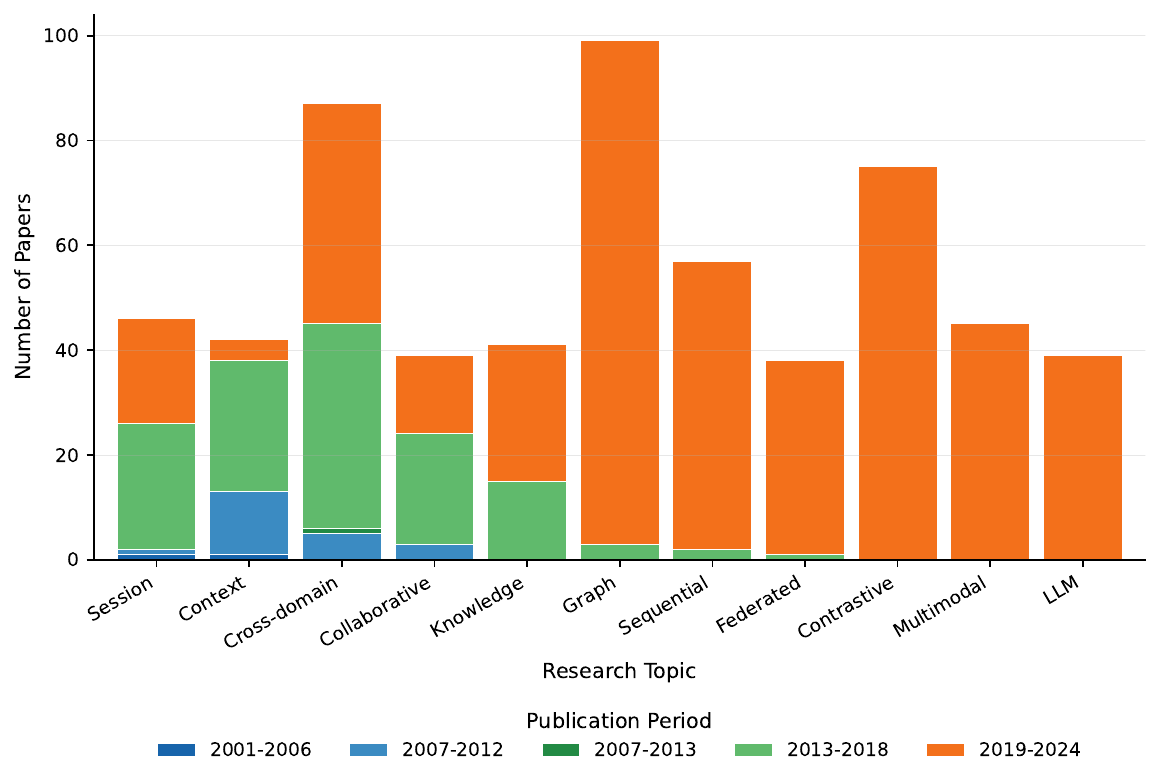}
    \caption{Distribution of papers across research topics and publication periods. Topics are ordered by temporal dominance, i.e., according to the period in which each topic becomes most prominent, from the earliest to the most recent interval.}
    \label{fig:paper_year}
\end{figure*}

For each survey listed in~\Cref{tab:surveys}, we retained papers that satisfied the following criteria: (i) they proposed a novel recommender system method or empirical study, (ii) they were relevant to the survey’s domain, (iii) they focused on recommendation tasks, (iv) they adopted an offline evaluation protocol, and (v) they were accessible to the authors. The latter criterion was necessary because a small number of papers could not be retrieved due to paywall restrictions.
This process resulted in a corpus of \textbf{517 unique papers} published between 2001 and 2024. \Cref{fig:paper_year} reports their temporal distribution and the contribution of each recommendation domain. For analyses stratified by domain, papers appearing in multiple surveys are counted once per domain.

\begin{figure}[t]
    \centering
    \includegraphics[width=\linewidth]{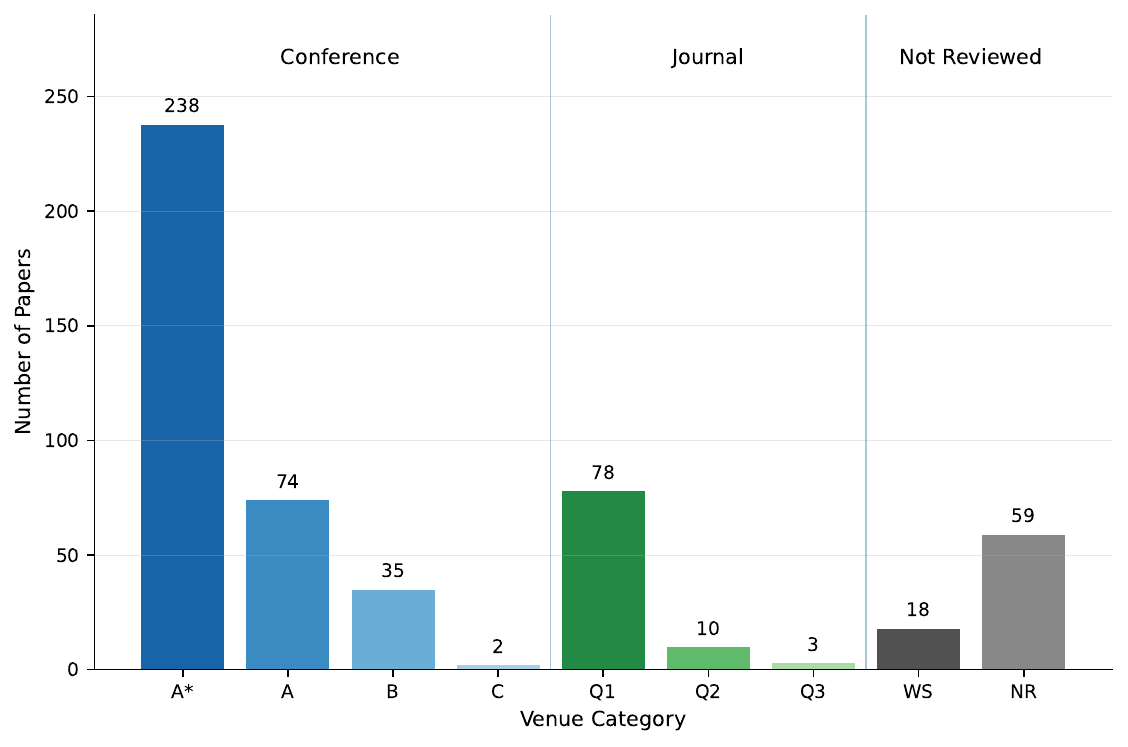}
    \caption{Distribution of papers across venue ranking categories.}
    \label{fig:papers_by_rank}
\end{figure}

To further characterise the quality and impact of the surveyed literature, we recorded conference rankings according to the CORE Ranking\footnote{\url{https://portal.core.edu.au/conf-ranks/}} and journal quartile rankings according to the Scimago Journal Rank (SJR)\footnote{\url{https://www.scimagojr.com/}}. For each venue, we considered the ranking corresponding to the year of publication. For multidisciplinary journals, we adopted the highest ranking within the Computer Science and Information Systems categories.
\Cref{fig:papers_by_rank} summarizes the distribution of papers across conference rankings (A*, A, B, C) and journal quartiles (Q1, Q2, Q3). The surveyed corpus is predominantly composed of papers published in top-ranked venues, with approximately 75\% of the literature appearing in A*/A conferences or Q1 journals. The figure also highlights the predominance of conference publications over journal publications, reflecting the publication culture of the recommender systems community.

Finally, we report workshop papers (WS) and papers grouped under the NR category, which includes non-ranked venues and non-peer-reviewed publications. The latter are particularly common in the LLM-based recommendation domain, reflecting the rapid evolution of this research area and the widespread dissemination of results through preprint platforms.

\section{Terminology and Conceptual Scope}
\label{sec:terminology}

Terminology in recommender systems research is often overloaded, with the same terms carrying different meanings across subfields and application domains. This is particularly evident for data-centric notions such as \emph{data processing}, \emph{preprocessing}, \emph{data preparation}, and \emph{feature extraction}, which are frequently used inconsistently or left implicit. To avoid ambiguity, this section introduces the terminology adopted throughout the survey and clarifies the conceptual boundaries of the topics under consideration.

\paragraph{Data selection.}
We use the term \emph{data selection} to refer to the choice of datasets, data sources, and benchmark collections used for experimentation. In recommender systems research, data selection determines the empirical context in which models are developed and evaluated, and therefore represents one of the earliest decisions in the experimental pipeline.

\paragraph{Data processing.}
We use the term \emph{data processing} as an umbrella concept encompassing all operations applied to data before model training and evaluation. Following common practice in the recommender systems literature, we treat \emph{data processing} and \emph{preprocessing} as interchangeable terms. This notion includes activities such as data cleaning, filtering, normalisation, feature extraction, and dataset splitting.

\paragraph{Data preparation.}
Within the broader data processing workflow, we use the term \emph{data preparation} to refer to the subset of operations that modify the structure, content, or representational form of datasets before model training. In the context of recommender systems, these operations may affect both interaction data and associated side information. A detailed discussion of data preparation practices is provided in~\Cref{sec:chapter4}.

\paragraph{Multimodal feature extraction.}
We use the term \emph{multimodal feature extraction} to denote the transformation of heterogeneous side information, such as text, images, audio, or structured knowledge, into machine-readable representations that can be consumed by recommendation models. In this survey, we focus on the generation, standardisation, sharing, and reuse of such representations as components of the data preparation pipeline. In contrast, model architectures designed to combine or exploit multiple modalities fall outside the scope of this discussion. Multimodal feature extraction techniques are discussed in detail in Section~\ref{subsec:feature_extraction}.

\paragraph{Evaluation paradigms.}
Throughout this survey, the term \emph{evaluation} refers to the experimental setting adopted to assess a recommender system. The literature commonly distinguishes between three evaluation paradigms: \emph{online evaluation}, \emph{user studies}, and \emph{offline evaluation}. Online evaluation assesses the quality of recommendations on a live platform by observing real user responses. User studies involve participants interacting with a recommender system under controlled conditions. Offline evaluation relies on historical interaction data and evaluates recommendation models without direct user involvement or live deployment.
Unless otherwise stated, the term \emph{evaluation} refers to offline evaluation, as this is the paradigm considered throughout the present survey. A more detailed discussion of evaluation paradigms and their implications for recommender systems research is provided in Chapter~\ref{sec:evaluation_paradigms}.

These definitions establish the terminology adopted throughout the remainder of the survey. Subsequent chapters build upon these definitions to analyse how data preparation and feature extraction practices influence the reproducibility, comparability, and validity of offline recommender systems research.

\section{Survey Context and Related Surveys}

The recommender systems literature includes numerous surveys covering both broad and specialised aspects of the field. Existing works can be broadly grouped into four categories: (i) general surveys of recommender systems, (ii) domain-specific studies that discuss data-related aspects, (iii) works focusing on evaluation and reproducibility, and (iv) studies investigating specific preprocessing techniques. While these contributions provide valuable insights, none offer a systematic and cross-domain analysis of data processing practices in recommender systems.

\paragraph{General recommender systems surveys.}
Several surveys aim to provide a comprehensive overview of the recommendation landscape. The \emph{Recommender Systems Handbook}~\citep{DBLP:reference/sp/2022rsh} remains one of the most influential references in the field, covering the fundamental concepts, algorithms, and evaluation methodologies of recommender systems. Similarly, surveys such as those by \citet{DBLP:journals/csur/ZangerleB23} and \citet{DBLP:journals/tors/BauerZS24} provide extensive analyses of evaluation methodologies and experimental protocols. However, their treatment of data-related aspects is largely limited to describing commonly used benchmark datasets, without a systematic discussion of data preparation and preprocessing practices.

\paragraph{Domain-specific studies on data processing.}
A smaller body of work has examined data-related issues more explicitly, although typically within specific recommendation domains. For example, \citet{DBLP:journals/data/ChaudharyC19} analysed the data pipeline from collection to transformation in e-commerce recommendation, while \citet{DBLP:conf/recsys/HidasiC23a} discussed methodological challenges associated with offline evaluation in sequential recommendation. Although these studies highlight the importance of data preparation, their scope remains limited to individual domains and does not provide a unified view across recommender systems research.

\paragraph{Evaluation and reproducibility studies.}
Several works have investigated how data-related decisions affect the reliability of offline evaluation. For instance, \citet{DBLP:journals/tois/JiS0L23} analysed the impact of data leakage on evaluation outcomes, while \citet{DBLP:journals/corr/abs-2211-01261} identified dataset preprocessing as one of the main threats to reliable offline experimentation and introduced \emph{TrainRec} to support reproducible research. Related efforts have also proposed frameworks, benchmarks, and best practices for improving reproducibility~\citep{DBLP:conf/sigir/AnelliBFMMPDN21, DBLP:conf/cikm/ZhaoHPYZLZBTSCX22, DBLP:conf/cikm/Ekstrand20}. These works recognise the importance of data processing but primarily address it from an implementation or evaluation perspective rather than as an object of systematic study.

The relationship between data processing and reproducibility has also been emphasised by broader methodological critiques. \citet{beel20224} argued that the lack of reliable evaluation practices, including inadequate data management and transformation procedures, contributes to a stagnation of progress in recommender systems research. Similarly, \citet{DBLP:reference/sp/JannachQC22} highlighted how inconsistent experimental protocols can create an illusion of progress in session-based recommendation, a phenomenon they describe as \emph{phantom progress}.

\paragraph{Task-specific preprocessing studies.}
Other studies have focused on preprocessing techniques designed to address specific objectives, such as debiasing~\citep{DBLP:journals/jiis/CarraroB22}, denoising and rating correction~\citep{DBLP:journals/jips/ToledoMG13}, fairness-aware preprocessing~\citep{DBLP:journals/tist/ZhaoWLCAD25}, and the mitigation of popularity bias~\citep{DBLP:journals/umuai/KlimashevskaiaJET24}. While highly relevant within their respective contexts, these approaches are tailored to particular modelling goals and therefore fall outside the general-purpose perspective adopted in this survey.

Despite these valuable contributions, no existing work provides a systematic, cross-domain characterisation of data processing practices in recommender systems. Most prior studies either focus on specific recommendation paradigms, discuss isolated aspects of the pipeline, or address data-related issues primarily through the lens of evaluation and reproducibility.

To the best of our knowledge, this is the first survey to provide a systematic, cross-domain characterisation of data processing practices in recommender systems. In addition to traditional preprocessing operations, our analysis explicitly considers multimodal feature extraction pipelines, which have become increasingly central to modern recommendation systems yet remain largely absent from previous surveys. By combining evidence from multiple recommendation domains, we provide a unified perspective that highlights both shared principles and domain-specific differences in data preparation practices. The growing number of workshops, tutorials, and reproducibility initiatives dedicated to data quality and preprocessing further underscores the relevance and timeliness of this topic for the recommender systems community.
\chapter{The Recommendation Pipeline}
\label{sec:pipeline}
This chapter introduces the experimental pipeline for evaluating offline recommender systems, with particular emphasis on the role of data throughout the process. In contemporary recommender systems research, models are rarely evaluated directly in production environments. Instead, they are typically trained and assessed using historical interaction data, making offline evaluation the dominant experimental paradigm. As a result, understanding how recommendation datasets are constructed, transformed, and partitioned is essential for interpreting experimental findings and ensuring reproducible research.

Rather than focusing on specific recommendation algorithms, we examine the sequence of steps through which raw platform data is transformed into datasets suitable for model training and evaluation. Starting from recorded user–item interactions, we discuss how different assumptions about user behaviour give rise to distinct representations of the same underlying data. We then introduce multimodal side information as an additional source of knowledge that can enrich user and item representations beyond those provided by interaction logs alone. Finally, we present the main stages of dataset preparation and splitting, highlighting the design choices that define the experimental conditions under which recommendation models are developed and evaluated.

The chapter begins by briefly discussing the main evaluation paradigms adopted in recommender systems and motivating the central role of offline evaluation in contemporary research. Building upon this perspective, we introduce the notion of interaction data, multimodal side information, dataset preparation, and dataset splitting, which together form the core components of the offline recommendation pipeline considered throughout this survey.

Overall, the goal of this chapter is to provide a conceptual overview of the data lifecycle underlying the evaluation of offline recommender systems. The concepts introduced here establish the foundation for the detailed analyses of data representation, preprocessing, multimodal enrichment, and evaluation protocols presented in the following chapters.

\section{Evaluation Paradigms in Recommender Systems}
\label{sec:evaluation_paradigms}
Recommender systems can be evaluated through different experimental paradigms, each providing a distinct perspective on recommendation quality and user behaviour. The literature commonly distinguishes between three main paradigms: online evaluation, user studies, and offline evaluation. Since the focus of this survey is on recommendation datasets and their processing procedures, it is important to clarify the role of these paradigms and motivate the centrality of offline evaluation in contemporary recommender systems research.

\paragraph{Online evaluation}
refers to experiments conducted within a live platform, typically through controlled A/B testing, where different recommendation strategies are exposed to subsets of real users~\citep{DBLP:journals/jss/QuinWGS24, DBLP:conf/www/OngTL24,DBLP:conf/www/BrennanCYLPMHPH25}. Because it directly measures user responses and downstream objectives such as engagement, conversion, or revenue, online evaluation is generally regarded as the most realistic assessment of recommendation quality. However, it requires access to a production environment, a sufficiently large user base, and substantial engineering and organisational effort. Moreover, experimental conditions are difficult to reproduce across platforms, making systematic comparison of recommendation algorithms particularly challenging. As a result, online evaluation remains largely inaccessible to the broader academic community.

\paragraph{User studies}
provide an alternative form of empirical assessment by placing users in controlled experimental settings. In this paradigm, participants interact with a recommender system or assess its outputs under predefined conditions, enabling the collection of qualitative and perceptual measures such as satisfaction, trust, transparency, or perceived usefulness~\citep{DBLP:conf/recsys/BurkeKE24, DBLP:journals/corr/abs-2404-19093,DBLP:journals/umuai/KnijnenburgWGSN12}. Although user studies offer a high degree of experimental control and can investigate aspects of the user experience that are not directly observable from behavioural logs, they are typically limited in scale and duration. Small sample sizes, laboratory effects, and participant bias may reduce their external validity, making them less suitable for large-scale benchmarking and comparative evaluation.

\paragraph{Offline evaluation}
relies on historical interaction data collected by a platform. In this setting, recommendation models are trained and evaluated by simulating user behaviour on logged interactions, typically by withholding a portion of the observed data for validation and testing. The underlying assumption is that historical interactions provide a meaningful proxy for user preferences, and that a model's ability to predict or rank withheld interactions correlates with its recommendation quality. Although this abstraction neglects important aspects of real-world deployment, such as exposure bias, feedback loops, and the influence of recommendations on future user behaviour, it enables controlled, repeatable, and cost-effective experimentation.

Despite its limitations, offline evaluation has become the dominant paradigm in recommender systems research. Its popularity stems from several practical advantages. First, it enables reproducibility, as experiments can be independently replicated using shared datasets and evaluation protocols. Second, it facilitates comparability, allowing different recommendation approaches to be assessed under identical conditions and according to common metrics. Finally, it supports large-scale experimentation with relatively limited infrastructure requirements, making it particularly well-suited to academic research.

Among the three paradigms, offline evaluation is therefore the one most closely connected to the study of recommendation datasets. Since all offline experiments ultimately rely on historical interaction data and their associated representations, the construction, processing, and partitioning of such datasets become fundamental components of the experimental pipeline. For this reason, the remainder of this chapter focuses on the data abstractions and transformations that underpin the evaluation of offline recommender systems.

\section{Interactions and Data for Offline Evaluation}
\label{sec:interactions_data}

Offline evaluation is fundamentally grounded in historical interaction data, which constitute the primary empirical evidence available to recommender systems. Regardless of the application domain or the recommendation model being considered, offline experiments rely on recorded user behaviour to infer preferences, train models, and assess performance.

We refer to an \emph{interaction} as any recorded event that establishes an observable relation between a user and an item within a recommendation system. Ratings, clicks, purchases, reviews, and other user actions are all examples of interactions captured by a platform. Formally, an interaction can be represented as a tuple

\[
r = (u,i,\tau),
\]

where $u$ denotes a user, $i$ an item, and $\tau$ represents the set of attributes associated with the interaction event. Typical examples of such attributes include the action performed by the user (e.g., rating, click, purchase), the timestamp of the event, and contextual information recorded at the time of the interaction. Depending on the application domain and the available data, some of these attributes may be absent, implicit, or aggregated. A recommendation dataset can therefore be viewed as a collection of observed interactions

\[
\mathcal{D} = \{r_1,r_2,\ldots,r_n\},
\]

where each interaction corresponds to a user--item event recorded by the platform during a given observation period. In offline evaluation, this interaction log serves as the primary source of information for developing and assessing recommendation models.

Interactions may encode different types of preference signals depending on the action component contained in $\tau$. A widely adopted distinction is between \emph{explicit} and \emph{implicit} signals. Explicit signals arise when users directly express an opinion about an item through actions such as ratings, likes, or reviews. In these cases, the interaction itself contains an explicit indication of user preference. Implicit signals, by contrast, are inferred from observed behaviours such as clicks, views, purchases, or add-to-cart actions. While these interactions are abundant and naturally collected by most platforms, they provide only indirect evidence of preference. Moreover, the absence of an interaction does not necessarily indicate negative preference, but rather a lack of observable evidence. The same interaction log can be interpreted under different structural assumptions depending on the recommendation task under consideration. For example, interactions may be treated as independent observations, organised into temporal sequences, or grouped into sessions that capture short-term user intent. Although these representations originate from the same underlying data, they induce different modelling and evaluation settings. We revisit these perspectives in greater detail in~\Cref{sec:data_taxonomy}.

The interaction log, therefore, represents the core component of any recommendation dataset. In practice, however, recommendation systems often exploit additional information beyond observed interactions. The next section introduces these complementary sources of information.

\section{Augmenting Recommendation Data}
\label{sec:augmenting_data}

The interaction dataset introduced in the previous section provides the minimum information required to train and evaluate recommendation models. However, user-item interactions alone often offer only a partial view of user preferences and item characteristics. Moreover, recommendation datasets are typically sparse, recording only a small fraction of the interactions that could potentially occur \citep{DBLP:conf/aaai/HeM16, DBLP:conf/icdm/KangFWM17}. For these reasons, many recommendation systems exploit additional sources of information describing users, items, or their relationships. These complementary data sources, commonly referred to as \emph{side information} \citep{DBLP:conf/recsys/FangS11, DBLP:conf/recsys/VasileSC16, DBLP:conf/www/LiuWTYHL19}, enrich the representation of the dataset and provide signals that are not directly observable from interaction logs alone.

Formally, let $\mathcal{U}$ and $\mathcal{I}$ denote the sets of users and items introduced previously, and let $\mathcal{E} = \mathcal{U} \cup \mathcal{I}$ be the set of entities in the system. Side information can be represented as a collection of modality-specific representation functions

\begin{equation}
\phi_m : \mathcal{E} \rightarrow \mathcal{Z},
\label{eq:modality-source-function}
\end{equation}

where $\mathcal{Z}$ denotes the representation space associated with modality $m$. These functions provide additional descriptions of users or items beyond the interaction log itself. Examples include demographic information associated with users, textual descriptions and images associated with items, audio tracks, videos, or external knowledge sources describing semantic relations between entities.

Due to the heterogeneous nature of these information sources, the recommendation literature commonly refers to them as \emph{multimodal side information} \citep{DBLP:journals/tors/MalitestaCPMNS25, DBLP:journals/csur/LiuHXZGWLT25, DBLP:journals/corr/abs-2302-04473, DBLP:journals/corr/abs-2502-15711, DBLP:conf/kdd/LiuZYDD0ZZD24, DBLP:journals/inffus/PanPCC26}. Following this terminology, we define the set of admissible modalities as

\[
\mathcal{M} = \{img, txt, aud, vid, kg, \dots\},
\]

where each modality corresponds to a specific type of entity description. For example, $img$ and $txt$ may represent visual and textual item content, while $kg$ denotes information derived from knowledge graphs encoding semantic relationships among entities. In practice, each modality is associated with a dedicated representation function $\phi_m$ that transforms raw information into a machine-readable representation. The construction and use of such representations will be discussed in later chapters.

At this point, it is important to distinguish multimodal side information from the attributes that belong to interaction events. As discussed in the previous section, interactions may include attributes such as timestamps, action types, ratings, or contextual conditions that are recorded as part of the interaction itself. These elements, therefore, belong to the interaction log. By contrast, side information exists independently of any specific interaction and provides descriptive information about the entities involved in the system.

A useful example highlighting this distinction concerns user reviews \citep{DBLP:conf/cikm/AnelliDNSFMP22, DBLP:conf/www/0012OM21}. Although reviews are often treated as textual content associated with items, they are more accurately interpreted as attributes of the interaction itself, since they are generated as part of a specific user-item event. In this sense, reviews belong to the interaction log rather than to the set of entity descriptions. By contrast, information such as a product description, an image, or an external knowledge graph exists independently of any particular interaction and can therefore be naturally considered side information.

Taken together, the interaction log $\mathcal{D}$ and the set of multimodal side information sources define the informational environment available to a recommendation model. We denote this augmented representation as

\[
\mathcal{D}_{\mathcal{M}} = \mathcal{D} \cup \mathcal{Z},
\]

which extends the traditional interaction-based view of recommendation data by incorporating heterogeneous descriptions of users and items.

\section{Dataset Preparation}

Before a recommendation dataset can be used for model training and evaluation, it typically undergoes a series of transformations aimed at improving data quality, adapting the data to a specific recommendation task, and integrating auxiliary information. Collectively, these procedures constitute the \emph{dataset preparation} stage of the recommendation pipeline.

Formally, given an augmented dataset $\mathcal D_{\mathcal M}$, dataset preparation can be viewed as a transformation

\[
T : \mathcal D_{\mathcal M}
\rightarrow
\widetilde{\mathcal D}_{\mathcal M},
\]

where $\widetilde{\mathcal D}_{\mathcal M}$ denotes the processed dataset used for subsequent training and evaluation.

Broadly speaking, dataset preparation can be divided into two complementary components: (i) transformations applied to the interaction log and (ii) transformations applied to multimodal side information. While the former operate directly on the observed user-item interactions, the latter focus on processing the auxiliary information associated with users and items. The following discussion provides a high-level overview of these operations, which will be analysed in greater detail later in the survey (\Cref{sec:chapter4}).

Interaction-log transformations can be further categorized into \emph{dataset-level} and \emph{representation-dependent} operations. Dataset-level transformations include all preprocessing procedures that can be applied independently of the recommendation task or the adopted data representation. Typical examples include filtering operations, such as $k$-core pruning, which removes users and items with fewer than $k$ interactions, as well as interaction-value transformations such as rating normalisation, binarisation of explicit signals, or filtering based on temporal constraints and action types.

Representation-dependent transformations, by contrast, depend on the structural interpretation adopted for the interaction log. As discussed previously, the same interaction data may be viewed as independent observations, temporal sequences, sessions, or other task-specific structures. Consequently, interactions may be reorganised into user-specific sequences, grouped into sessions, represented as graphs, or enriched with contextual information according to the requirements of the recommendation task and the underlying learning paradigm.

Beyond interaction logs, multimodal side information also requires dedicated preparation procedures before it can be incorporated into recommendation models. Although the specific processing steps vary across modalities, they generally involve two stages. First, raw information is cleaned, standardized, and, when necessary, reduced. Second, the processed content is transformed into machine-readable representations that can be exploited by recommendation models. These representations are typically obtained through modality-specific feature extractors that map textual, visual, auditory, or relational information into a common representational space.

Taken together, these preparation procedures transform raw interaction logs and auxiliary information into a structured dataset suitable for recommendation tasks. Once the dataset has been prepared, the next step consists of partitioning it into training, validation, and test sets to enable offline experimentation.
\section{Dataset Splitting}

Once a recommendation dataset has been prepared, it must be partitioned into training, validation, and test sets to enable offline experimentation. Dataset splitting is a fundamental component of the evaluation protocol, as it determines which interactions are available during model training and which are reserved for performance assessment. Consequently, the adopted splitting strategy directly influences the statistical properties of the resulting data partitions and may substantially affect the observed performance of recommendation models.

The literature proposes a variety of splitting strategies, which can be broadly categorized into three groups: (i) \textit{random} splitting, (ii) \textit{time-aware} splitting, and (iii) \textit{benchmark-specific} splitting protocols.

In the \textit{random} setting, interactions are partitioned independently of their temporal order. Splitting may be performed globally over the entire interaction log or separately for each user. In a global random split, all interactions are randomly assigned to the training, validation, and test sets according to predefined proportions (e.g., 80\%, 10\%, and 10\%). In a user-level random split, the same procedure is applied independently to each user's interaction history. Random splitting is simple and widely adopted, but it may introduce information leakage by allowing future interactions to appear in the training set while earlier interactions are assigned to the test set.

In the \textit{time-aware} setting, interactions are partitioned according to their chronological order. As in the random case, splitting can be performed globally or at the user level. Interactions are sorted according to their timestamps and divided using temporal boundaries, ensuring that training interactions always precede validation and test interactions. By preserving the temporal structure of user behaviour, time-aware splitting provides a more realistic approximation of real-world deployment and is therefore commonly adopted in sequential and next-item recommendation tasks.

Finally, several datasets and benchmarking frameworks employ \textit{benchmark-specific} splitting protocols. These protocols may reproduce historically adopted splits, follow dataset-specific conventions, or combine multiple splitting strategies. Although such protocols often facilitate direct comparison with prior work, they may also introduce hidden assumptions that are not immediately apparent from the dataset itself.

Dataset splitting influences not only the composition of the training, validation, and test sets, but also the evaluation setting in which recommendation models are assessed. We discuss this aspect in the next section. The splitting approaches introduced here will be analysed in greater detail later in the survey (\Cref{ch:chapter5}), where we examine their formal definitions, practical implementations, and implications for experimental reproducibility and comparability.

\section{Transductive and Inductive Evaluation}

Beyond determining which interactions belong to the training, validation, and test sets, dataset splitting also defines the assumptions under which recommendation models are evaluated. In particular, the adopted evaluation protocol determines whether the recommendation task is framed as \emph{transductive} or \emph{inductive}.

In a transductive setting, models are trained and evaluated on a fixed set of users and items. Consequently, every user and item appearing during validation or testing must also be observed during training. By contrast, inductive evaluation assesses a model's ability to generalize to previously unseen users, items, or both. While transductive evaluation remains the dominant setting in the recommendation literature, inductive scenarios are increasingly relevant in real-world applications characterized by continuously evolving users and item catalogues.

The availability of multimodal side information can partially bridge the gap between these two settings. Even when historical interactions are unavailable, textual, visual, auditory, or knowledge-based descriptions may provide informative representations for unseen entities, enabling recommendation models to generate predictions beyond the observed interaction graph \citep{DBLP:conf/mm/WeiWN0HC19, DBLP:conf/mm/Zhang00WWW21, DBLP:conf/aaai/GuoL0WSR24}. Having introduced the main components of the offline recommendation pipeline, we now conclude this chapter by briefly formalizing the recommendation task learned from the resulting dataset.

\section{The Recommendation Task}

The interaction data, multimodal side information, and evaluation assumptions introduced in the previous sections define the informational environment available to a recommender system. Given the processed and partitioned recommendation dataset
$\widetilde{\mathcal D}_{\mathcal M}$, the objective of a recommender system is to learn a scoring function

\[
\rho : \widetilde{\mathcal D}_{\mathcal{M}} \rightarrow \mathbb{R},
\]

which exploits the dataset's information to estimate the relevance of potential user-item interactions. Depending on the recommendation task and the adopted modelling assumptions, the scoring function may rely solely on interaction data or incorporate additional multimodal information associated with users and items.

Let $\bm{\Theta}_{\rho}$ denote the set of parameters defining the recommendation model. In a learning-based setting, the recommendation task can be formulated as the estimation of the model parameters from the available data:

\[
\hat{\bm{\Theta}}_{\rho}
=
\argmax_{\bm{\Theta}_{\rho}}
P\!\left(
\bm{\Theta}_{\rho}
\mid
\widetilde{\mathcal D}_{\mathcal{M}}
\right).
\]

The specific form of the scoring function, the model architecture, and the optimisation procedure depend on the recommendation approach under consideration. Since the focus of this survey is on recommendation datasets and their processing procedures rather than on recommendation algorithms themselves, we do not discuss model-specific methodological aspects further. The following chapters analyse the recommendation dataset in greater detail, examining its possible representations, the preprocessing operations applied during dataset preparation, and the splitting strategies used to support offline experimentation.
\chapter{Recommendation Datasets}
\label{ch:datasets}

In the context of offline evaluation for recommender systems, a dataset is not merely a collection of interactions; rather, it constitutes the empirical foundation upon which recommendation models are trained, evaluated, and compared. Its structure, scale, temporal characteristics, available side information, and underlying assumptions define the experimental environment and directly influence the validity, comparability, and generalizability of reported results. Consequently, recommendation experiments remain inherently bounded by the properties of the selected datasets, even when multiple domains or multimodal sources of information are considered.

Over the past two decades, the recommender systems community has developed a rich ecosystem of benchmark datasets spanning a wide range of application domains, interaction structures, and information modalities. These datasets differ substantially in terms of their size, density, temporal characteristics, available side information, and intended use cases, reflecting the diversity of recommendation scenarios investigated in the literature. Consequently, understanding the characteristics of recommendation datasets is essential not only for selecting appropriate benchmarks but also for interpreting experimental findings and identifying potential sources of bias.

This chapter presents a systematic overview of recommendation datasets used in offline evaluation. We first discuss the dataset as a structured experimental object and identify the key dimensions through which recommendation datasets can be characterised. We then examine the domains represented by existing benchmarks, introduce a taxonomy of recommendation datasets, review representative multimodal resources, and conclude with an empirical analysis of dataset usage patterns in the recommender systems literature.

\section{The Dataset as a Structured Experimental Object}

In the recommender-systems literature, datasets are commonly described in concrete and operational terms, typically as collections of user actions or historical interaction records generated by real users (e.g., \cite{DBLP:conf/um/GurbanovRP16}; \cite{DBLP:journals/ijhci/Burke12}). The \textit{Recommender Systems Handbook}~\cite{DBLP:reference/sp/2022rsh} similarly characterises datasets as collections of users’ choices or rated items, with the expectation that they approximate real user behaviour as closely as possible. These definitions effectively capture the empirical nature of recommendation data and emphasise the importance of realistic observations for experimental research.

However, within the context of offline evaluation, such descriptions remain primarily focused on what datasets contain rather than on the role they play in the experimental process. Recommendation datasets are not merely repositories of observations. Their structure, available information, and inherent assumptions determine the conditions under which recommendation models are trained, evaluated, and compared. Consequently, the dataset itself becomes an integral component of the experimental setting rather than a neutral input to it.

To better capture this perspective, it is useful to consider broader conceptualisations of datasets developed outside the recommender systems literature. In the data-management field, \cite{DBLP:journals/vldb/ChapmanSKKIKG20} define a dataset as \emph{``a collection of related observations organised and formatted for a particular purpose’’}. This definition introduces a key conceptual element: datasets are inherently purpose-driven artefacts. Their structure is not accidental but reflects the objectives for which they are created and used.
When applied to recommender systems, this notion suggests that recommendation datasets should not be viewed solely as collections of user–item interactions. Rather, they should be understood as artefacts specifically organised to support the training and evaluation of recommendation models. Under this interpretation, the dataset implicitly encodes the assumptions, constraints, and observational scope of the evaluation scenario.

Building on that formalisation, a recommendation dataset can be viewed as a concrete representation of user–item interaction data, along with the information needed to model and evaluate recommendation tasks. In particular, recommender systems introduce a relational structure involving users and items, interactions as the primary observational units, and potentially additional information associated with these entities. These characteristics motivate the following domain-specific definition.

\begin{tcolorbox}[colback=gray!5!white,
colframe=gray!75!black,
title=Definition 1: Recommendation Dataset,
fonttitle=\bfseries]

A \textbf{recommendation dataset} is a structured collection of users, items, and user–item interactions, optionally enriched with interaction attributes and side information describing users, items, or interactions, organised to support the training and evaluation of recommendation models.

\end{tcolorbox}

This definition emphasises that recommendation datasets are not merely collections of records but structured experimental artefacts designed to support a specific evaluation objective. The formal structure underlying recommendation data was introduced in~\Cref{sec:interactions_data}, where interactions and their associated attributes were defined as the fundamental observational units of recommender systems. Building on that formalisation, a recommendation dataset can be viewed as an empirical realisation of a user–item interaction process, together with the information required to represent and study it within an offline evaluation framework.

\subsection{Beyond Interaction Data: Side Information}

Interactions constitute the core observational units of recommendation datasets. However, the information available in modern recommendation benchmarks often extends beyond interaction records alone. Many datasets include additional sources of information describing users or items, thereby providing complementary knowledge that cannot be directly inferred from user–item interactions.

The motivation for incorporating such information stems from the intrinsic limitations of interaction data. User–item interactions are typically sparse and only partially informative: they indicate that an interaction occurred, but they do not fully explain the factors that influenced user behaviour. For example, a user may repeatedly interact with a movie because of its genre, cast, visual style, popularity, or personal circumstances at the time of consumption. While the interaction log records the observed behaviour, many of the factors underlying that behaviour remain unobserved.

To mitigate these limitations, recommendation datasets are often enriched with additional information describing users or items. Such information can provide complementary perspectives on the entities involved in the recommendation process and support richer representations than those obtainable from interaction data alone. Depending on the application domain, these additional sources may include textual descriptions, images, audio signals, demographic attributes, relational knowledge structures, or user-generated content.

The distinction between interaction data and additional information is not always clear-cut. In some cases, the same data source may play different roles depending on how it is represented and used within the recommendation pipeline. User-generated reviews provide a notable example. When considered as explicit feedback associated with a user–item interaction, a review forms part of the interaction record itself. Conversely, when its textual content is processed to extract semantic representations, the review can be treated as an additional source of information describing users, items, or both. This observation suggests that the notion of side information depends not only on the origin of the data but also on its functional role within the dataset representation.

\begin{tcolorbox}[colback=gray!5!white,
colframe=gray!75!black,
title=Definition 2: Side Information,
fonttitle=\bfseries]

\textbf{Side information} refers to any information associated with users or items that complements interaction data by providing additional descriptive, relational, or semantic knowledge about the entities involved in the recommendation process.

\end{tcolorbox}

Side information can therefore originate from heterogeneous data sources and can be associated with different entities within the dataset. Depending on the entity being described, the information can be broadly categorised as user-side or item-side. In practice, such information is typically derived from raw data that must be processed before being incorporated into recommendation models. The extraction of representations from heterogeneous sources such as text, images, audio, or knowledge graphs, therefore, constitutes an important stage of the recommendation pipeline, as discussed in~\Cref{sec:augmenting_data}.

\section{Recommendation Domains and Task Formulations}

Recommendation datasets originate from specific application domains, such as music streaming services, movie platforms, e-commerce websites, social networks, or location-based systems. A recommendation domain can be understood as the application context from which users, items, interactions, and auxiliary information are collected. As a consequence, domains largely determine the types of entities involved, the semantics of interactions, and the modalities of information available within a dataset. For example, music datasets typically contain listening events and artist-related information, movie datasets often include ratings and content descriptors, while e-commerce datasets may provide purchases, product metadata, images, and textual descriptions.

While domains determine \emph{what} information is available, they do not determine \emph{how} that information is used within an experimental setting. The interpretation of interaction data depends on the recommendation task being studied and on the assumptions adopted by the recommendation model. Consequently, the same dataset may support multiple recommendation tasks, each emphasising different aspects of the underlying interaction data.

As discussed in~\Cref{sec:pipeline}, recommendation tasks emerge from specific structural interpretations of user–item interactions. The same interaction records can be represented as entries of a user–item matrix in collaborative filtering, as edges in a user–item graph in graph-based recommendation, as temporally ordered interaction sequences in sequential recommendation, or as short-term interaction sessions in session-based recommendation. These representations do not alter the underlying observations; rather, they provide different perspectives on the same data and support different modelling assumptions.

This distinction highlights an important property of recommendation datasets: tasks are not intrinsic to datasets themselves. Instead, they arise from the way interaction data is structured and interpreted within the experimental pipeline. As a consequence, a single dataset may be reused across multiple recommendation paradigms. Well-known benchmarks such as MovieLens~\citep{DBLP:journals/tiis/HarperK16} have been extensively studied in collaborative filtering, graph-based, and sequential settings despite originating from a single application domain.

Although domain and task formulation are conceptually distinct, empirical associations between them can often be observed in the literature. Certain domains naturally lend themselves to particular recommendation settings because of the behavioural patterns they generate and the information they provide. Music recommendation, for instance, is frequently investigated through sequential or session-based approaches due to the inherently temporal nature of listening behaviour, whereas point-of-interest recommendation commonly incorporates contextual information related to time and location. Similarly, domains such as fashion, social media, and short-video platforms often motivate multimodal recommendation approaches that exploit visual and textual information alongside interaction data.

These associations, however, should not be interpreted as strict relationships between domains and tasks. Rather, they reflect common research practices and modelling preferences observed within the recommender systems literature. Understanding this distinction is essential when analysing recommendation datasets, since datasets can be characterised simultaneously by the domains from which they originate and by the task formulations they support.

This distinction motivates the taxonomy introduced in the next section, where recommendation datasets are characterised according to multiple dimensions, including interaction structure, side information, and dataset organisation paradigms.

\section{A Taxonomy of Recommendation Datasets}
\label{sec:data_taxonomy}

Building upon the interaction-based definition introduced in Section~\ref{sec:interactions_data}, recommendation datasets can be analysed along multiple complementary dimensions, depending on how interactions are represented, which information is available, and how data are organised.

The proposed taxonomy comprises three main dimensions. \emph{Structural representations} describe how interaction records are organised and interpreted, for instance, as graphs, sequences, sessions, or context-dependent interactions. \emph{Information sources} describe the types of information available within a dataset, including both interaction data and side information. Finally, \emph{dataset organisation} captures how data are arranged at a global level, including the scope of the domains and the distribution of data across storage or computational units.

\begin{figure*}[t]
\centering
\includegraphics[width=\linewidth]{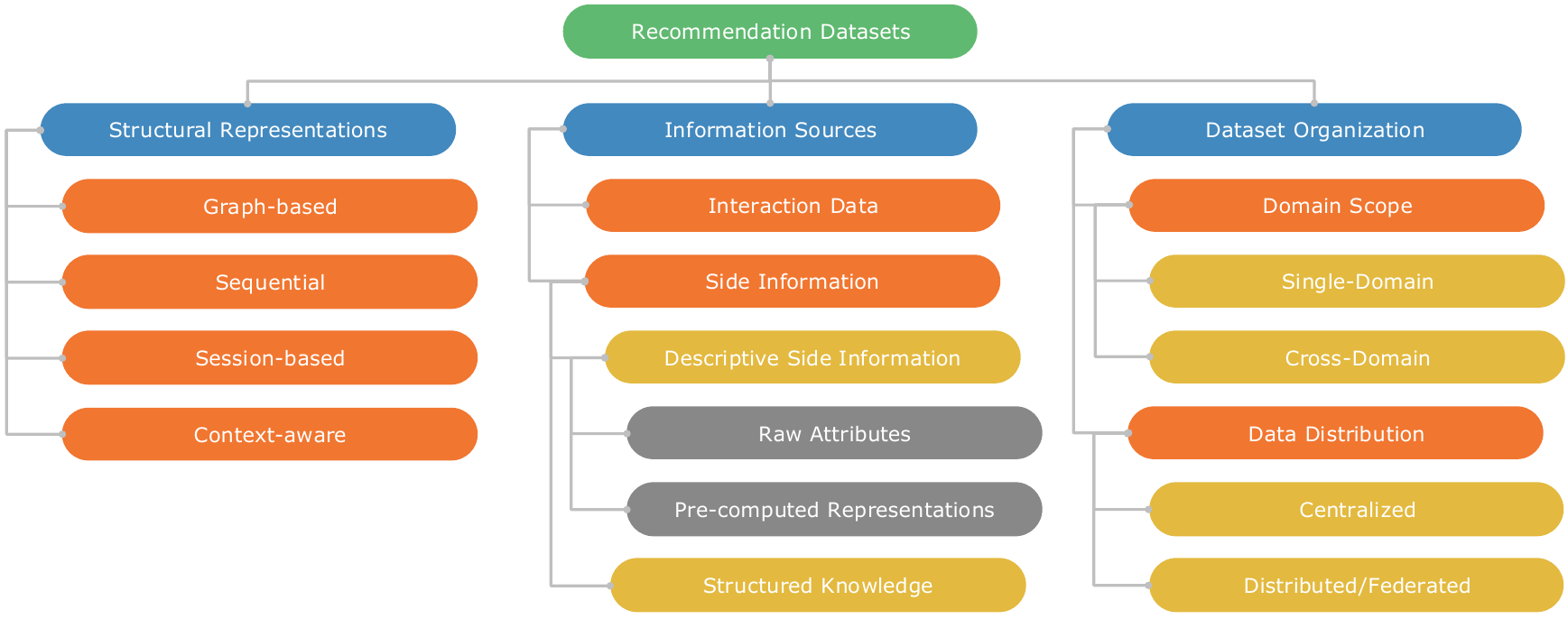}
\caption{Taxonomy of recommendation datasets. Datasets can be characterized along three complementary dimensions: structural representations, information sources, and dataset organization.}
\label{fig:taxonomy_recommendation_datasets}
\end{figure*}

These dimensions are complementary rather than mutually exclusive. A dataset may simultaneously admit a sequential representation, include descriptive side information, and be organised in a cross-domain or federated setting. Figure~\ref{fig:taxonomy_recommendation_datasets} summarises the proposed taxonomy.
The remainder of this section examines each dimension in detail.

\subsection{Structural Representations}

Structural representations provide alternative views of the interaction dataset $\mathcal{D}$ introduced in Section~\ref{sec:interactions_data} by selecting, organising, and emphasising specific components of the interaction tuple $r=(u,i,\tau)$.

\subsubsection{Graph-Based Representation}
\label{sec:graph_repr}
The graph-based view relies on the relational information contained in user–item interactions. Under this interpretation, the dataset is reduced to its core connectivity structure, where each interaction $r=(u,i,\tau)$ contributes only the pair $(u,i)$, while the remaining attributes in $\tau$ are ignored or optionally incorporated as edge features.

This interpretation induces a bipartite graph $G=(\mathcal{E},\mathbf{A})$, where the node set is $\mathcal{E}=\mathcal{U}\cup\mathcal{I}$ and each observed interaction generates an edge connecting a user node to an item node. The connectivity structure of the graph can be represented through the adjacency matrix
\[
\mathbf{A}\in\mathbb{R}^{(|\mathcal{U}|+|\mathcal{I}|)\times(|\mathcal{U}|+|\mathcal{I}|)},
\]

defined as

\[
\mathbf{A} =
\begin{bmatrix}
\mathbf{0} & \mathbf{R} \\
\mathbf{R}^{\top} & \mathbf{0}
\end{bmatrix},
\]

where $\mathbf{R}\in\mathbb{R}^{|\mathcal{U}|\times|\mathcal{I}|}$ denotes the user–item interaction matrix.

By focusing exclusively on relationships between users and items, this representation highlights the structural properties of the interaction network and forms the basis of methods that exploit connectivity patterns. Representative examples include random-walk-based recommenders such as $RP^3_{\beta}$\citep{DBLP:journals/tiis/PaudelCNB17} and graph neural approaches such as NGCF\citep{DBLP:conf/sigir/Wang0WFC19} and LightGCN~\citep{DBLP:conf/sigir/0001DWLZ020}.

\subsubsection{Sequential Representation}
\label{sec:sequential_repr}

Sequential recommendation exploits the temporal ordering of interactions. Consequently, in addition to user and item identities, this view requires information that allows interactions to be arranged chronologically. Such ordering may be explicitly provided through timestamps or implicitly derived from the data collection process.

For each user $u$, the corresponding interaction history is obtained from the subset

\[
\mathcal{D}_u = { (u,i,\tau) \in \mathcal{D} },
\]

containing all events associated with $u$. By ordering these events according to the information encoded in $\tau$, the user’s history can be represented as a sequence

\[
s_u = \langle i_1, i_2, \dots, i_k \rangle.
\]

The dataset is therefore represented as a collection of user sequences

\[
\mathcal{S}^{\text{seq}} = \{ s_u \mid u \in \mathcal{U} \}.
\]

Under this interpretation, users are implicitly characterised by the evolution of their interaction histories rather than by explicit profiles. This enables models to capture temporal dependencies in behaviour, including short-term intent and changes in preferences over time.

Such a representation naturally supports next-item prediction and underlies a broad family of sequential recommendation models, including convolutional architectures such as Caser~\citep{DBLP:conf/wsdm/TangW18} and self-attention-based approaches such as SASRec~\citep{DBLP:conf/icdm/KangM18}.

\subsubsection{Session-Based Representation}
\label{sec:session_repr}

Session-based recommendation focuses on short interaction sequences associated with sessions rather than persistent user identities. In this setting, the interaction attributes $\tau$ must encode session membership, allowing observations to be partitioned into distinct sessions. Consequently, the information required for this representation consists of item identities together with a session identifier, without relying on long-term user profiles.

Let $\mathcal{D}_s \subseteq \mathcal{D}$ denote the set of interactions belonging to a session $s$. Each session can then be represented as an ordered sequence.

\[
s = \langle i_1, i_2, \dots, i_k \rangle,
\]

obtained by arranging the interactions in $\mathcal{D}_s$ according to their occurrence.

The dataset can therefore be viewed as a collection of sessions.

\[
\mathcal{S}^{\text{sess}} = \{ s_1, s_2, \dots, s_n \}.
\]

From this perspective, session-based recommendation can be interpreted as a specialisation of sequential recommendation in which the unit of analysis shifts from users to sessions. Whereas user-level sequential models aim to capture long-term preference dynamics, session-based approaches focus on short-term interaction patterns and are particularly suitable when persistent user identities are unavailable, anonymous, or unreliable.

This formulation naturally supports next-item prediction within a session and forms the basis of a wide range of recommendation models, including recurrent architectures such as GRU4Rec~\citep{DBLP:journals/corr/HidasiKBT15} and graph-based session models such as SR-GNN~\citep{DBLP:conf/aaai/WuT0WXT19}.

\subsubsection{Context-Aware Representation}
\label{sec:context_repr}
Context-aware recommendation explicitly incorporates contextual attributes associated with interaction events. In this setting, the interaction attributes $\tau$ are required to encode variables describing the circumstances under which interactions occur, allowing recommendations to be conditioned on the surrounding context.

Interactions can therefore be represented as tuples $(u,i,c)$, where $c$ denotes contextual variables derived from $\tau$. The dataset can thus be viewed as a collection of context-dependent interactions

\[
\mathcal{D}_c = \{ (u,i,c) \mid (u,i,\tau)\in\mathcal{D} \}.
\]

Typical contextual variables include temporal conditions, spatial location, device type, weather, or other situational factors. By incorporating such signals, recommendation models can capture variations in user preferences across different circumstances and adapt their predictions accordingly.

This representation forms the basis of a broad family of context-aware recommendation methods, including factorisation-based approaches such as Factorisation Machines~\citep{DBLP:conf/icdm/Rendle10} and neural extensions such as Neural Factorisation Machines~\citep{DBLP:conf/sigir/0001C17}.

\subsection{Information Sources}

From the perspective of the proposed taxonomy, recommendation datasets comprise two complementary sources of information: the interaction data and the side information associated with dataset entities. While interaction data constitute the empirical basis of recommendation experiments, side information provides complementary knowledge describing users or items. Together, these two sources determine the information available for model development and evaluation.

\subsubsection{Interaction Data}

Interaction data constitute the primary source of information in recommendation datasets and form the empirical basis of offline recommendation experiments. As introduced in Section~\ref{sec:interactions_data}, each interaction is represented as a tuple $r=(u,i,\tau)$, where $u$ and $i$ identify the interacting user and item. At the same time, $\tau$ contains the attributes associated with that specific interaction event.

These attributes describe properties of the interaction itself, such as its time, context, or other event-specific characteristics, and should therefore be distinguished from side information, which instead provides complementary knowledge about users or items independently of any particular interaction.

\subsubsection{Side Information}
Side information includes any information associated with users or items that is not part of the interaction event itself~\citep{DBLP:journals/ecra/SunGYFGZB19, DBLP:journals/tors/MalitestaCPMNS25}. Its role is to enrich the entities appearing in the interaction log with additional content or knowledge that supports a richer characterisation of the recommendation environment.

From a dataset-centric perspective, side information can be broadly divided into two categories. The first, referred to as \emph{descriptive side information}, provides attributes describing individual users or items. The second, referred to as \emph{structured knowledge}, connects dataset entities through explicit semantic relationships.

\paragraph{Descriptive Side Information.}
Descriptive side information consists of attributes associated with users or items. Typical examples include tabular metadata, profile information, textual descriptions, images, or audio content associated with dataset entities~\citep{DBLP:conf/kdd/LiuZYDD0ZZD24}. Such information may originate from the platform itself, such as product images or item descriptions, or may be linked from external resources. Unlike structured knowledge, it primarily describes individual entities rather than modelling relationships among them.

Although the underlying information remains the same, datasets may expose it at different levels of processing. Some preserve the original content, whereas others provide representations that have already been extracted from that content.

\emph{Raw Attributes.}
The dataset provides directly observable and human-interpretable information, such as metadata fields, textual descriptions, images, or audio files. Feature extraction and representation learning are therefore left to subsequent stages of the recommendation pipeline.

\emph{Pre-computed Representations.}
Alternatively, descriptive information may be released in an already processed form, typically as embeddings, dense feature vectors, or modality-specific descriptors. In this case, the distinction lies not in the nature of the information itself, but in how it is represented: feature extraction has already been performed, allowing these representations to be directly employed in reproducible recommendation experiments without additional preprocessing.

\paragraph{Structured Knowledge.}

Structured knowledge consists of explicit relational information connecting users or items to a broader network of concepts and entities. Unlike descriptive side information, which characterises individual entities through their attributes, structured knowledge introduces semantic relationships among entities.

Typical examples include knowledge graphs, ontologies, or structured taxonomies linking items to categories, brands, creators, semantic concepts, or real-world entities~\citep{DBLP:journals/tkde/GuoZQZXXH22, DBLP:journals/eswa/ShaoLB21, DBLP:conf/i-semantics/NoiaMORZ12}. Through these associations, dataset entities become embedded within a richer semantic structure that extends beyond the interaction log.

The defining characteristic of this category is therefore not the presence of additional attributes, but the explicit representation of relationships. Such knowledge can support recommendation models by providing semantic connections that are not directly observable from user-item interactions alone.

\subsection{Dataset Organisation}

Beyond their structural representation and available information sources, recommendation datasets can also differ in how they are organised at a global level. From a dataset-centric perspective, this organisational dimension can be characterised along two complementary aspects: the \emph{domain scope} covered by the dataset and the \emph{distribution of the data} across storage or computational units. While the notion of domain refers to the application context from which interactions are collected (e.g., movies, music, books, or e-commerce), domain scope concerns whether the dataset covers a single domain or multiple interconnected domains.

These aspects are independent and therefore not mutually exclusive. 
For example, a dataset may be centralized and single-domain, centralized and cross-domain, or distributed in a federated setting while still covering one or multiple domains.

\subsubsection{Domain Scope}

A first organisational aspect concerns the scope of the application domains represented in the dataset. 
In the simplest case, a recommendation dataset is \emph{single-domain}, meaning that interactions are collected within one application domain only, such as movies, music, books, or e-commerce. 
In this setting, users, items, and interactions all belong to the same recommendation environment.

By contrast, \emph{cross-domain} datasets involve multiple domains connected through shared entities or explicit correspondences, enabling information to be transferred across recommendation settings. Such connections may arise from shared users, aligned items, common attributes, or other forms of cross-domain linkage. Consequently, cross-domain datasets are not merely collections of multiple domains, but datasets in which meaningful relationships exist across domains.

This distinction is particularly relevant because single-domain and cross-domain datasets support different forms of transfer, generalisation, and integration across recommendation settings. 
From a dataset perspective, however, the key difference lies simply in whether the recorded interactions belong to one domain or to several interconnected domains.

\subsubsection{Data Distribution}

A second organisational aspect concerns how the dataset is arranged across storage or computational units. 
In the most common case, recommendation datasets are \emph{centralised}, meaning that all interactions and any associated information are collected and stored in a single accessible repository.

In other settings, data may instead be \emph{distributed}, so that different portions of the dataset are held separately across multiple nodes, clients, or devices. 
A particularly important instance of this setting is \emph{federated} organisation, in which interaction data are partitioned across decentralised clients and remain locally stored, while only intermediate model updates or aggregated information are shared during training.

From a dataset-centric perspective, the key distinction is therefore whether the dataset is available as a single centralised collection or as a set of distributed local partitions. 
This organisational property has important consequences for privacy, accessibility, and experimental design, even when the underlying interaction data and side information remain otherwise comparable.

%sec:selected_papers
\section{Recommendation Dataset Usage in the Literature}
The previous sections introduced a formal definition and a structured taxonomy of recommendation datasets. While this conceptual perspective provides a systematic understanding of their characteristics, it does not capture how datasets are actually adopted by the research community. Understanding these practices is equally important, as widely used datasets often become \emph{de facto} benchmarks, influencing experimental protocols, enabling fair model comparisons, and shaping the research questions addressed by subsequent work. Conversely, datasets that receive limited adoption may remain comparatively underexplored despite their potential relevance. Examining dataset usage, therefore, provides valuable insights into how benchmark choices influence the evolution of recommender systems research.

To investigate these aspects, we analyse a large collection of datasets extracted from recommender systems publications. Specifically, for each paper selected in our study, as reported in \Cref{sec:selected_papers}, we annotated every dataset used and recorded whether the authors referenced it through a bibliographic citation, a link to the data source, or both. We then manually traced each dataset to its original source and, when available, its original publication to assess the correctness and completeness of dataset references.

Based on this curated corpus, our analysis examines recommendation dataset usage from four complementary perspectives. First, we investigate dataset popularity by analysing how frequently individual datasets appear in the literature, highlighting the coexistence of a small number of widely adopted benchmarks and a broad long-tailed distribution of rarely used datasets. Second, we examine how datasets are distributed across recommendation domains and task-oriented research areas, identifying both well-established and underexplored settings. Third, we analyse dataset provenance and dissemination, distinguishing contributions from academia, industry, and community-driven initiatives while discussing the challenges associated with each source. Finally, we assess dataset accessibility and referencing practices to evaluate reproducibility and examine the consequences of incomplete, incorrect, or outdated dataset references.

\subsection{Dataset Frequency and Long-Tail Distribution}
We begin our empirical analysis by examining how frequently datasets are used in the recommender systems literature. Here, each occurrence denotes the use of a dataset in a single publication, capturing dataset \emph{usage} rather than merely counting distinct datasets.

To ensure a consistent analysis, we adopt the following conventions. Different versions of the same dataset (e.g., yearly releases or benchmark variants) are treated as distinct entries, as they correspond to different data instances. In contrast, different preprocessing or filtering strategies applied to the same underlying dataset are not distinguished, since they originate from the same data source.
When the specific version of a dataset is not explicitly reported in the original paper, we retain the generic dataset name as provided by the authors. However, in cases where sufficient information is available (e.g., the number of users or items), we infer the most likely dataset version and annotate it accordingly.

Overall, our analysis spans $517$ papers and $1497$ paper--dataset pairs, involving $270$ distinct datasets. On average, each paper evaluates approximately $2.9$ datasets, while each distinct dataset appears in approximately $5.5$ paper--dataset pairs. These figures indicate substantial dataset diversity, while also showing that a subset of datasets is reused across multiple studies.

\begin{figure*}[htbp]
    \centering
    \includegraphics[width=\textwidth]{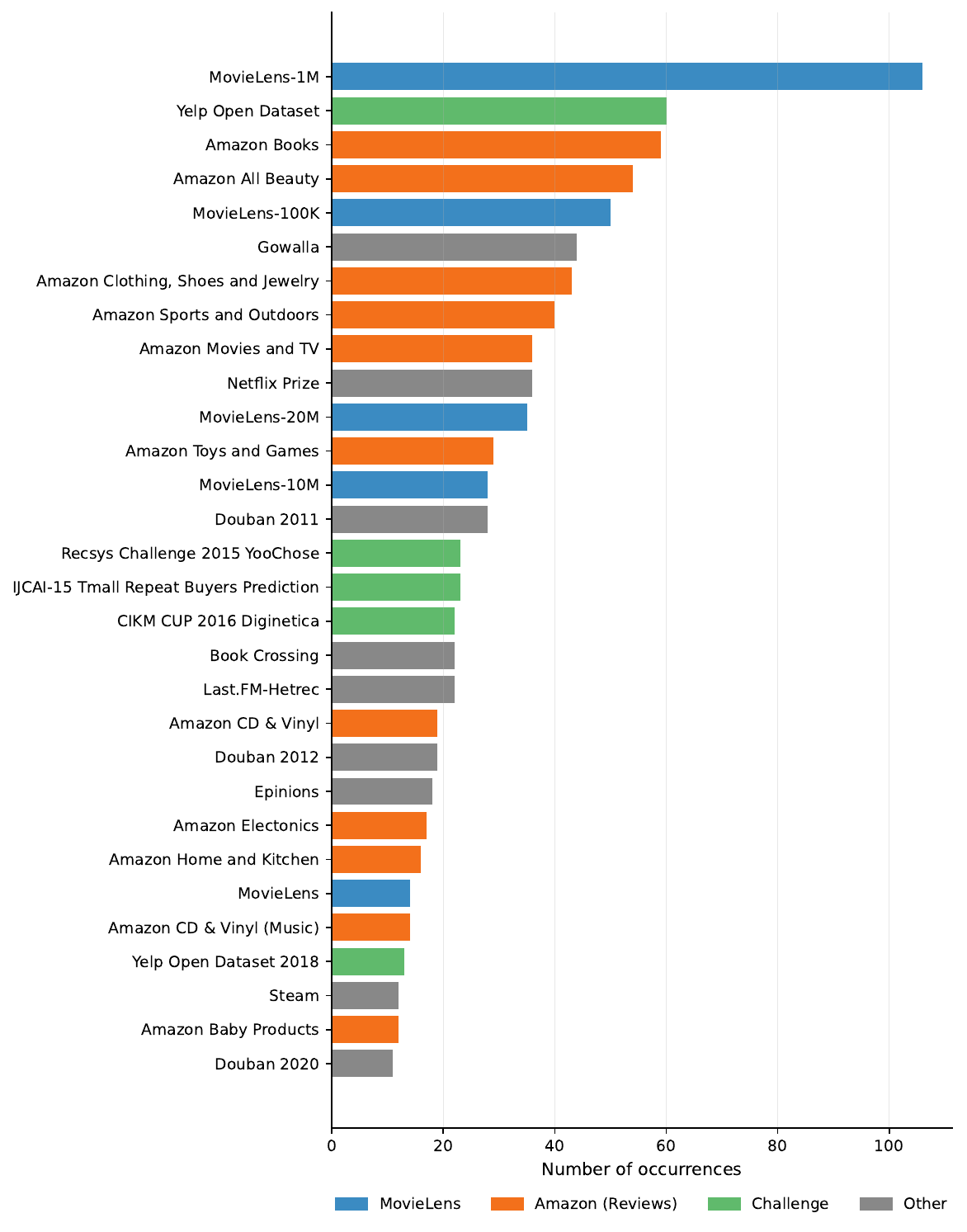}
    \caption{Most frequently used datasets in the analyzed literature. 
The figure shows the top 30 datasets ranked by the number of occurrences across publications, considering only datasets that appear in at least two studies. 
Different versions of the same dataset are treated as distinct entries, while variations due to preprocessing are not distinguished. 
The distribution reveals a strong concentration of empirical evaluation around a small number of widely reused benchmarks, which dominate the experimental landscape.
}
    \label{fig:datasets_top30}
\end{figure*}

Figure~\ref{fig:datasets_top30} reports the frequency of dataset usage across the analysed literature. For readability, we restrict the visualisation to datasets that appear in at least two publications and report only the top 30 most frequently used datasets. This choice focuses on datasets that exhibit a minimum level of reuse while avoiding visual clutter from a large number of rarely used datasets.

Despite this filtering, the distribution remains highly skewed. A small number of datasets dominate the literature, appearing in a large number of publications, while the remaining datasets are used only a few times.
This concentration of empirical evaluation around a limited set of benchmarks has important implications. On the one hand, the presence of widely adopted datasets facilitates comparison across models and supports cumulative progress. On the other hand, the limited diversity of commonly used datasets may bias evaluation toward specific domains, data characteristics, and interaction patterns, potentially limiting the generalizability of experimental findings.

In Figure~\ref{fig:datasets_top30}, datasets are also categorised according to their origin, distinguishing between datasets derived from major benchmark families (e.g., MovieLens~\citep{DBLP:journals/tiis/HarperK16} and Amazon Reviews~\citep{DBLP:conf/www/HeM16}), datasets released in the context of evaluation challenges, and other datasets. The figure shows that a large portion of the most frequently used datasets belong to these categories. In particular, it is worth noting that one of the most widely used datasets (Yelp~\footnote{\url{https://www.yelp.com/dataset/.}}) originates from an industrial challenge. Furthermore, the presence of an entry labelled simply as ``MovieLens'' highlights cases where the dataset version is not explicitly specified in the original paper and cannot be unambiguously inferred. We will further discuss dataset creation, versioning, and traceability in the following sections.

To complement the analysis of the most frequently used datasets, we further investigate the overall distribution of dataset usage in order to highlight its long-tail behaviour.

\begin{figure*}[htp]
    \centering
    \includegraphics[width=\textwidth]{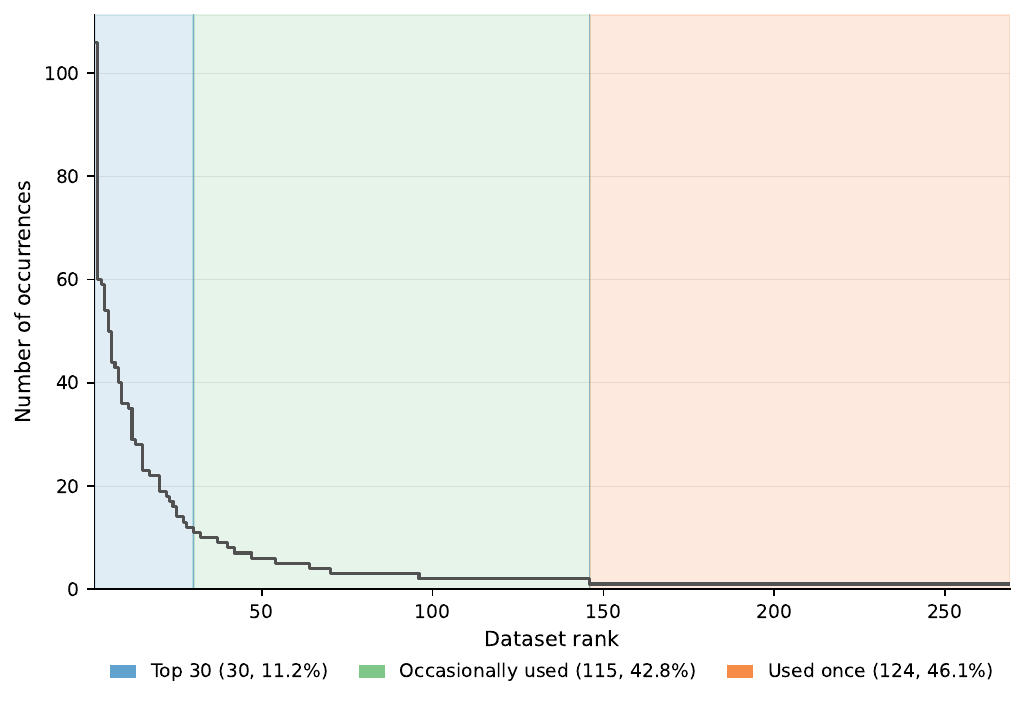}
    \caption{Rank-frequency distribution of dataset usage in the analyzed literature. 
Datasets are ordered by decreasing frequency of occurrence across publications. 
The distribution exhibits a pronounced long-tail behavior: a small fraction of datasets is extensively reused (top 30), while a large portion appears only occasionally or in a single study. 
In particular, 46.1\% of the datasets are used only once, highlighting a substantial degree of fragmentation in experimental practices.
}
    \label{fig:datasets_long_tail}
\end{figure*}

Figure~\ref{fig:datasets_long_tail} reports the rank-frequency distribution considering all identified datasets. The figure clearly shows a highly skewed distribution, where a small number of datasets are repeatedly used across many studies, while a large fraction appears only rarely. In particular, a substantial portion of datasets is used in a single publication, accounting for 46.1\% of the total. This observation reveals a marked fragmentation in dataset usage within the literature.

While the reasons behind the limited reuse of specific datasets cannot be directly inferred from frequency statistics alone, several recurring patterns emerge from a closer inspection of the data. First, some datasets (e.g., \textit{Amadeus CME}, \textit{AlipayMiniApp}, \textit{NAVER}) are proprietary industrial datasets that are not publicly shared, inherently limiting their reuse. Second, other datasets (e.g., \textit{8TRACKS}, \textit{Vine}, \textit{Allegro}) are accessible only upon request, which introduces additional barriers to adoption. Third, in some cases (e.g., \textit{Art of the Music}, \textit{RecSys Challenge 2016}, \textit{Exact Street2Shop}), although a source is reported, the corresponding links are no longer accessible, preventing reproducibility. Finally, certain datasets (e.g., \textit{ASSISTments}, \textit{Telefonica}, \textit{WSRec}) are tied to highly specific domains, such as learning environments, click data in telecommunication services, and web services, which may limit their applicability in broader recommendation scenarios.

Although these factors do not provide an exhaustive explanation of the observed long-tail distribution, they offer concrete evidence of structural barriers affecting dataset reuse. These aspects will be further investigated in the following sections, where we analyse dataset availability, origin, and reproducibility in greater detail.

\subsection{Domain Distribution of Recommendation Datasets}

While the previous subsection examined how frequently individual datasets are reused, analysing their distribution across recommendation domains provides a complementary perspective on the empirical focus of recommender systems research. Although many recommendation algorithms are designed to be largely domain-agnostic, the benchmarks adopted by the community ultimately determine which application scenarios receive the greatest experimental attention.

Since no standardised taxonomy of recommendation domains exists, we manually assigned each dataset to a domain family according to its primary application context. Closely related domains (e.g., e-commerce, marketplace, and retail) were grouped together to reduce fragmentation and facilitate high-level comparisons across the literature.

Table~\ref{tab:domain_usage} summarises dataset usage across recommendation domains. For each domain, we report the number of occurrences, the number of distinct datasets, and the average usage (occurrences per dataset), capturing both domain popularity and the degree of benchmark reuse.

\begin{table}[t]
\centering
\caption{Dataset usage across domains. ``Occurrences'' denotes the total number of dataset usages in the analyzed literature, while ``Distinct'' reports the number of unique datasets within each domain. ``Avg. Usage'' is defined as the ratio between occurrences and distinct datasets, capturing the average reuse of datasets within a domain. Domains are sorted by decreasing number of occurrences. Darker shades indicate higher values of Avg. Usage.
}
\label{tab:domain_usage}
\small
\begin{tabularx}{\textwidth}{lccc}
\toprule
\textbf{Domain} & \textbf{Occurrences} & \textbf{Distinct} & \textbf{Avg. Usage} \\
\midrule
Movies & 361 & 25 & \cellcolor{gray!80}14.44 \\
E-commerce / Marketplace & 263 & 54 & \cellcolor{gray!40}4.87 \\
Fashion & 138 & 14 & \cellcolor{gray!60}9.86 \\
POI / Travel / Tourism & 125 & 31 & \cellcolor{gray!40}4.03 \\
Music & 119 & 24 & \cellcolor{gray!40}4.96 \\
Multi Domain & 110 & 7 & \cellcolor{gray!80}15.71 \\
Books & 101 & 12 & \cellcolor{gray!60}8.42 \\
Social / Friendship & 62 & 19 & \cellcolor{gray!40}3.26 \\
Games & 60 & 6 & \cellcolor{gray!80}10.00 \\
Video / Short Video & 45 & 20 & \cellcolor{gray!40}2.25 \\
News & 27 & 13 & \cellcolor{gray!40}2.08 \\
Food & 20 & 9 & \cellcolor{gray!40}2.22 \\
Other & 15 & 8 & \cellcolor{gray!20}1.88 \\
Scholar & 8 & 4 & \cellcolor{gray!40}2.00 \\
Advertising / Marketing & 7 & 6 & \cellcolor{gray!20}1.17 \\
Financial / Insurance & 5 & 5 & \cellcolor{gray!20}1.00 \\
Apps / Mobile & 3 & 3 & \cellcolor{gray!20}1.00 \\
Conversational & 2 & 2 & \cellcolor{gray!20}1.00 \\
Courses / Learning & 2 & 2 & \cellcolor{gray!20}1.00 \\
\bottomrule
\end{tabularx}
\end{table}

The results reveal a markedly uneven distribution across domains. The \emph{Movies} domain dominates the literature in terms of occurrences while also exhibiting one of the highest average usage values. Despite comprising 25 distinct datasets, empirical evaluation is concentrated around a small number of benchmarks, most notably the MovieLens family (e.g., MovieLens-1M and MovieLens-100K) and Amazon Movies and TV. This finding extends the observations of the previous subsection, showing that the concentration of benchmark usage is particularly pronounced within a limited number of application domains.

The \emph{E-commerce / Marketplace} domain represents a contrasting pattern. It ranks second in terms of occurrences while exhibiting the largest number of distinct datasets. This reflects both the widespread adoption of established benchmarks (e.g., YooChoose\footnote{\url{https://www.kaggle.com/datasets/chadgostopp/recsys-challenge-2015}} and Tmall\footnote{\url{https://tianchi.aliyun.com/dataset/dataDetail?dataId=42}}) and the availability of numerous domain-specific subsets derived from Amazon Reviews and industrial platforms such as Taobao.

A different behaviour is observed for the \emph{Multi-domain} category, where a relatively small number of datasets accounts for a disproportionately high level of reuse. Although limited in number, these datasets have become standard benchmarks for specific research problems, particularly cross-domain recommendation. Representative examples include the Douban datasets, which are widely adopted in this research area.

Overall, three distinct patterns emerge. Some domains, including Movies, Fashion, and Games, attract considerable research attention but rely on a relatively small number of widely reused benchmarks. Others, such as E-commerce, Point of Interest, and Music, combine high research interest with greater dataset diversity. Finally, several application domains remain represented by only a handful of datasets, indicating that they have received comparatively limited attention from the research community. The \emph{Other} category groups domains represented by a single dataset.

These findings suggest that empirical evaluation is concentrated not only around a small set of benchmark datasets, but also around a limited number of application domains. As a result, the evolution of recommender systems research has been shaped primarily by a relatively narrow set of evaluation scenarios, while many application contexts remain comparatively underexplored.

In the next subsection, we move beyond application domains and investigate how dataset usage is distributed across different research areas, providing a complementary perspective on benchmark adoption in the recommender systems literature.

\subsection{Benchmark Adoption Across Research Topics}
\begin{figure*}[!ht]
    \centering
    \includegraphics[width=\textwidth]{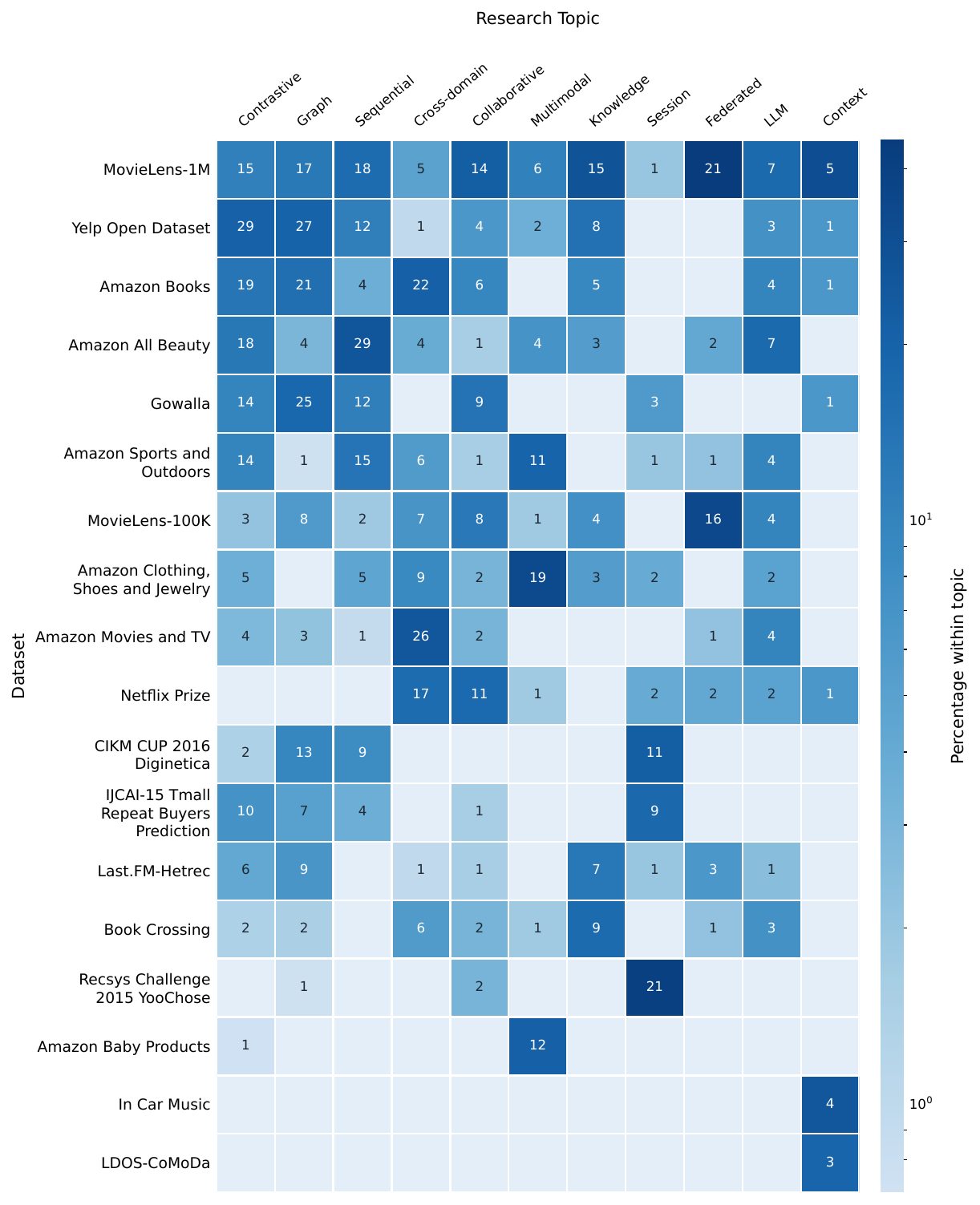}
    \caption{Benchmark adoption across recommender systems research topics. Rows correspond to datasets (ordered by overall popularity) and columns to research topics. Cell colors represent the percentage of occurrences of each dataset within a topic, while numbers denote absolute occurrence counts. The visualization illustrates both the broad reuse of general-purpose benchmarks and the emergence of task-specific benchmark ecosystems.}
    \label{fig:datasets_topic}
\end{figure*}
While recommendation domains describe the application context of a dataset, research topics capture the methodological perspective from which datasets are employed. To investigate how benchmark adoption varies across research topics, \Cref{fig:datasets_topic} reports a heatmap of dataset usage. Each cell shows the percentage of occurrences of a dataset within a given research topic, while the annotated values indicate the corresponding number of occurrences. By normalising occurrences within each topic, the visualisation enables meaningful comparisons across research areas with different publication volumes. Datasets are ordered according to their overall popularity.
The most evident pattern is that benchmark adoption is highly topic-dependent. While some datasets are consistently reused across multiple research areas, others are strongly associated with specific methodological paradigms. For instance, the \textit{Yelp Open Dataset} is extensively used in graph-based, contrastive, and sequential recommendation, whereas it appears only rarely in cross-domain, session-based, and federated recommendation.
Federated recommendation provides a particularly interesting example. Most studies rely on MovieLens datasets, despite the fact that these benchmarks were not originally designed for federated evaluation. This suggests that, in the absence of dedicated datasets, researchers tend to adapt well-established benchmarks rather than introduce new evaluation collections. Conversely, when task-specific datasets are available, they quickly become the preferred choice. For example, \textit{YooChoose} and \textit{Diginetica} dominate the session-based recommendation literature.
A similar phenomenon can be observed in several specialised research areas, where niche datasets become closely associated with a particular task. For instance, \textit{InCarMusic}\citep{DBLP:conf/ecweb/BaltrunasKLMRALS11} and \textit{LDOS-CoMoDa}\citep{kovsir2011database} are used almost exclusively in context-aware recommendation. More generally, benchmark selection is also influenced by the type of auxiliary information required by different recommendation paradigms, an aspect that we examine in greater detail in the following sections.
At the same time, the heatmap highlights the existence of broadly reusable benchmarks. Most notably, \textit{MovieLens-1M} appears consistently across a wide range of research topics, including collaborative, sequential, federated, and knowledge-aware recommendation. This confirms its role as a general-purpose benchmark that has remained relevant despite substantial differences between recommendation paradigms. Amazon datasets also exhibit relatively broad adoption, although with greater variability across topics.
Interestingly, no single benchmark clearly dominates the emerging area of LLM-based recommendation. Instead, several datasets are used with comparable frequency, suggesting that the community has not yet converged on a shared evaluation standard. This behaviour contrasts with more mature research areas, where benchmark selection has progressively consolidated around a small number of widely adopted datasets.
Overall, the analysis demonstrates that benchmark adoption is shaped by two complementary factors: the overall popularity of a dataset and the methodological requirements of individual research topics. Mature research areas tend to converge toward a small set of widely reused benchmarks, whereas specialised or emerging paradigms either rely on dedicated datasets or adapt existing benchmarks to new evaluation settings. These findings further emphasise the importance of dataset availability in shaping experimental practice and motivate the analysis of dataset provenance and accessibility presented in the next subsection.

\subsection{Dataset Provenance, Accessibility, and Reproducibility}

The analyses presented in the previous sections revealed that benchmark adoption in recommender systems is highly uneven. While a small number of datasets are repeatedly reused across research topics, many others appear only sporadically or remain confined to specific application domains. Although methodological requirements partly explain these trends, they do not provide the full picture. The way datasets are created, disseminated, documented, and maintained also plays a fundamental role in determining whether they become widely adopted benchmarks or remain isolated resources.

These aspects directly affect reproducibility. Modern recommender systems are evaluated through complex experimental pipelines involving data collection, preprocessing, model training, and evaluation. Within this pipeline, the dataset constitutes the empirical foundation upon which every subsequent step depends. Consequently, its provenance, accessibility, and correct identification are essential prerequisites for reproducible and comparable research.

Building upon the curated collection of 270 datasets assembled in this survey, we therefore investigate three complementary aspects that directly influence dataset reuse. We first analyse the provenance of recommendation datasets, examining the different actors involved in their creation and dissemination. We then study dataset accessibility, evaluating how frequently datasets remain publicly available over time. Finally, we investigate dataset referencing practices and discuss a set of recommendations aimed at improving reproducibility in future research.

\subsubsection{Dataset Provenance}

\begin{table}[t]
\centering
\caption{Distribution of datasets by origin, distinguishing between academic, industrial, and challenge-based sources.}
\label{tab:dataset_origin}
\begin{tabular}{lcc}
\toprule
\textbf{Origin} & \textbf{Count} & \textbf{Percentage} \\
\midrule
Academic & 156 & 59.5\% \\
Industry & 72 & 27.5\% \\
\midrule
Academic Challenge & 23 & 8.8\% \\
Industry Challenge & 11 & 4.2\% \\
\bottomrule
\end{tabular}
\end{table}

Recommendation datasets originate from a heterogeneous ecosystem involving academic institutions, industrial organisations, challenge initiatives, and community-driven projects. Understanding these different sources is important because they influence not only the characteristics of the released data but also their availability, maintenance, and long-term reuse.

In this survey, we classify datasets into four categories according to the primary actor responsible for their creation and dissemination. Academic datasets are proposed and released by universities or research institutions. Industrial datasets originate from companies or industrial collaborations in which the industry plays the leading role. Challenge datasets are released in the context of competitive evaluation campaigns, either academic or industrial. Finally, community datasets are created independently by practitioners or enthusiasts outside formal research initiatives.

Table~\ref{tab:dataset_origin} summarises the resulting distribution. Academic datasets account for the majority of the resources considered in our survey (59.5\%), followed by industrial datasets (27.5\%). Challenge datasets represent a smaller but still significant portion, comprising 8.8\% of academic challenges and 4.2\% of industrial challenges.

\subsubsection{Dataset Accessibility}

\begin{table}[t]
\centering
\caption{Dataset availability and link quality.}
\label{tab:dataset_links}
\begin{tabular}{llcc}
\toprule
\textbf{Group} & \textbf{Category} & \textbf{Count} & \textbf{Percentage} \\
\midrule
\multirow{4}{*}{Dataset availability}
 & Link available & 186 & 70.2\% \\
 & Not shared & 30 & 11.3\% \\
 & Proprietary & 47 & 17.7\% \\
 & Unknown & 2 & 0.8\% \\
\midrule
\multirow{5}{*}{Link quality}
 & Working & 154 & 82.8\% \\
 & Broken link & 23 & 12.4\% \\
 & Closed (challenge) & 4 & 2.2\% \\
 & After request & 2 & 1.1\% \\
 & Redistribution & 3 & 1.6\% \\
\bottomrule
\end{tabular}
\end{table}

Beyond their provenance, the long-term accessibility of recommendation datasets plays a crucial role in enabling reproducible research. A dataset that cannot be retrieved, or whose source is no longer available, cannot be reliably reused to reproduce published experiments, regardless of the quality of the accompanying methodology.

To assess this aspect, we verified the accessibility of every dataset in our curated collection and reported the results in \Cref{tab:dataset_links}. For each one, we identified its original source whenever possible and recorded whether it remained publicly accessible at the time of our analysis.

Our results show that only 70.2\% of datasets are currently associated with a publicly accessible source, whereas the remaining 29.8\% cannot be directly retrieved. Among these inaccessible datasets, 17.7\% correspond to proprietary industrial resources that were never intended for public release, while 11.3\% are datasets that, despite not being inherently proprietary, are no longer shared with the research community. The remaining 0.8\% could not be conclusively classified.

Accessibility is not only determined by whether a dataset is public, but also by whether its original source remains available over time. Among datasets with an identified source, 82.8\% provide working links, whereas 12.4\% point to webpages or repositories that are no longer accessible. The remaining cases correspond to datasets available only upon request (1.1\%), links to closed evaluation challenges (2.2\%), or alternative, non-official mirrors (1.6\%).

These findings suggest that dataset accessibility is inherently dynamic. Even datasets that were originally released as public resources may gradually become unavailable because repositories disappear, websites are discontinued, or challenge infrastructures are decommissioned. Consequently, reproducibility should not be regarded as a static property of a dataset, but as one that depends on its long-term preservation and maintenance.

\subsubsection{Dataset Referencing}

Dataset accessibility alone is insufficient to guarantee reproducibility. Even when data remain publicly available, experiments can only be reproduced if the adopted dataset can be unambiguously identified. This depends on how datasets are referenced within scientific publications.

For each dataset occurrence in our survey, we therefore annotated whether the corresponding publication explicitly reported (i) a reference paper describing the dataset and (ii) a direct link to its source.

Our analysis shows that 78\% of datasets are associated with a reference publication. Such references provide a stable mechanism for identifying datasets, understanding their construction, and distinguishing between different releases or variants.

However, referencing practices remain heterogeneous. In many cases, datasets are cited only through their name, without a bibliographic reference or an official source. In others, references point to secondary publications rather than to the original dataset paper, making it more difficult to trace the dataset's provenance and identify the exact version employed.

These issues become particularly relevant for benchmark families that have evolved through multiple releases. Collections such as MovieLens or Amazon Reviews comprise several versions with different characteristics, yet publications do not always specify which release was used. Similar ambiguities also arise when datasets undergo custom preprocessing while retaining their original name. Such practices complicate the comparison of experimental results across studies, even when researchers nominally rely on the same benchmark.

\subsubsection{Lessons Learned for Reproducible Dataset Usage}

The analyses presented in this section highlight that reproducibility in recommender systems depends not only on the availability of algorithms and evaluation protocols, but also on the quality of dataset documentation and dissemination. Although most studies rely on publicly known benchmarks, our findings reveal that ambiguities in dataset identification, inaccessible sources, broken links, and inconsistent referencing remain common throughout the literature.

Based on these observations, we identify four simple practices that can substantially improve the transparency and reproducibility of dataset-based experiments.

First, datasets should always be identified using their official name, avoiding abbreviations or alternative names that may introduce ambiguity. This is particularly important for benchmark families that comprise multiple datasets or variants.

Second, whenever available, the original publication introducing the dataset should be explicitly cited. Beyond acknowledging the creators, the reference paper provides the most reliable description of the dataset construction process, its characteristics, and its intended use, facilitating both interpretation and reproducibility.

Third, authors should provide a direct link to the official dataset source rather than relying solely on bibliographic references or secondary repositories. Whenever possible, links should point to stable repositories maintained by the dataset creators, reducing the risk of future ambiguity and simplifying dataset retrieval.

Finally, when multiple releases or versions of a dataset exist, the exact version adopted in the experiments should always be specified. Differences between releases may affect the number of users, items, interactions, or available side information, making version identification essential for meaningful comparisons across studies.

Taken together, these practices require little additional effort from authors, yet they considerably improve the traceability and long-term reproducibility of experimental research. More broadly, our analysis suggests that reproducibility should not be viewed solely as a property of recommendation algorithms, but also as a consequence of how datasets are documented, maintained, and shared within the research community.

\subsection{Auxiliary Information Sources in Recommendation Datasets}
\label{sec:auxiliary_information}

The previous sections focused on recommendation datasets as collections of user-item interactions. However, modern recommendation models increasingly rely on auxiliary information that complements the interaction log with additional signals describing users, items, or their surrounding semantic environment. Such information has become a key component of contemporary recommendation research, particularly in multimodal and knowledge-aware recommendation.

To complement the dataset-centric perspective developed throughout this chapter, we analyse the auxiliary information sources provided by representative benchmark datasets adopted in these research areas. Following the methodology adopted throughout this survey, the analysed benchmarks were identified from the primary studies included in the selected multimodal and knowledge-aware recommendation surveys. Although not exhaustive, this collection provides a representative snapshot of the auxiliary information currently available in recommendation benchmarks.

Table~\ref{tab:auxiliary_sources} summarises the analysed datasets together with the auxiliary information they provide. In addition to identifying the available modalities, the table highlights how each source of auxiliary information is made available to researchers. In particular, we distinguish between raw data included in the original dataset release, pre-computed feature representations, official links to external resources, and external enrichments introduced after the original dataset release.

A first observation concerns the diversity of auxiliary information available across recommendation benchmarks. Across the analysed datasets, we identify six recurring categories of auxiliary information:

\begin{itemize}
    \item \textbf{Textual information}, including titles, descriptions, tags, categories, and other textual metadata associated with users or, more commonly, items;

    \item \textbf{Visual information}, typically consisting of product images, photographs, movie posters, or video keyframes;

    \item \textbf{Audio information}, derived from original audio content or from audio tracks associated with multimedia items;

    \item \textbf{Video information}, available in datasets containing or linking multimedia content;

    \item \textbf{Reviews}, i.e., user-generated textual content that, although originating from interaction data, is frequently exploited to derive user and item representations in review-aware recommendation models;

    \item \textbf{External knowledge graphs}, which connect dataset entities to semantic resources such as Freebase, DBpedia, or Microsoft Satori.
\end{itemize}

A second observation concerns the entities described by auxiliary information. Most auxiliary sources are associated with items rather than users. Images, textual descriptions, categorical metadata, and semantic annotations naturally characterise items and are therefore directly released with many recommendation benchmarks. In contrast, user-side information is generally limited to basic profile attributes, such as gender, age group, or other demographic characteristics, when available. Rich user descriptions are rarely released explicitly due to privacy constraints. Instead, they are commonly constructed by recommendation models from the user's interaction history, for example, by aggregating review text or the multimodal representations of previously consumed items. Consequently, while auxiliary information is predominantly item-centric at the dataset level, user representations are often derived indirectly from item-side information during model training.

A third observation concerns how auxiliary information is released. While textual information is almost always distributed as raw content, such as titles, descriptions, or metadata, reflecting both its central role and the relative ease with which textual resources can be shared, visual, audio, and semantic information follow more heterogeneous release strategies.

Knowledge graphs provide a particularly interesting example. Rather than being distributed with the original dataset release, they are often introduced in subsequent studies that align benchmark entities with external semantic resources such as Freebase, DBpedia, or Microsoft Satori. This practice illustrates that recommendation datasets are not static artefacts but evolve over time through community-driven enrichments. The MovieLens family represents perhaps the clearest example of this evolution, having progressively expanded from a benchmark dataset of user-item interactions into an ecosystem enriched with movie posters, trailers, and multiple knowledge graph alignments that support different recommendation paradigms.

A different trend emerges for datasets originating from multimedia platforms. Instead of distributing raw video or audio content, benchmarks such as TikTok (ICME 2019) provide pre-computed visual and audio representations extracted from the original videos. Sharing feature representations rather than raw multimedia content substantially reduces storage requirements and simplifies dataset dissemination while providing standardised inputs for downstream models. At the same time, this choice constrains the representation space available to researchers, since the extracted features cannot be recomputed using alternative encoders or preprocessing pipelines.

These release strategies also reveal that auxiliary information is often derived from other available sources rather than acquired independently. For example, visual features may be extracted from videos through representative keyframes, while textual, visual, and audio embeddings are commonly generated from raw content using pre-trained models. Consequently, the same modality may be distributed either as original data or as a processed representation, depending on the design choices adopted by the dataset creators or by subsequent benchmark extensions.

Overall, the analysed benchmarks show that recommendation datasets should no longer be viewed as static collections of interactions complemented by auxiliary information. Instead, they increasingly evolve as data ecosystems, where new modalities, semantic resources, and feature representations are progressively integrated by both dataset creators and the research community. This evolution mirrors the growing importance of content-aware, multimodal, and knowledge-enhanced recommendation models, for which auxiliary information has become an integral component of both model design and experimental evaluation.
\begin{table*}[hbtp]
\centering
\caption{Representative benchmark datasets and their auxiliary information. Symbols denote how each modality is released: \raw~raw data; \feat~pre-computed representations; \link~official external resources; \enrich~external enrichments introduced by subsequent studies. Multiple symbols indicate multiple release strategies. Datasets marked with * aggregate closely related versions.}
\label{tab:auxiliary_sources}
\small
\begin{tabular}{lcccccc}
\toprule
\textbf{Dataset} &
\textbf{Text} &
\textbf{Image} &
\textbf{Audio} &
\textbf{Video} &
\textbf{Reviews} &
\textbf{KG} \\
\midrule

Alibaba iFashion      & \cmark        & \cmark          & --            & --      & --      & -- \\
Allrecipes.com        & \cmark        & \cmark          & --            & --      & --      & -- \\
Amadeus CME           & \feat         & --              & --            & --      & --      & \cmark \\
Amazon Reviews*       & \cmark\feat   & \link\feat      & --            & --      & \cmark  & \enrich \\
Bing News             & \cmark        & --              & --            & --      & --      & \cmark \\
BookCrossing          & \cmark        & \cmark          & --            & --      & --      & \enrich \\
ChinaMM 2018 Kwai     & \cmark        & \cmark          & --            & --      & --      & -- \\
CNRec                 & \cmark        & --              & --            & --      & --      & \enrich \\
Dianping              & --            & \cmark          & --            & --      & \cmark  & \cmark \\
Douban 2012           & --            & --              & --            & --      & --      & \enrich \\
ESWC 2014 DBbook      & \cmark        & --              & --            & --      & --      & \enrich \\
Exact Street2Shop     & \cmark        & \link           & --            & --      & --      & -- \\
Fashion-136K          & \cmark        & \link           & --            & --      & --      & -- \\
ICME 2019 TikTok      & \feat         & \feat           & \feat         & --      & --      & -- \\
IntentBooks           & \cmark        & \cmark          & --            & --      & --      & \cmark \\
Kwai                  & \cmark        & \cmark          & --            & --      & --      & -- \\
Last.FM*              & \cmark        & --              & --            & --      & --      & \enrich \\
Meituan               & \cmark        & \cmark          & --            & --      & --      & -- \\
Microsoft MMRec       & \cmark        & \cmark          & --            & --      & --      & -- \\
MovieLens*            & \cmark        & \link\external  & \enrich       & \link   & --      & \enrich \\
Netflix               & \cmark        & \enrich         & --            & --      & --      & -- \\
RecSys 2013 Yelp      & \cmark        & \cmark          & --            & --      & --      & \enrich \\
Taobao                & \cmark        & \cmark          & --            & --      & --      & -- \\
Toffee*               & \cmark        & \enrich         & \enrich       & \cmark  & --      & -- \\
V-MIND                & \cmark        & \cmark          & --            & --      & --      & -- \\
What-to-Wear          & \cmark        & \cmark          & --            & --      & --      & -- \\
Yelp Open Dataset*    & \cmark        & \cmark          & --            & --      & \cmark  & \enrich \\

\bottomrule
\end{tabular}
\end{table*}

Conceptualising datasets as measurement instruments clarifies the central methodological tension addressed in this survey: 
\chapter{Data Preparation}
\label{sec:chapter4}

Recommendation datasets, in their raw form, consist of interaction logs and associated side information collected from real-world systems. Before these data can be used for model training and offline evaluation, they undergo a sequence of transformations that modify their statistical properties, determine the information available to recommendation models, and ultimately define the effective benchmark on which algorithms are evaluated. Despite this central role, data preparation is often treated as a secondary implementation detail, with preprocessing choices reported only in part or inherited from previous work, making it difficult to separate the contribution of the recommendation model from that of the underlying data-processing pipeline.

In this chapter, we collectively refer to these transformations as \emph{data preparation}. Consistent with the terminology introduced in Section~\ref{sec:terminology}, data preparation comprises all transformations applied to the interaction log and its associated side information before dataset splitting. Within the broader data processing pipeline, it corresponds to the stage that modifies the dataset's structure, content, or representation, whereas the construction of training, validation, and test partitions is addressed separately in \Cref{ch:chapter5}.

Rather than presenting these transformations as an isolated collection of techniques, this chapter develops a unified view of data preparation. We first introduce a formal view in which each preparation step is modelled as an operator acting on either the interaction log or the associated side information. This formalisation naturally leads to a taxonomy that organises data preparation according to the dataset component being transformed and the criterion driving each operation. The resulting taxonomy is then used to analyse the principal families of interaction-log and side-information transformations, and to examine how these operations are employed in the recommendation literature through a large-scale empirical analysis.

\section{A Formal View of Data Preparation}
\label{sec:formal_view}
The wide variety of data preparation techniques adopted in recommender systems makes it difficult to reason about them as isolated procedures. Operations such as interaction filtering, feedback transformation, sequence truncation, and multimodal content processing differ substantially in purpose and implementation, yet they all share a common objective: transforming the raw dataset prior to model training and evaluation. To provide a unified perspective, here we introduce the formal framework used throughout the remainder of the chapter. We first define the objects on which data preparation operates and then model the corresponding preparation steps as transformation operators. This formalisation provides the foundation for the taxonomy developed in the next section.

\paragraph{Interaction log.}
We adopt the notation introduced in \Cref{sec:pipeline}. An interaction is represented as the tuple $r=(u,i,\tau)$, where $u\in\mathcal{U}$ denotes a user, $i\in\mathcal{I}$ an item, and $\tau\in\mathcal{T}$ denotes the set of attributes associated with the interaction event, including the interaction type, timestamp, and contextual information. The observed interaction log is therefore the finite set

\[
\mathcal{D}=\{r_1,r_2,\ldots,r_n\}
\subseteq
\mathcal{U}\times\mathcal{I}\times\mathcal{T},
\]

which serves as the input to the interaction-log preparation operators introduced below.
\paragraph{Interaction-log preparation operators.}

Each data preparation step is modelled as a parameterised operator

\[
P_\theta:\mathcal{D}\mapsto\tilde{\mathcal{D}},
\qquad
\theta\in\Theta,
\]

where $\theta$ denotes the parameters governing the transformation, examples include the activity threshold in $k$-core filtering, the decision threshold used for feedback binarisation, or the temporal boundaries defining a time window.

In practice, data preparation is implemented as a sequence of such operators. Given a pipeline
$\{P^{(1)}_\theta,\ldots,P^{(K)}_\theta\}$, the resulting interaction log is

\[
\tilde{\mathcal{D}}
=
(P^{(K)}_\theta
\circ
P^{(K-1)}_\theta
\circ
\cdots
\circ
P^{(1)}_\theta)
(\mathcal{D}).
\]

This formulation emphasises that the dataset used in an experiment is not the raw interaction log itself, but the result of an explicit sequence of transformation choices. Different preparation pipelines applied to the same raw interaction log may therefore yield substantially different experimental datasets, even when the underlying data source is identical.

\paragraph{Side-information preparation operators.}

The operator $P$ models transformations acting on the interaction log. Recommendation datasets, however, often include auxiliary side information associated with users or items, such as textual descriptions, images, audio, or knowledge graphs. Let $\Phi=\{\phi_s\}$ denote the collection of such information. Since these data are external to the interaction log, their preparation is modelled separately through the operator

\[
\Psi:\Phi\mapsto Z,
\]

where $Z$ denotes the representations obtained from the original side information.

Unlike interaction-log transformations, which operate on structured interaction tuples, side-information transformations act on heterogeneous raw content. Consequently, the operator $\Psi$ naturally decomposes into two conceptually distinct stages. The first standardises raw modality-specific content into a canonical representation, while the second extracts fixed-dimensional features suitable for recommendation models. These stages are examined later in the chapter through a modality-oriented perspective.

Together, the operators $P$ and $\Psi$ provide a unified formalisation of data preparation by capturing its two complementary dimensions: transformations of the interaction log and transformations of auxiliary side information. The following section organises these transformations into a taxonomy that structures the remainder of the chapter.

%Data Transformations
%│
%├── Interaction Log Transformations
%│   │
%│   ├── Dataset-Level Transformations
%│   │   ├── Support-Driven Transformations
%│   │   └── Attribute-Driven Transformations
%│   │
%│   └── Representation-Dependent Transformations
%│       ├── Sequential-Based Transformations
%│       ├── Session-Based Transformations
%│       ├── Graph-Based Transformations
%│       └── Context-Based Transformations
%│
%└── Side Information Transformations
%    │
%    ├── Multimodal Data Preparation
%    │   ├── Text
%    │   ├── Image
%    │   ├── Audio
%    │   ├── Video
%    │   └── Knowledge Graph
%    │
%    └── Feature Extraction
%        ├── Hand-Crafted Features
%        ├── Frozen Pre-trained Encoders
%        ├── End-to-End Learned Encoders
%        └── Dataset-Shipped Features

\section{Taxonomy of Data Transformations}
\label{sec:taxonomy_transformations}

\begin{figure}[tbhp]
    \centering
    \includegraphics[width=\linewidth]{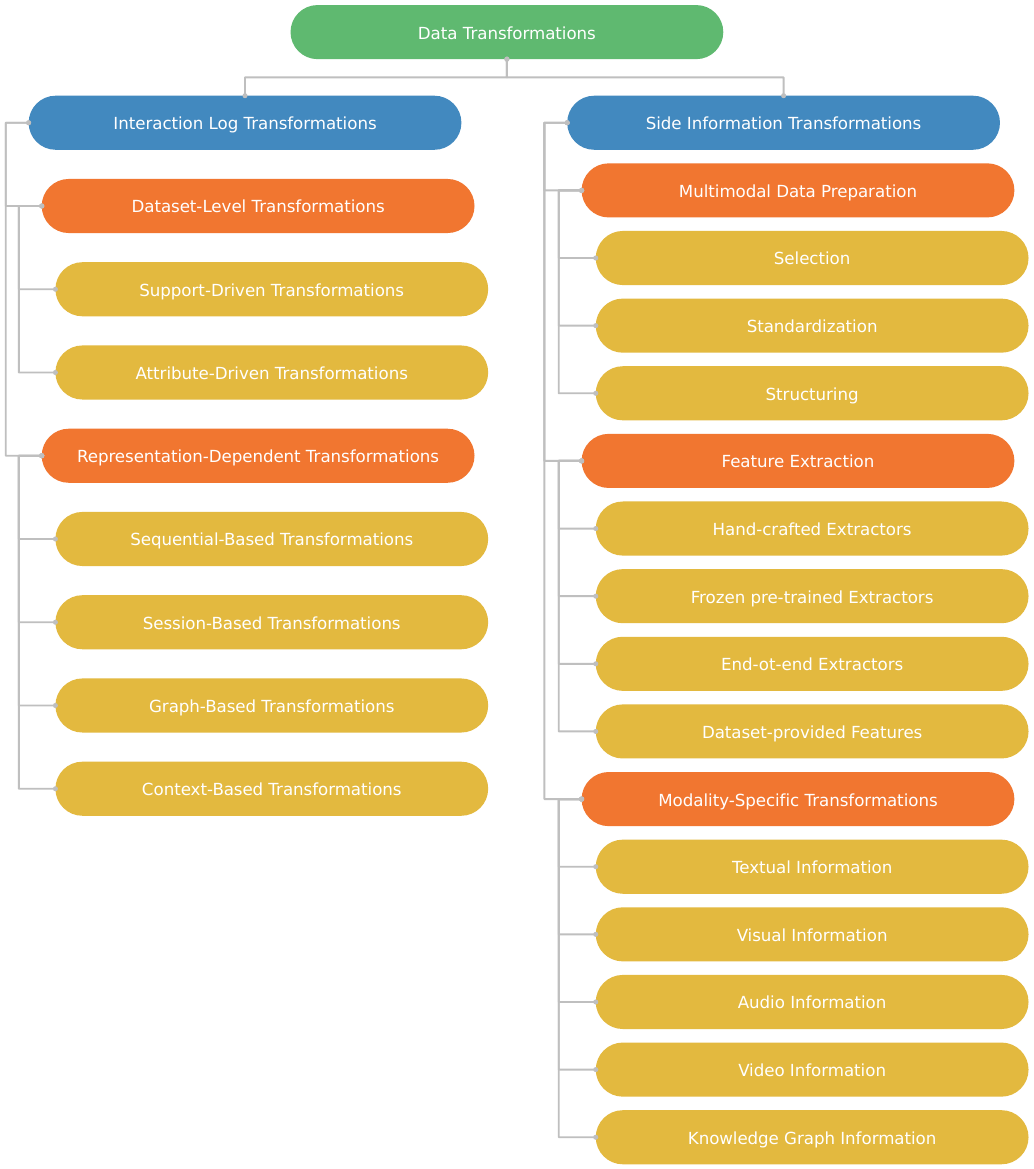}
    \caption{Taxonomy of data preparation transformations.}
    \label{fig:prep-taxonomy}
\end{figure}

The formal framework introduced in the previous section naturally gives rise to the taxonomy shown in Figure~\ref{fig:prep-taxonomy}. The taxonomy organises data preparation transformations according to the object they act upon and the criterion that defines each family of operations.

\paragraph{Interaction-log transformations.}
Interaction-log transformations, represented by the operator $P$, can be further distinguished according to whether they are defined directly on the interaction log or require a specific structural representation of it. Dataset-level transformations operate directly on $\mathcal{D}$ without assuming any structural interpretation beyond the interaction tuples themselves. Representation-dependent transformations, in contrast, become meaningful only after the interaction log has been interpreted under a particular structural representation, such as a sequence, a collection of sessions, a graph, or a context-aware interaction model.

Dataset-level transformations can, in turn, be classified by the criterion that drives them. Support-driven transformations rely on properties of the users and items appearing in the interaction log, such as interaction frequency or entity availability. In some cases, this criterion may also depend on the availability of associated side information (e.g., retaining only interactions involving items with auxiliary content). Although parameterised by side information, these transformations still operate by filtering the interaction log and therefore remain within the scope of the operator $P$. Attribute-driven transformations instead use the values stored in the interaction attributes $\tau$, such as ratings, timestamps, interaction types, or contextual variables.

\paragraph{Side-information transformations.}
Side-information transformations, represented by the operator $\Psi$, follow a different organisation because they operate on heterogeneous content rather than interaction tuples. In the multimodal setting considered in this survey, they naturally decompose into two stages. The first, referred to as \emph{multimodal data preparation}, standardises raw modality-specific content into a canonical representation. The second performs \emph{feature extraction}, transforming the prepared content into feature representations suitable for recommendation models. Feature extraction is further organised into distinct families of extractors, discussed in~\Cref{subsec:feature_extraction}, whereas both stages are analysed separately for each modality.

The remainder of this chapter follows this organisation. We first examine interaction-log transformations, considering dataset-level and representation-dependent operations, before turning to multimodal side-information preparation and feature extraction.

\section{Interaction-Log Transformations}
\label{sec:interaction_log}

Interaction-log transformations comprise all data preparation operations that act directly on the interaction log through the operator $P$. As introduced in the taxonomy, these transformations can be divided into two broad families. Dataset-level transformations are defined directly on the interaction log and therefore apply independently of any particular structural representation of the data. Representation-dependent transformations, in contrast, become meaningful only after the interaction log has been interpreted under a specific representation, such as a sequence, a graph, or a collection of sessions. The following sections examine these two families in turn.

\subsection{Dataset-Level Transformations}
\label{sec:dataset_level_trans}

\begin{table}[t]
\centering
\caption{Overview of dataset-level transformation techniques associated with the corresponding taxonomy categories. Each technique refers to its detailed description in the text.}
\label{tab:dataset-level}

\small
\resizebox{\textwidth}{!}{%
\begin{tabular}{@{}lll@{}}
\toprule
\textbf{Family} &
\textbf{Subfamily} &
\textbf{Technique} \\
\midrule

\multirow{7}{*}{\textbf{Support-driven}}
& \multirow{3}{*}{Activity-based filtering}
& \hyperref[par:user-filtering]{User filtering} \\

&
& \hyperref[par:item-filtering]{Item filtering} \\

&
& \hyperref[par:iterative-kcore]{Iterative $k$-core filtering} \\

\cmidrule(lr){2-3}

&
\multirow{2}{*}{User and item subsampling}
& \hyperref[par:user-item-subsampling]{User subsampling} \\

&
& \hyperref[par:user-item-subsampling]{Item subsampling} \\

\cmidrule(lr){2-3}

&
\multirow{2}{*}{Other}
& \hyperref[par:external-filtering]{External-criterion filtering} \\

&
& \hyperref[par:deduplication]{Interaction deduplication} \\

\cmidrule(lr){1-3}

\multirow{4}{*}{\textbf{Attribute-driven}}
& \multirow{2}{*}{Feedback transformations}
& \hyperref[par:explicit-implicit]{Explicit-to-implicit conversion} \\

&
& \hyperref[par:rating-normalization]{Rating normalization} \\

\cmidrule(lr){2-3}

&
Temporal transformations
& \hyperref[par:time-window]{Time-window selection} \\

\cmidrule(lr){2-3}

&
Interaction-type transformations
& \hyperref[par:action-filtering]{Action filtering} \\

\bottomrule
\end{tabular}
}
\end{table}

Dataset-level transformations operate directly on the interaction log without assuming any structural interpretation beyond the interaction tuples themselves. They modify the composition or interpretation of the observed interactions using criteria that can be evaluated directly from the interaction log. As introduced in the taxonomy, these transformations can be grouped into two categories. Support-driven transformations are based on properties of the users and items appearing in the interaction log, whereas attribute-driven transformations rely on the values associated with each interaction event, such as ratings, timestamps, interaction types, or contextual attributes. Table~\ref{tab:dataset-level} provides an overview of the dataset-level transformation techniques, linking the taxonomy introduced in Section~\ref{sec:taxonomy_transformations} to the specific operations discussed throughout this section.

\subsubsection{Support-Driven Transformations}
\label{sec:support_driven}

Support-driven transformations comprise all dataset-level operations whose selection criterion depends on the entities participating in the interaction log rather than on the values associated with individual interaction events. Their defining characteristic is that they operate on properties of the interaction log's support, such as user or item activity, entity availability, or the presence of associated external resources. Although these criteria are defined over users or items, the transformation's output is always a modified interaction log obtained by retaining or discarding interactions. Consequently, support-driven transformations also affect the associated interaction attributes, since removing an interaction necessarily removes its corresponding values in $\tau$.

Support-driven transformations encompass a broad family of operators that differ in the criterion used to select which entities to retain. The most common strategies include filtering entities based on their activity in the interaction log, selecting subsets of users or items for experimentation, retaining only entities that satisfy specific availability constraints, and removing redundant interactions. We discuss these families in turn.

\paragraph{Activity-based filtering.}
The most common support-driven transformations filter the interaction log according to the observed activity of users or items. Their underlying assumption is that entities associated with only a few interactions provide insufficient evidence for learning reliable preference patterns. Removing these low-activity entities reduces the long tail of extremely sparse users and items, producing a denser interaction log that is often considered more suitable for offline evaluation.

Activity-based filtering can be applied independently to users or items, or jointly to both. The latter case corresponds to the well-known iterative $k$-core filtering procedure, which is by far the most widely adopted support-driven transformation in recommender systems research. 

\paragraph{User filtering.}
\label{par:user-filtering}
The simplest form of activity-based filtering removes interactions involving users whose activity falls below a predefined threshold. Let the degree of a user in the interaction log be defined as

\[
\deg_u(u,\mathcal{D})
=
\left|
\left\{
(u',i',\tau')\in\mathcal{D}
\mid
u'=u
\right\}
\right|.
\]

Given a minimum activity threshold $k\in\mathbb{N}$, the corresponding filtering operator retains only interactions involving users whose degree is at least $k$:

\[
P^{\mathrm{u\mbox{-}core}}_{k}(\mathcal{D})
=
\{
(u,i,\tau)\in\mathcal{D}
\mid
\deg_u(u,\mathcal{D})\ge k
\}.
\]

This operation removes all users whose activity falls below the specified threshold, together with their associated interactions, while leaving the remaining interaction records unchanged.

\paragraph{Item filtering.}
\label{par:item-filtering}
The same principle can be applied to items. Let

\[
\deg_i(i,\mathcal{D})
=
\left|
\left\{
(u',i',\tau')\in\mathcal{D}
\mid
i'=i
\right\}
\right|
\]

denote the degree of item $i$. The corresponding operator is

\[
P^{\mathrm{i\mbox{-}core}}_{k}(\mathcal{D})
=
\{
(u,i,\tau)\in\mathcal{D}
\mid
\deg_i(i,\mathcal{D})\ge k
\}.
\]

This operator removes interactions involving items whose observed activity falls below the specified threshold and is commonly employed to discard items for which only limited interaction evidence is available.

\paragraph{Iterative $k$-core filtering.}
\label{par:iterative-kcore}
Applying user and item filtering only once does not guarantee that both activity constraints are simultaneously satisfied. Removing low-activity users may reduce the degree of some items below the threshold, while removing low-activity items may, in turn, reduce the degree of additional users. Consequently, recommender systems typically adopt an iterative $k$-core procedure that alternates user and item filtering until convergence.

Formally,

\[
P^{\mathrm{iter}}_{k}(\mathcal{D})
=
\lim_{t\rightarrow\infty}
\left(
P^{\mathrm{u\mbox{-}core}}_{k}
\circ
P^{\mathrm{i\mbox{-}core}}_{k}
\right)^t
(\mathcal{D}).
\]

The iterative procedure terminates once no additional users or items violate the activity threshold, yielding the maximal interaction log in which every retained user and every retained item satisfy the minimum degree constraint. From a graph-theoretic perspective, this corresponds to extracting the maximal $k$-core of the bipartite interaction graph induced by the interaction log.

\paragraph{User and item subsampling.}
\label{par:user-item-subsampling}
Whereas activity-based filtering retains or removes entities based on their observed activity, subsampling explicitly selects a subset of users or items before constructing the interaction log for experimentation. The objective is typically to reduce the dataset's scale, facilitate controlled experiments, or construct benchmark variants with specific statistical properties.

User subsampling selects a subset $\mathcal{U}'\subseteq\mathcal{U}$ and retains only interactions involving users in the selected set:

\[
P^{\mathrm{u\mbox{-}sample}}_{\mathcal{U}'}
(\mathcal{D})
=
\{
(u,i,\tau)\in\mathcal{D}
\mid
u\in\mathcal{U}'
\}.
\]

Similarly, item subsampling selects a subset $\mathcal{I}'\subseteq\mathcal{I}$ and is defined as

\[
P^{\mathrm{i\mbox{-}sample}}_{\mathcal{I}'}
(\mathcal{D})
=
\{
(u,i,\tau)\in\mathcal{D}
\mid
i\in\mathcal{I}'
\}.
\]

The subsets $\mathcal{U}'$ and $\mathcal{I}'$ may be obtained through different sampling strategies, including random selection, activity-based sampling, popularity-based sampling, or application-specific criteria. User and item subsampling can also be combined, yielding an interaction log restricted simultaneously to selected subsets of users and items.

\paragraph{External-criterion filtering.}
\label{par:external-filtering}
Support-driven transformations may also retain or remove interactions based on properties external to the interaction log itself. Rather than relying on the observed activity of users or items, these transformations select entities satisfying an application-specific criterion defined outside $\mathcal{D}$. Typical examples include retaining only items for which auxiliary information (e.g., textual descriptions, images, or knowledge-graph entities) is available, restricting the dataset to entities that also appear in another domain or benchmark, or satisfying any other requirement imposed by the experimental setting.

Let $\mathcal{E}\subseteq\mathcal{U}\cup\mathcal{I}$ denote the subset of entities satisfying the external selection criterion. The corresponding filtering operator can be expressed as

\[
P^{\mathrm{ext}}_{\mathcal{E}}(\mathcal{D})
=
\{
(u,i,\tau)\in\mathcal{D}
\mid
u\in\mathcal{E}
\;\vee\;
i\in\mathcal{E}
\}.
\]

The criterion defining $\mathcal{E}$ is application-dependent and does not need to originate from the interaction log itself. In practice, the filtering condition is most commonly applied to items, reflecting the greater availability of auxiliary resources on the item side. However, the same formulation naturally extends to any externally defined subset of users or items, regardless of the source of the criterion.

\paragraph{Interaction deduplication.}
\label{par:deduplication}
Some recommendation datasets contain multiple interaction events associated with the same user–item pair. Although these events represent distinct observations, some recommendation settings require each user–item pair to appear only once in the interaction log. In these cases, the repeated interactions are consolidated into a single representative event according to a predefined selection policy.

Let $\mathrm{sel}_{\pi}(u,i,\mathcal{D})$ denote a selection function that returns one interaction among all occurrences of the pair $(u,i)$ in $\mathcal{D}$ according to a policy $\pi$, such as retaining the earliest interaction, the most recent one, or an aggregated representative event. The corresponding operator is
\[
P^{\mathrm{dedup}}_{\pi}(\mathcal{D})
=
\left\{
\mathrm{sel}_{\pi}(u,i,\mathcal{D})
\;\middle|\;
\exists\,\tau:(u,i,\tau)\in\mathcal{D}
\right\}.
\]
Unlike activity-based and external-criterion filtering, duplicate removal is driven by the multiplicity of interactions associated with the same user–item pair rather than by properties of the users or items themselves.

\subsubsection{Attribute-Driven Transformations}
\label{sec:attribute_driven}

Unlike support-driven transformations, which retain or remove interactions according to properties of the participating users or items, attribute-driven transformations operate on the values stored in the interaction attributes $\tau$. Rather than modifying which interactions are observed, these transformations modify how interactions are represented or interpreted. Consequently, the support of the interaction log generally remains unchanged, whereas the semantic interpretation or attribute values associated with each interaction are modified.

Since the interaction tuple introduced in Chapter~\ref{sec:pipeline} comprises multiple attributes, attribute-driven transformations can be naturally organised according to the component of $\tau$ on which they operate. Throughout this section, we distinguish three families: \emph{feedback transformations}, which modify the representation of user preference signals; \emph{temporal transformations}, which operate on timestamps and temporal information; and \emph{interaction-type transformations}, which select or reinterpret different categories of user actions.

\paragraph{Feedback transformation: Explicit-to-implicit.}
\label{par:explicit-implicit}
One of the most common feedback transformations is converting explicit feedback into implicit feedback. This operation is widely adopted when recommendation models designed for implicit interactions are applied to datasets originally collected with explicit ratings. Rather than preserving different levels of preference expressed through explicit ratings, the transformed interaction log retains only the information that an interaction occurred, treating observed interactions as positive evidence of user preference.

The most common implementation of this transformation is rating binarisation. Given an interaction $(u,i,\tau)$ with rating attribute $v(\tau)$, a threshold $\delta$ is introduced to distinguish positive from non-positive feedback. Interactions whose rating satisfies the threshold are retained and assigned a binary value:

\[
P^{\mathrm{bin}}_{\delta}(\mathcal{D})
=
\{
(u,i,\tau')
\mid
(u,i,\tau)\in\mathcal{D},
\;
v(\tau)\ge\delta
\},
\]

where $\tau'$ is obtained from $\tau$ by replacing the original rating with the binary value $v(\tau')=1$.

Although explicit-to-implicit conversion is driven by the interaction attributes, it simultaneously modifies both the interaction attributes and the interaction log's support. Ratings below the threshold are not converted into explicit negative interactions; instead, they are removed entirely and become unobserved. This distinction reflects the semantics of implicit-feedback recommendation, where the absence of an interaction does not imply negative preference but merely the lack of observed positive evidence.

\paragraph{Feedback transformation: Rating normalisation.}
\label{par:rating-normalisation}
Unlike explicit-to-implicit conversion, which changes the semantics of the feedback signal, rating normalisation preserves the ordinal relationships among ratings while transforming their numerical scale. This operation is commonly adopted when combining datasets collected under different rating scales or when recommendation models assume normalised input values.

Let $v(\tau)$ denote the rating associated with interaction $(u,i,\tau)$, with observed range $[v_{\min},v_{\max}]$. A linear normalisation maps each rating into a target interval $[\alpha,\beta]$ according to

\[
v'(\tau)
=
\alpha
+
(\beta-\alpha)
\frac{v(\tau)-v_{\min}}
{v_{\max}-v_{\min}}.
\]

The corresponding preparation operator is

\[
P^{\mathrm{norm}}_{[\alpha,\beta]}(\mathcal{D})
=
\{
(u,i,\tau')
\mid
(u,i,\tau)\in\mathcal{D}
\},
\]

where $\tau'$ is obtained from $\tau$ by replacing the original rating $v(\tau)$ with the normalised value $v'(\tau)$.

Unlike explicit-to-implicit conversion, rating normalisation does not modify the support of the interaction log. All interactions are preserved, and only the numerical representation of the feedback signal is transformed. Depending on the dataset and recommendation setting, the normalisation parameters may be computed globally over the entire interaction log or separately for each user.

\paragraph{Temporal transformation: Time-window selection.}
\label{par:time-window}
Temporal information is a fundamental attribute of recommendation interactions. A common transformation is to restrict the interaction log to a specified time interval, either to focus on a particular observation period or to exclude interactions deemed obsolete or not representative of the target recommendation scenario.

Given a temporal interval $[t_{\min},t_{\max}]$, the corresponding operator retains only interactions whose timestamp falls within the selected window:

\[
P^{\mathrm{time}}_{[t_{\min},t_{\max}]}(\mathcal{D})
=
\{
(u,i,\tau)\in\mathcal{D}
\mid
t_{\min}\leq t(\tau)\leq t_{\max}
\}.
\]

Time-window selection modifies the support of the interaction log by removing interactions outside the specified interval while leaving the remaining interaction attributes unchanged. Typical applications include restricting the analysis to recent periods, removing interactions collected during a platform's early stages, or constructing datasets corresponding to specific seasons or events.

\paragraph{Interaction-type transformation: Action filtering.}
\label{par:action-filtering}
Many recommendation datasets record multiple categories of user actions, such as clicks, purchases, likes, reviews, or bookmarks. Since these actions convey different levels of user intent, it is often desirable to retain only a subset of interaction types relevant to the recommendation task under consideration.

Let $\mathcal{A}$ denote the set of possible action types and let $a(\tau)\in\mathcal{A}$ be the action associated with interaction $(u,i,\tau)$. Given a subset of actions $A\subseteq\mathcal{A}$, the corresponding operator is

\[
P^{\mathrm{action}}_{A}(\mathcal{D})
=
\{
(u,i,\tau)\in\mathcal{D}
\mid
a(\tau)\in A
\}.
\]

Action filtering modifies the interaction log's support by retaining only interactions associated with the selected action types. This transformation is particularly common in multi-behaviour recommendation datasets, where different interaction types carry different semantic meanings and predictive value.

\subsection{Representation-Dependent Transformations}
\label{sec:repr_dependent}

\begin{table}[t]
\centering
\caption{Overview of representation-dependent transformation techniques associated with the corresponding taxonomy categories. Each technique refers to its detailed description in the text.}
\label{tab:representation-dependent}

\small
\resizebox{\textwidth}{!}{%
\begin{tabular}{@{}lll@{}}
\toprule
\textbf{Family} &
\textbf{Subfamily} &
\textbf{Technique} \\
\midrule

\multirow{2}{*}{\textbf{Sequential-based}}
& \multirow{2}{*}{Sequence transformations}
& \hyperref[par:sequence-truncation]{Sequence truncation} \\

&
& \hyperref[par:sequence-length]{Sequence-length filtering} \\

\cmidrule(lr){1-3}

\textbf{Session-based}
& Session transformations
& \hyperref[par:session-length]{Session-length filtering} \\

\cmidrule(lr){1-3}

\textbf{Graph-based}
& Connectivity-based transformations
& \hyperref[par:connectivity-filtering]{Connectivity-based filtering} \\

\cmidrule(lr){1-3}

\multirow{2}{*}{\textbf{Context-based}}
& \multirow{2}{*}{Context transformations}
& \hyperref[par:context-selection]{Context selection} \\

&
& \hyperref[par:context-discretization]{Context discretization} \\

\bottomrule
\end{tabular}
}
\end{table}

Unlike dataset-level transformations, which are defined directly on the interaction log, representation-dependent transformations become meaningful only after the interaction log has been interpreted under a specific structural representation. Their applicability, therefore, depends not on the interaction tuples themselves but on higher-level structures derived from them, such as ordered sequences, sessions, graphs, or contextual interaction representations.

As discussed in \Cref{ch:datasets}, the same interaction log may be represented in different ways depending on the recommendation task under consideration. Each representation introduces structural objects that are absent from the original interaction log, thereby enabling preparation operators that cannot be defined directly on $\mathcal{D}$. Consequently, representation-dependent transformations are naturally organised by the adopted representation rather than by the criterion driving the transformation. Table~\ref{tab:representation-dependent} provides an overview of the representation-dependent transformation techniques, organizing them according to the structural representation on which they operate and linking each category with the specific operations discussed in the following sections.

\subsubsection{Sequential Representations}

In sequential recommendation, the interaction log is interpreted as a collection of chronologically ordered user interaction sequences, as introduced in Section~\ref{sec:sequential_repr}. Formally, the sequential representation of the interaction log is

\[
\mathcal{S}^{\mathrm{seq}}
=
\{s_u\}_{u\in\mathcal{U}},
\]

where each sequence

\[
s_u
=
\langle
r_1,
r_2,
\ldots,
r_{|s_u|}
\rangle
\]

contains the interactions associated with user $u$, ordered according to their timestamps. Unlike dataset-level transformations, preparation operators in this setting act on complete interaction sequences rather than directly on individual interaction tuples.

\paragraph{Sequence truncation.}
\label{par:sequence-truncation}

User interaction histories may become arbitrarily long, increasing both computational cost and the amount of historical information considered by sequential recommendation models. Therefore, a common preparation step retains only the most recent $L$ interactions of each user sequence.

Given a maximum sequence length $L$, the corresponding operator is

\[
P^{\mathrm{trunc}}_{L}
(\mathcal{S}^{\mathrm{seq}})
=
\{
s'_u
\}_{u\in\mathcal{U}},
\]

where

\[
s'_u
=
\begin{cases}
\langle
r_{|s_u|-L+1},
\ldots,
r_{|s_u|}
\rangle,
&
|s_u|>L,
\\[1ex]
s_u,
&
\text{otherwise.}
\end{cases}
\]

Sequence truncation is only meaningful once the interaction log has been represented as an ordered sequence. Unlike activity-based filtering, which removes users or items according to their interaction frequency, sequence truncation preserves every user while discarding only the oldest interactions beyond the specified sequence length.

\paragraph{Sequence-length filtering.}
\label{par:sequence-length}

Another common transformation removes user sequences that are shorter than a minimum length, under the assumption that very short histories provide insufficient sequential context for predicting future interactions.

Given a minimum sequence length $\ell$, the corresponding operator is

\[
P^{\mathrm{seq\mbox{-}len}}_{\ell}
(\mathcal{S}^{\mathrm{seq}})
=
\{
s_u
\in
\mathcal{S}^{\mathrm{seq}}
\mid
|s_u|
\geq
\ell
\}.
\]

Although conceptually related to user activity filtering, this transformation is defined on the sequential representation rather than on the interaction log itself. Its criterion is the length of the ordered interaction sequence, making it inherently dependent on the sequential interpretation of the data.

\subsubsection{Session Representations}
In session-based recommendation, the interaction log is represented as a collection of sessions, as introduced in Section~\ref{sec:session_repr}. Formally, the session representation is

\[
\mathcal{S}^{\mathrm{sess}}
=
\{s_1,s_2,\ldots,s_m\},
\]

where each session

\[
s
=
\langle
r_1,r_2,\ldots,r_{|s|}
\rangle
\]

is an ordered sequence of interactions grouped according to a session identifier or a temporal proximity criterion. Preparation operators, therefore, act on complete sessions rather than on individual interactions.

\paragraph{Session-length filtering.}
\label{par:session-length}

A common preparation step removes sessions whose length falls outside a predefined range. Very short sessions often provide insufficient context to predict the next interaction, whereas extremely long sessions may reflect anomalous user behaviour or data-collection artefacts.

Given minimum and maximum admissible session lengths $\ell_{\min}$ and $\ell_{\max}$, the corresponding operator is

\[
P^{\mathrm{sess\mbox{-}len}}_{[\ell_{\min},\ell_{\max}]}
(\mathcal{S}^{\mathrm{sess}})
=
\{
s\in\mathcal{S}^{\mathrm{sess}}
\mid
\ell_{\min}
\le
|s|
\le
\ell_{\max}
\}.
\]

Unlike sequence-length filtering, this transformation operates on sessions, which are local interaction contexts rather than complete user histories. Consequently, its applicability depends on the session-based interpretation of the interaction log.

\subsubsection{Graph Representations}

In graph-based recommendation, the interaction log is represented as a graph, as introduced in Section~\ref{sec:graph_repr}. In the simplest setting, this representation is a bipartite graph

\[
\mathcal{G}=(\mathcal{V},\mathcal{E}),
\]

where the node set $\mathcal{V}=\mathcal{U}\cup\mathcal{I}$ contains users and items, and each observed interaction in $\mathcal{D}$ induces an edge in $\mathcal{E}$. Data preparation, therefore, operates on the graph structure rather than directly on the interaction log.

It is important to distinguish between transformations that are genuinely graph-specific and dataset-level transformations that merely admit a graph-theoretic interpretation. For example, the activity-based filtering operators introduced in Section~\ref{sec:support_driven} correspond to removing user or item nodes whose degree falls below a specified threshold. In contrast, duplicate removal corresponds to collapsing multiple edges between the same pair of nodes into a single edge. Although these operations can be naturally described on the graph representation, their definition depends only on the interaction log, and therefore, they remain dataset-level transformations.

Graph-based transformations, by contrast, rely on structural properties that are defined only once the graph representation has been constructed.

\paragraph{Connectivity-based filtering.}
\label{par:connectivity-filtering}

One of the most common graph-specific transformations removes nodes or edges according to connectivity properties of the interaction graph. A typical example consists of retaining only the largest connected component, thereby discarding isolated nodes and disconnected graph fragments.

Let $\mathcal{G}_{\mathrm{lcc}}$ denote the largest connected component of $\mathcal{G}$,

\[
\mathcal{G}_{\mathrm{lcc}}
=
\arg\max_{\mathcal{G}'\subseteq\mathcal{G},
\;
\mathcal{G}'\text{ connected}}
|\mathcal{V}(\mathcal{G}')|,
\]

where $\mathcal{V}(\mathcal{G}')$ denotes the node set of the subgraph $\mathcal{G}'$. The corresponding preparation operator is

\[
P^{\mathrm{lcc}}(\mathcal{G})
=
\mathcal{G}_{\mathrm{lcc}}.
\]

Connectivity-based filtering removes graph components that are disconnected from the main interaction network, ensuring that all retained nodes belong to a single connected graph. Unlike activity-based filtering, which is driven solely by local interaction counts, this transformation depends on global connectivity properties that are defined only on the graph representation.

\subsubsection{Context-Aware Representations}

In context-aware recommendation, interactions are represented together with contextual variables describing the circumstances under which they occurred, as introduced in Section~\ref{sec:context_repr}. Formally, the context-aware representation is

\[
\mathcal{D}_{c}
=
\{
(u,i,c)
\mid
(u,i,\tau)\in\mathcal{D}
\},
\]

where $c\in\mathcal{C}$ denotes the contextual information extracted from the interaction attributes, such as temporal, spatial, or device-related variables. Data preparation, therefore, operates on contextual representations rather than on the original interaction tuples.

\paragraph{Context selection.}
\label{par:context-selection}

Context-aware datasets often contain multiple contextual variables, not all of which are necessarily relevant for the recommendation task. A common preparation step, therefore, selects only a subset of the available contextual dimensions.

Given a subset of contextual variables $C\subseteq\mathcal{C}$, the corresponding operator is

\[
P^{\mathrm{ctx\mbox{-}select}}_{C}
(\mathcal{D}_{c})
=
\{
(u,i,c')
\mid
(u,i,c)\in\mathcal{D}_{c}
\},
\]

where $c'$ denotes the restriction of $c$ to the variables contained in $C$.

Context selection reduces the dimensionality of the contextual representation while preserving the underlying interaction log. Unlike attribute-driven filtering, which operates directly on the interaction attributes, this transformation is defined only after the contextual representation has been constructed.

\paragraph{Context discretization.}
\label{par:context-discretization}
Many contextual variables are continuous, such as timestamps, geographic coordinates, or environmental measurements. A common preparation step, therefore, converts these variables into discrete categories better suited to downstream recommendation models.
Let

\[
\beta:\mathcal{C}\rightarrow\mathcal{B}
\]

denote a discretisation function mapping continuous contextual variables to a set of discrete bins $\mathcal{B}$. The corresponding preparation operator is

\[
P^{\mathrm{ctx\mbox{-}disc}}_{\beta}
(\mathcal{D}_{c})
=
\{
(u,i,c')
\mid
(u,i,c)\in\mathcal{D}_{c}
\},
\]

where $c'$ is obtained by replacing each continuous contextual variable with its discretised counterpart according to $\beta$.

Typical discretisation strategies include partitioning timestamps into time-of-day or day-of-week categories, grouping locations into geographical regions, or constructing application-specific intervals for continuous contextual measurements. Like context selection, this transformation depends on the context-aware representation and cannot be defined directly on the interaction log.

\section{Side-Information Transformations}
\label{sec:side_info}

Unlike interaction-log transformations, which operate on the interaction log $\mathcal{D}$, side-information transformations operate on the auxiliary information $\Phi$ associated with users or items. As introduced in Section~\ref{sec:formal_view}, their objective is to transform heterogeneous side information into representations that recommendation models can exploit. Although the proposed formal framework applies to side information in general, this section focuses on the multimodal setting, where $\Phi$ comprises modality-specific content such as text, images, audio, video, and knowledge graphs.

From a methodological perspective, side-information transformations naturally decompose into two conceptually distinct stages. The first, referred to as \emph{data preparation}, operates on the original modality while preserving its semantic nature, producing a canonical representation suitable for subsequent processing. The second, referred to as \emph{feature extraction}, transforms the prepared content into a different representation, typically a numerical embedding, that can be consumed by the recommendation model.

Formally, let $c_x^{(m)}$ denote the raw content associated with entity $x\in\mathcal{X}$ in modality $m\in\mathcal{M}_x\subseteq\mathcal{M}$, where $\mathcal{M}$ denotes the set of supported modalities. Data preparation applies a modality-specific operator $\pi_m$ to obtain a prepared representation $\widetilde{c}_x^{(m)}$, which is subsequently transformed by a feature extractor $\varphi_m$ into the final representation $z_x^{(m)}$. The overall side-information transformation is therefore expressed as

\[
\Psi_m
=
\varphi_m
\circ
\pi_m.
\]

This decomposition provides the conceptual foundation adopted throughout the remainder of this chapter. We first formalise the notions of data preparation and feature extraction, introducing a taxonomy of the transformations that characterise each stage. We then instantiate these concepts across the main categories of side information used in recommender systems, highlighting how the same preparation and extraction principles apply to different modalities while giving rise to modality-specific processing pipelines.

\subsection{Data Preparation}
\label{sec:side_preparation}

Data preparation comprises the transformations that operate on raw side information before feature extraction. Unlike feature extraction, whose objective is to map the original content into a numerical representation, data preparation preserves the nature of the underlying modality while transforming it into a canonical form suitable for subsequent processing. Its purpose is therefore to reduce unnecessary variability, improve consistency across data instances, and organise the available information without altering its semantic content.

Formally, let $c_x^{(m)}$ denote the raw content associated with entity $x$ in modality $m$. Data preparation is modelled by a modality-specific operator

\[
\pi_m :
\mathcal{C}^{(m)}
\rightarrow
\widetilde{\mathcal{C}}^{(m)},
\]

which transforms the raw content into its prepared counterpart

\[
\widetilde{c}_x^{(m)}
=
\pi_m\!\left(c_x^{(m)}\right),
\]

where both $c_x^{(m)}$ and $\widetilde{c}_x^{(m)}$ belong to the same modality. For example, textual preparation transforms raw textual documents into standardised textual inputs, visual preparation transforms raw images into standardised images, and audio preparation transforms raw audio recordings into standardised audio signals. The prepared content, therefore, constitutes the canonical input to the subsequent feature extraction stage.

Although the specific operations performed during preparation vary across modalities, they are driven by a common set of design decisions. In the following, we organise these decisions into three complementary categories: \emph{Selection}, \emph{standardisation}, and \emph{Structuring}.

\paragraph{Selection.}
\emph{What information enters the pipeline?}
Selection determines which portions of the available side information are retained for subsequent processing. Auxiliary information associated with the same entity is often heterogeneous and redundant, requiring the preparation pipeline to define which assets, fields, or fragments constitute the effective input to the recommendation pipeline.

The selection strategy depends on the underlying modality. For textual information, the choice may involve titles, descriptions, metadata, or reviews. For visual content, it determines whether to retain a single representative image or multiple images. Analogous decisions arise for audio recordings, video clips, and knowledge graphs, where only a subset of the available information may be preserved.

Selection, therefore, defines the information boundary of the side-information pipeline: content discarded at this stage cannot be

\paragraph{Standardisation.}
\emph{In which format is that information represented?}
standardisation transforms the selected content into a consistent representation while preserving its original modality. Its objective is to eliminate variability that is unrelated to the semantic information conveyed by the data, ensuring that different instances of the same modality are represented in a uniform format before feature extraction.

The specific operations depend on the underlying modality. For textual information, standardisation typically includes character-encoding normalisation, removal of markup, and harmonisation of whitespace and punctuation. Visual data processing encompasses operations such as image decoding, colour-space conversion, resizing, and pixel normalisation. Analogous procedures are applied to audio, video, and knowledge graphs, where differences in sampling rates, encoding formats, or identifier conventions are resolved to obtain a consistent representation.
Standardisation, therefore,e reduces accidental variability introduced during data collection or integration while preserving the semantic content of the original modality. As a result, subsequent feature extraction can focus on modelling the information contained in the data rather than compensating for inconsistencies in its representation.

\paragraph{Structuring.}
\emph{How is that standardised information organised for feature extraction?}
Structuring determines how standardised content is organised before feature extraction. Unlike standardisation, which aims to make individual data instances consistent, structuring defines how the available information is arranged in the form expected by the downstream feature extractor.

The organisation's strategy depends on the modality and on the assumptions of the selected extraction pipeline. In textual information, structuring includes decisions such as field concatenation, tokenisation, truncation, and padding. For visual and temporal media, it may involve organising images into patches, segmenting audio into fixed-length windows, or partitioning videos into clips or frame sequences. Similar decisions arise for knowledge graphs, where subgraphs or neighbourhoods may be constructed before representation learning.

Structuring, therefore, establishes the computational unit on which feature extraction operates. While it does not alter the semantic content of the selected information, it determines how that information is presented to the feature extractor and, consequently, influences the representations ultimately learned.

\subsection{Feature Extraction}
\label{subsec:feature_extraction}

Whereas data preparation standardises and organises side information while preserving its original modality, feature extraction transforms the prepared content into a numerical representation that can be consumed by recommendation models. The objective is no longer to modify the content itself, but to encode its informative properties into a representation suitable for downstream learning. As a result, feature extraction determines the information ultimately made available to the recommender and therefore represents one of the principal sources of variability across recommendation pipelines.

Formally, let $\widetilde{c}_x^{(m)}$ denote the prepared content associated with entity $x$ in modality $m$. Feature extraction is modelled by the modality-specific operator

\[
z_x^{(m)}
=
\varphi_m\!\left(\widetilde{c}_x^{(m)}\right),
\]

where $\varphi_m$ maps the prepared modality-specific content into a feature representation $z_x^{(m)}$. Unlike the preparation operator $\pi_m$, which preserves the original modality while improving its consistency, $\varphi_m$ changes the representation space itself, producing numerical vectors that recommendation algorithms can subsequently process.

Although the implementation of $\varphi_m$ depends on the considered modality, the underlying design question is common to all side-information sources: \emph{how should the prepared content be represented?} Rather than classifying feature extractors according to the specific architectures employed (e.g., BERT~\citep{DBLP:conf/naacl/DevlinCLT19}, ResNet~\citep{DBLP:conf/cvpr/HeZRS16}, or CLIP~\citep{DBLP:conf/icml/RadfordKHRGASAM21}), we organise them according to how the representation function is obtained. This perspective captures the methodological choices that most directly influence computational cost, experimental reproducibility, and the degree of control retained by the experimenter. We distinguish four families of feature extractors: hand-crafted, frozen pre-trained, end-to-end pre-trained, and dataset-provided.

\paragraph{Hand-crafted extractors.}

Hand-crafted extractors derive feature representations through manually designed algorithms that summarise specific properties of the input content. Rather than learning representations from data, these methods rely on predefined descriptors engineered for a particular modality or application domain. Consequently, the extraction function is completely specified before training and remains independent of the recommendation task.

\paragraph{Frozen pre-trained extractors.}

Frozen pre-trained extractors obtain feature representations from models trained on external tasks and datasets. During recommendation, the extractor is used only for inference, while its parameters remain fixed. This strategy has become the dominant paradigm in multimodal recommendation, as it enables the incorporation of rich semantic representations without the computational cost of end-to-end optimisation.

\paragraph{End-to-end trained extractors.}

End-to-end trained extractors learn the representation function jointly with the recommendation objective. Unlike frozen models, the feature extractor is optimised together with the recommender, allowing the learned representations to adapt to the target recommendation task. This increased flexibility is achieved at the cost of higher computational requirements, more complex optimisation, and reduced transferability across datasets.

\paragraph{Dataset-provided features.}

Some benchmark datasets directly distribute pre-computed feature representations together with the raw side information. In this setting, feature extraction is effectively performed by the dataset creators rather than by the experimenter, who directly consumes the provided representations. Although this strategy improves computational efficiency and facilitates reproducibility, it also limits control over the extraction process and ties the evaluation to a specific representation distributed with the dataset.

Although these four families are defined according to how the extraction function is obtained, they also exhibit different practical trade-offs in terms of computational cost, experimenter control, and reproducibility. Table~\ref{tab:family_axes} summarises the characteristics typically associated with each family.

\begin{table}[t]
\centering
\small
\caption{Typical practical trade-offs across the four families of feature extractors. Individual implementations may deviate from these general characteristics.}
\label{tab:family_axes}
\resizebox{\textwidth}{!}{%
\begin{tabular}{@{}lccccc@{}}
\toprule
\textbf{Family} &
\makecell{Encoder\\training cost} &
\makecell{Per-study\\extraction cost} &
\makecell{Experimenter\\control} &
\makecell{Cross-study\\comparability} &
\makecell{Reproducibility} \\
\midrule
Hand-crafted        & None       & Low    & Full   & Low    & High \\
Frozen pre-trained  & Amortized  & Low    & Medium & High   & Medium \\
End-to-end trained  & Per study  & High   & Full   & Low    & Low \\
Dataset-provided    & By dataset & None   & None   & Exact  & High \\
\bottomrule
\end{tabular}
}
\end{table}

The following subsections illustrate how these four families are instantiated across the principal sources of side information used in recommender systems, namely, textual, visual, audio, video, and knowledge-graph information.

\subsection{Modality-Specific Transformations}
\label{sec:modality_specific}

% ============================================================
% MODALITY-SPECIFIC TRANSFORMATIONS
% ============================================================

\begin{table}[t]
\centering
\caption{Overview of modality-specific side-information transformations. Each technique refers to its detailed description in the text.}
\label{tab:modality-specific-transformations}

\small
\resizebox{\textwidth}{!}{%
\begin{tabular}{@{}lll@{}}
\toprule
\textbf{Modality} &
\textbf{Stage} &
\textbf{Technique} \\
\midrule

% ------------------------------------------------------------
% TEXTUAL
% ------------------------------------------------------------

\multirow{6}{*}{\textbf{Textual}}
& \multirow{3}{*}{Data preparation}
& \hyperref[par:text-preparation]{Text-field selection and combination} \\

&
& \hyperref[par:text-preparation]{Text normalization and cleaning} \\

&
& \hyperref[par:text-preparation]{Tokenization, truncation, and padding} \\

\cmidrule(lr){2-3}

&
\multirow{3}{*}{Feature extraction}
& \hyperref[par:text-feature-extraction]{Sparse lexical representations} \\

&
& \hyperref[par:text-feature-extraction]{Pre-trained language-model representations} \\

&
& \hyperref[par:text-feature-extraction]{Task-specific or dataset-provided representations} \\

\cmidrule(lr){1-3}

% ------------------------------------------------------------
% VISUAL
% ------------------------------------------------------------

\multirow{6}{*}{\textbf{Visual}}
& \multirow{3}{*}{Data preparation}
& \hyperref[par:visual-preparation]{Single- or multi-image selection} \\

&
& \hyperref[par:visual-preparation]{Image decoding, resizing, and normalization} \\

&
& \hyperref[par:visual-preparation]{Image-set organization} \\

\cmidrule(lr){2-3}

&
\multirow{3}{*}{Feature extraction}
& \hyperref[par:visual-feature-extraction]{Hand-crafted visual descriptors} \\

&
& \hyperref[par:visual-feature-extraction]{Pre-trained visual representations} \\

&
& \hyperref[par:visual-feature-extraction]{Task-specific or dataset-provided representations} \\

\cmidrule(lr){1-3}

% ------------------------------------------------------------
% AUDIO
% ------------------------------------------------------------

\multirow{6}{*}{\textbf{Audio}}
& \multirow{3}{*}{Data preparation}
& \hyperref[par:audio-preparation]{Audio-stream or recording selection} \\

&
& \hyperref[par:audio-preparation]{Channel conversion, resampling, and spectral transformation} \\

&
& \hyperref[par:audio-preparation]{Fixed-duration segmentation} \\

\cmidrule(lr){2-3}

&
\multirow{3}{*}{Feature extraction}
& \hyperref[par:audio-feature-extraction]{Hand-crafted acoustic representations} \\

&
& \hyperref[par:audio-feature-extraction]{Pre-trained audio representations} \\

&
& \hyperref[par:audio-feature-extraction]{Task-specific or dataset-provided representations} \\

\cmidrule(lr){1-3}

% ------------------------------------------------------------
% VIDEO
% ------------------------------------------------------------

\multirow{6}{*}{\textbf{Video}}
& \multirow{3}{*}{Data preparation}
& \hyperref[par:video-preparation]{Frame or clip selection} \\

&
& \hyperref[par:video-preparation]{Frame decoding, resizing, and normalization} \\

&
& \hyperref[par:video-preparation]{Frame, clip, or sequence organization} \\

\cmidrule(lr){2-3}

&
\multirow{3}{*}{Feature extraction}
& \hyperref[par:video-feature-extraction]{Frame-based visual encoding} \\

&
& \hyperref[par:video-feature-extraction]{Pre-trained video encoding} \\

&
& \hyperref[par:video-feature-extraction]{Task-specific or dataset-provided representations} \\

\cmidrule(lr){1-3}

% ------------------------------------------------------------
% KNOWLEDGE GRAPH
% ------------------------------------------------------------

\multirow{6}{*}{\textbf{Knowledge graph}}
& \multirow{3}{*}{Data preparation}
& \hyperref[par:kg-preparation]{Knowledge-source and subgraph selection} \\

&
& \hyperref[par:kg-preparation]{Identifier and relation standardization} \\

&
& \hyperref[par:kg-preparation]{Triple, neighborhood, and subgraph construction} \\

\cmidrule(lr){2-3}

&
\multirow{3}{*}{Feature extraction}
& \hyperref[par:kg-feature-extraction]{Hand-crafted graph descriptors} \\

&
& \hyperref[par:kg-feature-extraction]{Pre-trained knowledge-graph embeddings} \\

&
& \hyperref[par:kg-feature-extraction]{Task-specific or dataset-provided representations} \\

\bottomrule
\end{tabular}
}
\end{table}

The preparation and feature-extraction principles introduced in the previous sections are common to all forms of side information. Their concreteness, however, depends on the characteristics of the underlying modality. In the following, we discuss the most common preparation and extraction strategies adopted for the principal sources of side information used in recommender systems, namely, textual, visual, audio, video, and knowledge-graph information. Table~\ref{tab:modality-specific-transformations} provides an overview of the modality-specific side-information transformations discussed in this section, organizing the corresponding data-preparation and feature-extraction techniques across the different modalities.

\subsubsection{Textual Information}
\label{subsec:textual_information}

Textual side information typically includes item titles, descriptions, metadata fields, captions, transcripts, or user-generated content such as reviews when these are treated as user- or item-level attributes rather than interaction attributes (see Section~\ref{sec:augmenting_data}). Among the different sources of side information adopted in recommender systems, textual information is arguably the most versatile, as it can describe both the semantic characteristics of items and the preferences or profiles of users.

\paragraph{Data preparation.}
\label{par:text-preparation}

The preparation of textual information instantiates the three preparation decisions introduced in Section~\ref{sec:side_preparation}.

Regarding \emph{selection}, textual preparation determines which textual fields are retained for subsequent processing. Depending on the available information, recommendation pipelines may rely on titles, descriptions, metadata, reviews, or combinations thereof. When multiple textual fields are available, the preparation pipeline must additionally specify whether they are processed independently or concatenated into a single document, as well as the ordering adopted and the use of separator tokens.

The second preparation decision concerns \emph{standardisation}. Typical operations include Unicode normalisation, removal of markup and malformed tokens, harmonisation of whitespace and punctuation, and other cleaning procedures aimed at eliminating artefacts introduced during data collection while preserving the lexical and semantic content of the original text.

Finally, \emph{structuring} organises the standardised text into the representation expected by the selected feature extractor. Classical pipelines generally require word-level tokenisation and may additionally employ stop-word removal, stemming, or lemmatization~\citep{DBLP:conf/emnlp/Camacho-Collados18, DBLP:journals/access/FarinaGKS22}. Transformer-based pipelines instead rely on subword tokenisation and require explicit decisions regarding maximum sequence length, truncation, and padding~\citep{DBLP:conf/naacl/DevlinCLT19}. Since these operations directly determine the input processed by the extractor, they should be reported together with the adopted extraction strategy.

\paragraph{Feature extraction.}
\label{par:text-feature-extraction}

Textual information is represented by feature extractors belonging to all four families introduced in Section~\ref{subsec:feature_extraction}.

Hand-crafted extractors were predominant in early recommender systems, where textual information was represented through sparse lexical descriptors such as bag-of-words, TF--IDF, and paragraph vectors~\citep{DBLP:conf/kdd/YingHCEHL18,DBLP:conf/mm/LiuCSWNK19}.

End-to-end trained extractors subsequently became widespread with the adoption of neural recommendation models. In these approaches, CNN-, RNN-, or hybrid architectures were jointly optimised with the recommendation objective to learn task-specific representations directly from textual descriptions or reviews~\citep{DBLP:conf/sigir/ChenCXZ0QZ19, DBLP:conf/kdd/ChenHXGGSLPZZ19, DBLP:conf/ijcai/Chen020, DBLP:journals/tcss/YangWLLGDW20, DBLP:journals/tomccap/WangDJJSN21}.

More recently, frozen pre-trained extractors have become the dominant paradigm. Transformer-based language models, particularly BERT~\citep{DBLP:conf/naacl/DevlinCLT19} and Sentence-BERT~\citep{DBLP:conf/emnlp/ReimersG19}, are now adopted by the majority of multimodal recommendation approaches~\citep{DBLP:conf/mm/LiuYLWTZSM21,DBLP:conf/mm/Zhang00WWW21,DBLP:conf/bigmm/VaswaniAA21,DBLP:conf/mm/MuZT0T22,DBLP:journals/corr/abs-2211-06924,DBLP:conf/www/WeiHXZ23,DBLP:conf/www/ZhouZLZMWYJ23}. Since these models are tightly coupled to their tokenization scheme and input format, reproducibility requires reporting not only the adopted checkpoint, but also the tokenizer, maximum sequence length, pooling strategy, and the textual fields provided as input.

Finally, some benchmark datasets distribute pre-computed textual representations together with the raw textual content, allowing experiments to directly consume the released embeddings without performing feature extraction as part of the experimental pipeline.

\subsubsection{Visual Information}
\label{subsec:visual_information}

Visual side information typically consists of images associated with items, such as product photographs, movie posters, book covers, or advertisement images. Compared with textual information, visual content is often represented as images with a common tensor-based input interface, leading to standardised preparation pipelines across recommendation datasets.

\paragraph{Data preparation.}
\label{par:visual-preparation}
Visual data preparation instantiates the three preparation decisions introduced in Section~\ref{sec:side_preparation}.

Regarding \emph{selection}, the preparation pipeline determines whether each item is represented by a single image or by multiple images. When multiple images are available, a representative image (e.g., the first or primary image) may be selected, or the entire image collection may be retained for subsequent aggregation. Since this choice influences both the item's semantic representation and the computational cost of feature extraction, it should be treated as an explicit preparation decision.

The second preparation decision concerns \emph{standardisation}. Raw images are first decoded into a common colour space, typically RGB, and subsequently resized to the input resolution required by the selected feature extractor~\citep{DBLP:conf/icml/RadfordKHRGASAM21, DBLP:conf/cvpr/HeZRS16}. Additional operations may include aspect-ratio correction via cropping or padding, and pixel normalisation using the statistics adopted during model pre-training (e.g., ImageNet mean and variance~\citep{DBLP:conf/nips/KrizhevskySH12}). These transformations reduce variability introduced by different image formats while preserving the semantic visual content.

Finally, \emph{structuring} organises the standardised visual information into the input expected by the downstream extractor. When a single image is associated with each item, the computational unit naturally corresponds to the entire image. Conversely, when multiple images are retained, the preparation pipeline must define how they are organised before extraction, for example, by preserving the complete image collection or by constructing fixed-size batches to be subsequently aggregated into a single item-level representation.

\paragraph{Feature extraction.}
\label{par:visual-feature-extraction}
Visual information has historically been represented using feature extractors from all four families introduced in Section~\ref{subsec:feature_extraction}.

Hand-crafted extractors were mainly adopted before the widespread use of deep learning, relying on manually designed descriptors such as colour histograms and SIFT-based features to characterise image appearance~\citep{DBLP:conf/iccv/SwainB90, DBLP:journals/ijcv/Lowe04}.

Frozen pre-trained extractors currently constitute the dominant paradigm in multimodal recommendation. Early approaches employed convolutional neural networks such as AlexNet~\citep{DBLP:conf/nips/KrizhevskySH12} and VGG~\citep{DBLP:journals/corr/SimonyanZ14a}, using intermediate or penultimate-layer activations as fixed visual representations~\citep{DBLP:conf/ijcai/ZhangWHHG17,DBLP:conf/kdd/YingHCEHL18,DBLP:conf/emnlp/WangNL18,DBLP:conf/sigir/ChenCXZ0QZ19,DBLP:conf/mm/DongSFJXN19,DBLP:conf/aaai/YangDW20}. More recent studies have largely converged toward stronger encoders, particularly ResNet-50 and Inception variants~\citep{DBLP:conf/mm/WeiWN0HC19,DBLP:journals/tois/ChengCZKK19,DBLP:conf/kdd/ChenHXGGSLPZZ19,DBLP:conf/cikm/SunCZWZZWZ20,DBLP:conf/ijcnn/Shen0LWC20,DBLP:journals/ipm/TaoWWHHC20,DBLP:journals/tcss/YangWLLGDW20,DBLP:journals/tmm/ZhanLASDK22,DBLP:conf/mm/ChenWWZS22}. More recently, vision-language foundation models such as CLIP~\citep{DBLP:conf/icml/RadfordKHRGASAM21} have started to appear in multimodal recommender systems~\citep{DBLP:conf/mir/LiuMSO022}, although their adoption remains considerably less widespread than in the broader computer vision literature.

End-to-end trained extractors are represented by recommendation models that jointly optimise convolutional visual encoders together with the recommendation objective. While common in earlier deep-learning architectures, this strategy has progressively been replaced by frozen pre-trained backbones as large-scale visual foundation models have become available.

Finally, dataset-provided representations are particularly common for visual information. The Amazon benchmark collections~\citep{DBLP:conf/sigir/McAuleyTSH15}, for example, distribute pre-computed CNN features together with the raw images, allowing experiments to directly consume visual embeddings without performing feature extraction as part of the experimental pipeline.

\subsubsection{Audio Information}
\label{subsec:audio_information}

Audio side information typically consists of music tracks, speech recordings, podcasts, or the audio stream extracted from videos. Compared with textual and visual information, audio presents an intrinsically temporal structure, requiring preparation pipelines that operate on continuous signals before feature extraction can be applied.

\paragraph{Data preparation.}
\label{par:audio-preparation}
Audio data preparation instantiates the three preparation decisions introduced in Section~\ref{sec:side_preparation}.

Regarding \emph{selection}, the preparation pipeline determines which audio content is associated with each entity. When the audio is embedded within a video file, the audio stream is first separated from the visual component so that the two modalities can be processed independently. If multiple audio recordings are available, the pipeline must additionally specify whether a single recording is selected or the complete collection is retained for subsequent aggregation.

The second preparation decision concerns \emph{standardisation}. Raw waveforms are converted into a consistent representation by resolving channel mismatches (typically through mono conversion), resampling the signal to the sampling rate expected by the downstream extractor, and applying the transformations required by the selected representation. For spectrogram-based pipelines, these operations include the computation of the Short-Time Fourier Transform, Mel-filterbank projection, and logarithmic compression, producing standardised log-Mel spectrograms suitable for subsequent processing~\citep{DBLP:conf/icassp/HersheyCEGJMPPS17}.

Finally, \emph{structuring} organises the standardised signal into the computational units processed by the feature extractor. Continuous audio recordings are typically partitioned into fixed-duration segments or patches, with explicit policies for handling recordings that are shorter or longer than the selected segment length~\citep{DBLP:conf/icassp/HersheyCEGJMPPS17}. Since these choices determine the temporal granularity of the extracted representations, they should be considered part of the preparation pipeline.

\paragraph{Feature extraction.}
\label{par:audio-feature-extraction}
The feature-extraction landscape for audio is considerably narrower than for textual or visual information.

Hand-crafted extractors appear only in a small number of early studies and are now largely absent from the multimodal recommendation literature.

Frozen pre-trained extractors constitute the dominant paradigm. VGGish~\citep{DBLP:conf/icassp/HersheyCEGJMPPS17} is by far the most frequently adopted audio encoder, particularly in multimedia and micro-video recommendation~\citep{DBLP:conf/mm/WeiWN0HC19,DBLP:journals/ipm/TaoWWHHC20,DBLP:conf/mm/LiuYLWTZSM21,DBLP:conf/mm/ChenWWZS22,DBLP:journals/tmm/WangWYWSN23}. More recent alternatives, including YAMNet~\citep{DBLP:conf/ijcnn/DrossosMGLV20}, OpenL3, AST, and music-specific encoders, have also been explored~\citep{DBLP:conf/recsys/OramasNSS17,DBLP:conf/bigmm/VaswaniAA21,DBLP:journals/eswa/LeiHZSZ21}, although they remain comparatively uncommon.

End-to-end trained audio encoders are only sporadically adopted, with the most recent recommender systems preferring frozen pre-trained representations over jointly optimised acoustic models.

Finally, dataset-provided representations are particularly common in short-video recommendation benchmarks. For example, the TikTok and Kwai datasets distribute pre-computed acoustic embeddings that downstream recommendation models directly consume without requiring feature extraction as part of the experimental pipeline~\citep{DBLP:conf/mm/WeiWN0HC19,DBLP:journals/tmm/WangWYWSN23}.

\subsubsection{Video Information}
\label{subsec:video_information}

Video side information consists of temporally ordered visual content and therefore combines both spatial and temporal information. Unlike static images, videos require preparation pipelines that determine not only how individual frames are processed but also how temporal information is sampled and organised before feature extraction.

\paragraph{Data preparation.}
\label{par:video-preparation}
Video data preparation instantiates the three preparation decisions introduced in Section~\ref{sec:side_preparation}.

Regarding \emph{selection}, the preparation pipeline determines which portions of the video are retained for subsequent processing. Since processing every frame is rarely computationally feasible, most recommendation pipelines sample either individual frames or short clips according to a predefined temporal policy, such as uniform sampling, fixed frame rates, or regularly spaced clips.

The second preparation decision concerns \emph{standardisation}. Sampled frames are decoded into a common visual representation and undergo the same normalisation procedures adopted for static images, including resizing, aspect-ratio correction when required, and pixel normalisation. When videos also contain an audio stream, the acoustic component is extracted and processed independently through the audio preparation pipeline.

Finally, \emph{structuring} determines the temporal organisation presented to the feature extractor. Depending on the adopted architecture, the prepared content may consist of independent frames, fixed-length clips, or ordered frame sequences. Consequently, the preparation pipeline should explicitly specify the clip duration, frame sampling strategy, temporal stride, and policies for handling videos of different lengths.

\paragraph{Feature extraction.}
\label{par:video-feature-extraction}
Video feature extraction is less standardised than audio extraction and encompasses several complementary strategies.

Hand-crafted extractors are virtually absent from modern recommendation pipelines.

Frozen pre-trained extractors constitute the predominant family. A common strategy applies image encoders such as VGG, ResNet, or Inception independently to sampled frames before aggregating the resulting representations through temporal pooling or attention mechanisms~\citep{DBLP:conf/ijcai/ZhangWHHG17,DBLP:conf/mm/WeiWN0HC19,DBLP:journals/tmm/SangXQMLW21}. Other approaches employ dedicated video encoders based on three-dimensional convolutional networks or temporal architectures that directly encode short video clips while modelling motion information~\citep{DBLP:journals/tmm/SangXQMLW21,DBLP:conf/sigir/WuWQZHX22}.

End-to-end trained video encoders are less common and generally appear in recommendation models that jointly optimise temporal representations together with the recommendation objective.

Dataset-provided video representations remain relatively uncommon compared with textual, visual, and audio information, with most studies performing feature extraction directly from the released video content.

\subsubsection{Knowledge Graph Information}
\label{subsec:kg_information}

Knowledge graphs provide structured semantic information describing entities and the relationships among them. In recommender systems, they are commonly used to enrich users and items with external knowledge derived from resources such as DBpedia~\footnote{https://www.dbpedia.org/}, Freebase~\citep{DBLP:conf/aaai/BollackerCT07}, Wikidata~\footnote{https://www.wikidata.org}, or domain-specific knowledge graphs. Unlike textual or visual information, which consists of unstructured content, knowledge graphs are naturally represented as collections of entities connected through typed relations.

\paragraph{Data preparation.}
\label{par:kg-preparation}
Knowledge-graph data preparation instantiates the three preparation decisions introduced in Section~\ref{sec:side_preparation}.

Regarding \emph{selection}, knowledge-graph preparation first determines the source of external knowledge adopted by the recommendation pipeline. Common choices include large public knowledge graphs such as DBpedia, Freebase, and Wikidata, as well as domain-specific or proprietary knowledge graphs maintained by individual organisations. Since these resources differ substantially in coverage, granularity, and semantic richness, the selected knowledge source directly determines the information available to the recommender.

Selection subsequently determines which portion of the knowledge graph is retained for recommendation. Rather than exploiting the entire graph, preparation typically extracts the subgraph associated with the users or items appearing in the interaction dataset. This process commonly involves entity linking between catalogue items and knowledge-graph entities, followed by the extraction of local neighbourhoods up to a predefined depth (e.g., one- or two-hop neighbourhoods). Additional selection decisions may further restrict the retained graph according to entity types, relation types, or frequency thresholds in order to remove semantically uninformative portions of the graph.

The second preparation decision concerns \emph{standardisation}. This stage aligns heterogeneous identifiers, removes duplicated entities and relations, resolves inconsistencies originating from multiple data sources, and converts the graph into a common representation. Depending on the downstream pipeline, entities may be linked to external knowledge bases through entity linking, while relations are represented using a consistent vocabulary of predicates.

Finally, \emph{structuring} organises the prepared graph into the computational units processed by the feature extractor. Depending on the adopted representation-learning approach, the preparation pipeline may construct collections of triples, extract local $k$-hop neighbourhoods around each entity, or generate task-specific subgraphs centred on users or items. These decisions determine the structural context available during representation learning and should therefore be regarded as part of data preparation.

\paragraph{Feature extraction.}
\label{par:kg-feature-extraction}
Knowledge-graph representations can be obtained through feature extractors belonging to all four families introduced in Section~\ref{subsec:feature_extraction}.

Hand-crafted extractors rely on manually designed graph descriptors, such as node degree, centrality measures, relation counts, or manually engineered semantic features. Although historically important, these approaches are now rarely adopted in modern knowledge-aware recommender systems.

Frozen pre-trained extractors obtain entity representations from knowledge graph embedding models trained independently of the recommendation task. Representative examples include translational approaches such as TransE~\citep{DBLP:conf/nips/BordesUGWY13}, as well as subsequent embedding models including TransR~\citep{DBLP:conf/aaai/LinLSLZ15}, ComplEx~\citep{DBLP:conf/icml/TrouillonWRGB16}, and RotatE~\citep{DBLP:conf/iclr/SunDNT19}. The resulting entity embeddings are subsequently incorporated into the recommender without further optimisation of the extraction model.

End-to-end trained extractors jointly optimise graph representations together with the recommendation objective. Rather than relying on fixed embeddings, these approaches learn task-specific representations through graph neural networks or knowledge-aware propagation mechanisms, allowing entity representations to adapt to the recommendation task during training.

Finally, some benchmark datasets directly provide precomputed knowledge graph embeddings along with the interaction data. In these settings, feature extraction is effectively externalised from the experimental pipeline, allowing recommendation models to directly consume the released representations while sacrificing control over the extraction process.

\subsection{Cross-Modal Considerations}
\label{subsec:cross_modal}
The previous sections discussed preparation and feature extraction independently for each modality. In practice, however, recommendation datasets often combine multiple sources of side information, requiring additional processing steps that concern the relationships among modalities rather than any individual modality. These considerations influence the amount of information available to the recommender, the consistency of the experimental pipeline, and ultimately the reproducibility of experimental results.

A first issue concerns \emph{cross-modal alignment}. For each entity $x$, the preparation pipeline must establish the set of available modalities
$\mathcal{M}_x \subseteq \mathcal{M}$ and define an unambiguous correspondence between the entity and its associated assets. In practice, this involves resolving inconsistent identifiers across heterogeneous data sources, removing dangling references (e.g., images without a corresponding catalogue item), and ensuring that all retained assets are correctly associated with the corresponding user or item. Since misaligned content cannot be meaningfully exploited during feature extraction, alignment constitutes a prerequisite for every multimodal recommendation pipeline.

A second consideration concerns \emph{modality coverage}. Different modalities are rarely available for exactly the same set of entities. Textual descriptions may cover the entire catalogue, whereas images, videos, or knowledge-graph entities may only be available for a subset of items. Consequently, the preparation pipeline must explicitly define how missing modalities are handled. Common alternatives include restricting the dataset to entities with complete modality coverage, retaining partially observed entities, or assigning fallback or imputed representations to the missing modalities. \citet{DBLP:conf/cikm/MalitestaRPNM24} systematically compare these strategies on several Amazon datasets, showing that silently discarding items with missing modalities may substantially overestimate the contribution of multimodal information, whereas graph-based imputation recovers much of the lost performance. Regardless of the adopted strategy, it should always be reported because it changes the effective benchmark and the population of evaluable entities.

A further consideration concerns \emph{multiple assets per modality}. Individual entities are frequently associated with multiple images, reviews, audio recordings, or video clips. The preparation pipeline must therefore specify whether a single representative asset is selected or whether all available assets are retained for subsequent aggregation. This decision determines both the quantity of information presented to the feature extractor and the semantics of the resulting representation, and should therefore be regarded as an explicit preparation choice rather than an implementation detail.

Finally, multimodal pipelines should preserve \emph{cross-modal consistency} throughout the experimental workflow. Whenever multiple assets belong to the same entity, they should all be assigned consistently to the same training, validation, or test partition to avoid information leakage across data splits~\citep{DBLP:conf/recsys/GusakVKVF25}. More generally, any transformation affecting modality availability or alignment should be applied before feature extraction, so that the extracted representations accurately reflect the information available at recommendation time.

\paragraph{Cross-modal feature extraction.}

Although feature extraction is commonly performed independently for each modality, recent representation models have demonstrated the possibility of learning a shared embedding space directly from multiple modalities. Examples include vision--language models such as CLIP~\citep{DBLP:conf/icml/RadfordKHRGASAM21} and multimodal transformers such as ViLBERT~\citep{DBLP:conf/nips/LuBPL19}, which jointly encode visual and textual content. From the perspective of the taxonomy introduced in this chapter, these models remain instances of frozen pre-trained extractors, the distinction being that the extraction function itself operates simultaneously on multiple modalities rather than on a single source.

Despite their growing popularity in adjacent fields, the surveyed recommendation literature remains largely dominated by modality-specific feature extraction followed by downstream fusion within the recommender. Consequently, multimodal extractors should be regarded as an alternative realisation of the feature-extraction stage rather than as a replacement for the distinction between preparation, extraction, and recommendation modelling adopted throughout this chapter.

%%%%
%%%%

%%%%%%%%%%%%%%
% Empirical Trends 
%%%%%%%%%%

\section{Empirical Trends in Data Preparation}
The preceding sections introduced a formal vocabulary for describing data-preparation choices in recommender-system pipelines and organised these choices into a set of transformation categories. This taxonomy identifies what can be done to interaction logs and side information; it does not, however, establish which transformations are actually adopted in the literature, how frequently they occur, or which combinations have become common practice.

This section, therefore, moves from conceptual characterisation to empirical observation. By coding the experimental pipelines of the surveyed studies according to the taxonomy introduced above, we examine how data preparation is implemented in practice across the literature. The analysis enables distinguishing broadly adopted conventions from uncommon alternatives, identifying recurrent combinations of processing decisions, and assessing the extent to which methodological choices are reported with sufficient detail for reproducibility.

The investigation is organised along the two components of the data-processing framework. The first part examines \emph{Interaction Log Transformations}, focusing on how feedback data are filtered, redefined, and structured. In particular, it analyses the use of filtering strategies, k-core pruning, binarisation, and the overall complexity of the resulting preprocessing pipelines.

The second part examines \emph{Side-Information Transformations}, with particular attention to the multimodal subset of the corpus. It analyses how auxiliary content is prepared and represented, including the families of feature extractors employed, the adoption of specific encoders such as BERT and ResNet, and the provenance of the data used to train these models.

Taken together, these analyses characterise the methodological profile of the surveyed literature: not only the transformations that are available in principle, but also the practices that have become routine, the alternatives that remain underexplored, and the reporting gaps that limit transparency and reproducibility.

\subsection{Empirical Trends in Interaction Log Transformations}
Having introduced a formal taxonomy of data preparation techniques, in this section, we analyse how these operations are actually employed in the literature, outlining emerging trends and patterns in recommender systems research. For each paper included in our study (Section~\ref{sec:selected_papers}) and for each dataset therein, we manually annotated the data preparation operations applied to the interaction logs. Our annotation focuses exclusively on the type of transformation, independent of its specific parameterisation (e.g., we record the use of user $k$-core filtering without considering the specific value of $k$). Furthermore, we restrict this analysis to transformations applied to the interaction log itself; data preparation steps that involve side information are excluded here and analysed separately in Section~\ref{sec:empirical_trends_mm}. 

\subsubsection{Reporting Practices and Reproducibility}

As a primary investigative step, we examine the extent to which authors explicitly report the data transformation operations applied to their datasets. Among the 517 papers collected in our study, we distinguish two reporting categories: Explicit, with 315 papers (approx. 61\%) clearly and explicitly describe the operations applied to the interaction log, including both the type of operation and, where applicable, its configuration. Implicit or Not Declared: 202 papers (approx. 39\%) in which data preparation steps are either entirely omitted or indirectly specified through references to prior work without a detailed description. This significant lack of explicit reporting raises serious concerns regarding the reproducibility and comparability of experimental results. Since interaction log transformations directly influence data sparsity, degree distributions, and user–item connectivity patterns, missing or incomplete information about data preparation steps can introduce non-trivial biases in empirical evaluations. These findings underscore the need for a systematic characterisation of data preparation practices. Consequently, we now move from reporting habits to the technical characteristics of the pipelines themselves, analysing which specific transformation techniques are most prevalent across the surveyed literature.

\subsubsection{Prevalence of Transformation Techniques}

In this section, we examine the distribution and frequency of specific data transformation techniques within the literature. Following the methodology used for dataset usage analysis, we define an occurrence as the application of a specific transformation within a single (paper, dataset) pair. This approach allows us to capture the granular diversity of data preparation pipelines, acknowledging that authors often apply different transformation strategies to different datasets within the same publication.

\begin{figure}[h]
    \centering    \includegraphics[width=\textwidth]{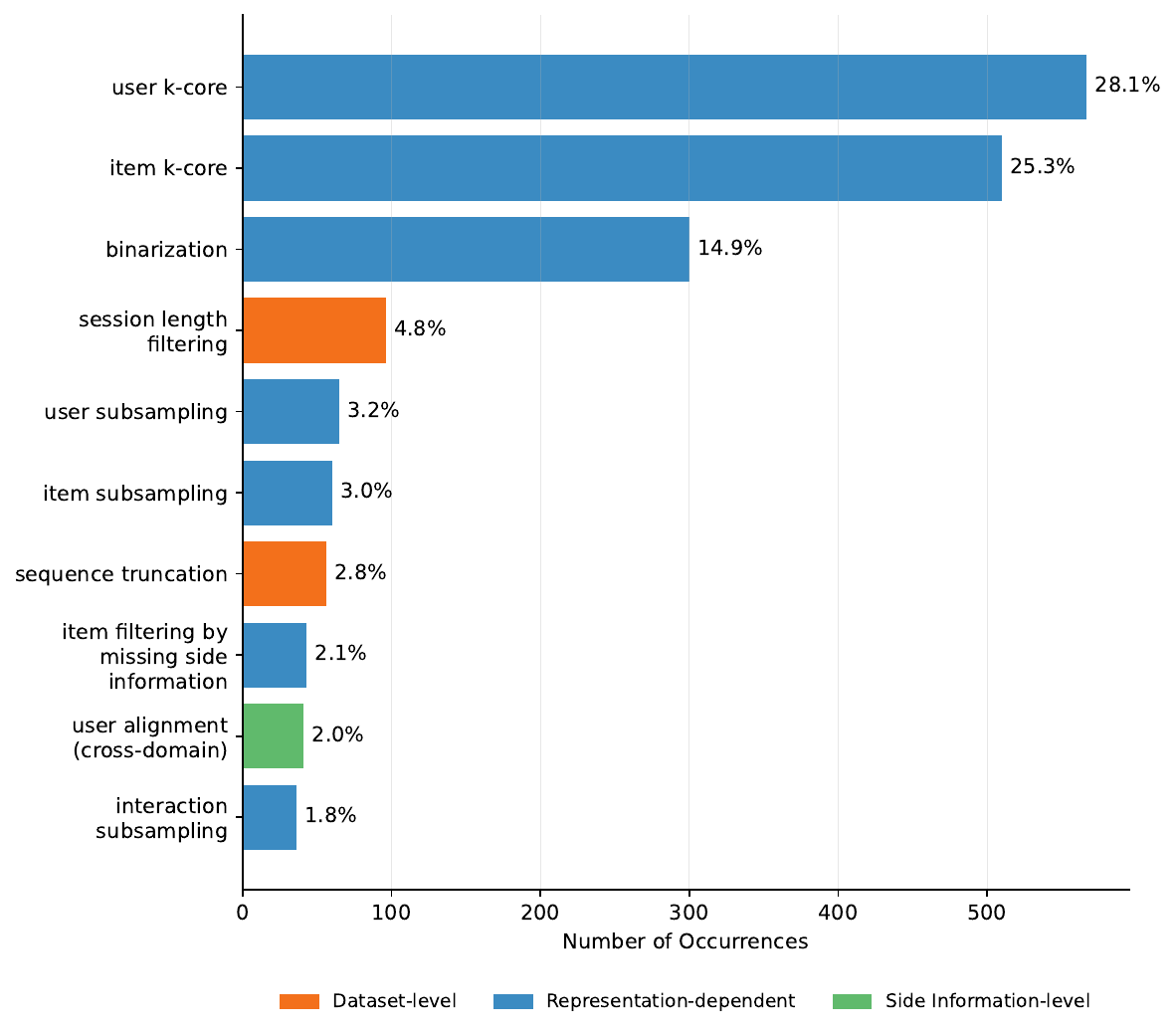}
    \caption{Distribution of the top 10 data transformation techniques applied to interaction logs. Occurrences are measured at the (paper, dataset) level to capture the specific preprocessing pipeline applied to each dataset within a study. Percentages indicate the relative prevalence of each technique across all annotated instances.}
    \label{fig:frequency}
\end{figure}

As illustrated in Figure \ref{fig:frequency}, the empirical landscape is heavily dominated by a small set of "standard" operations. Out of approximately 60 distinct techniques identified in our study, the top 10 account for the vast majority of applications, revealing a highly concentrated distribution. The most frequent transformations are user $k$-core (28.1\%) and item $k$-core (25.3\%) filtering. Together, these $k$-core variations represent over half of all observed transformations, confirming their role as the de facto standard for managing data sparsity and ensuring that users and items have sufficient interaction history for model training. 

However, a more critical look at these reporting practices reveals a significant methodological gap. While user and item $ k$-cores are ubiquitous, they are almost always reported as atomic, one-pass operations. Notably, the iterative $k$-core transformation, which is mathematically necessary to guarantee that all users and items satisfy the $k$ constraint simultaneously, is virtually absent from the top-tier results, falling deep into the long tail of our surveyed techniques. This lack of specificity is concerning: without an iterative approach, filtering users can drop items below the threshold and vice versa, leading to datasets that do not actually meet the intended core requirements. This ambiguity points to a standardisation bias in the field. By predominantly evaluating models on dense sub-graphs generated through these filters, the community may be inadvertently overlooking model performance on long-tail user behaviour and colder items, which are often the most challenging to recommend in real-world scenarios.

The third most common practice is binarisation, appearing in 14.9\% of cases. This technique is typically employed to adapt explicit feedback (such as ratings) into implicit signals, reflecting the broader research trend toward Top-N recommendation tasks. Beyond these dominant practices, we observe several specialised trends that reflect specific research niches. For instance, session length filtering (4.8\%) and sequence truncation (2.8\%)  are consistently linked to sequential and session-based recommendation architectures.  

The remainder of this top-10 list, which includes user alignment (2.0\%) for cross-domain tasks and various forms of subsampling (ranging from 1.8\% to 3.2\%), underscores that while the library of available techniques is vast, researchers rely on a very narrow subset for general-purpose preparation. It is also worth noting that an implicit pre-filtering effect might partially mask the frequency of these reported operations; many researchers utilise already-processed versions of popular datasets (e.g., the Amazon 5-core releases), where the transformation is a property of the data source itself rather than an explicit step in the experimental pipeline. Finally, the low frequency of item filtering due to missing side information (2.1\%)  highlights that log pruning driven by metadata availability remains a specialised practice, occurring almost exclusively in content-conscious or multimodal recommendation contexts.  

\subsubsection{Interplay Between Transformations}

\begin{figure}[!h]
    \centering   \includegraphics[width=\textwidth]{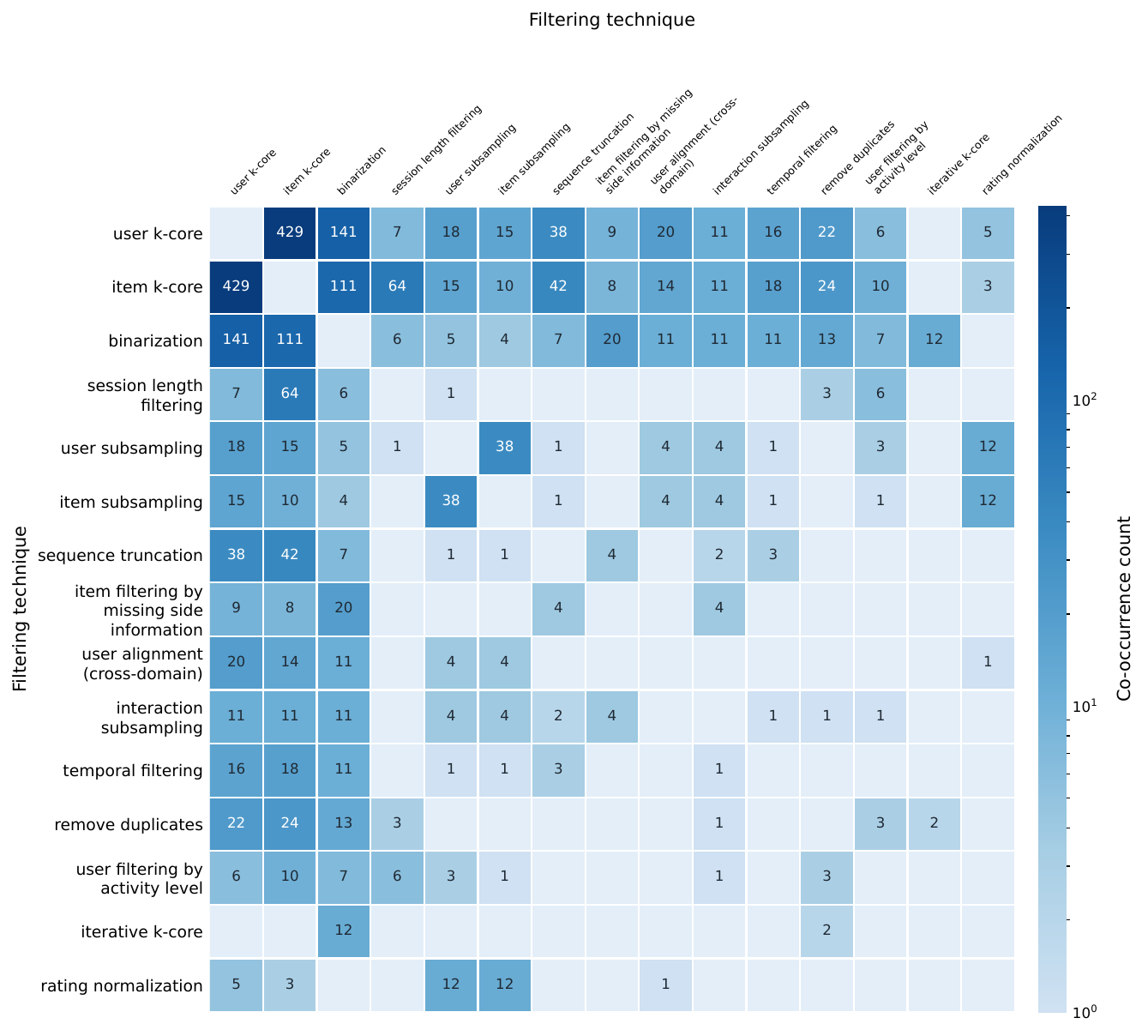}
    \caption{Co-occurrence heatmap of the top 15 data transformation techniques. Each cell indicates the number of times two distinct operations were applied to the same (paper, dataset) instance. The diagonal is intentionally left blank as it represents single-technique occurrences. The high values at the intersection of user and item k-core (429 instances) and binarisation (over 100 instances each) highlight the prevalence of a standardised data preparation pipeline in the literature.}
    \label{fig:co-occurrence}
\end{figure}

While identifying individual techniques is crucial, data preparation in recommender systems often involves a pipeline of multiple operations. To understand how these techniques are combined, we analyse the co-occurrence of the top 15 transformations within the same (paper, dataset) instance, as illustrated in Figure \ref{fig:co-occurrence}.

The heatmap reveals a high degree of coupling between specific operations, centred around a dominant standard pipeline. The most frequent configuration by a significant margin is the simultaneous application of user $k$-core and item $k$-core filtering, which co-occur in 429 instances. This combination is also frequently paired with binarisation, which co-occurs 141 times with user $k$-core and 111 times with item $k$-core. This pattern confirms that a vast portion of the literature follows a rigid, multi-stage data preparation strategy: first reducing the graph to a dense sub-structure from both dimensions to handle sparsity, and then converting the remaining interactions into implicit feedback for Top-N recommendation tasks.

However, a significant ambiguity emerges regarding the execution of these filters. While the simultaneous presence of user and item filtering is ubiquitous (429 instances), the specific use of an iterative $k$-core process is explicitly reported in almost no cases. In much of the surveyed literature, authors describe the removal of users and items below a certain threshold in generic terms, without specifying whether the process was performed iteratively to guarantee simultaneous core consistency. This lack of detail represents a substantial methodological gap: since a one-pass filter on users can drop item frequencies below the established threshold (and vice versa), the absence of explicit information regarding iteration makes it impossible to determine if the resulting datasets actually meet the intended core requirements. This underscores the reporting challenges identified earlier, as such ambiguity hinders exact experimental replication.

Beyond this central cluster, the matrix highlights several specialised research niches. For example, user subsampling and item subsampling show a strong mutual co-occurrence (38 instances), indicating they are almost always used together to downscale massive datasets. We also observe a notable link between sequence truncation and $k$-core filtering (38 instances with user $k$-core and 42 with item $k$-core), reflecting the requirements of sequential recommendation where minimum interaction frequencies are necessary to learn transition patterns. Finally, the sparse nature of the rest of the matrix, particularly for techniques like rating normalisation or temporal filtering, underscores that most papers adhere strictly to the dominant $k$-core/binarisation paradigm, with little variation in their data preparation logic.

\subsubsection{Structural Nature of Transformations}

To gain a deeper understanding of the research community's priorities, we categorise the identified transformation techniques according to their structural intent, following the proposed taxonomy. We distinguish between Dataset-level transformations, which modify global log properties, and Representation-dependent transformations, which adapt the data structure for specific architectures.

\begin{figure}[h]
    \centering   \includegraphics[width=\textwidth]{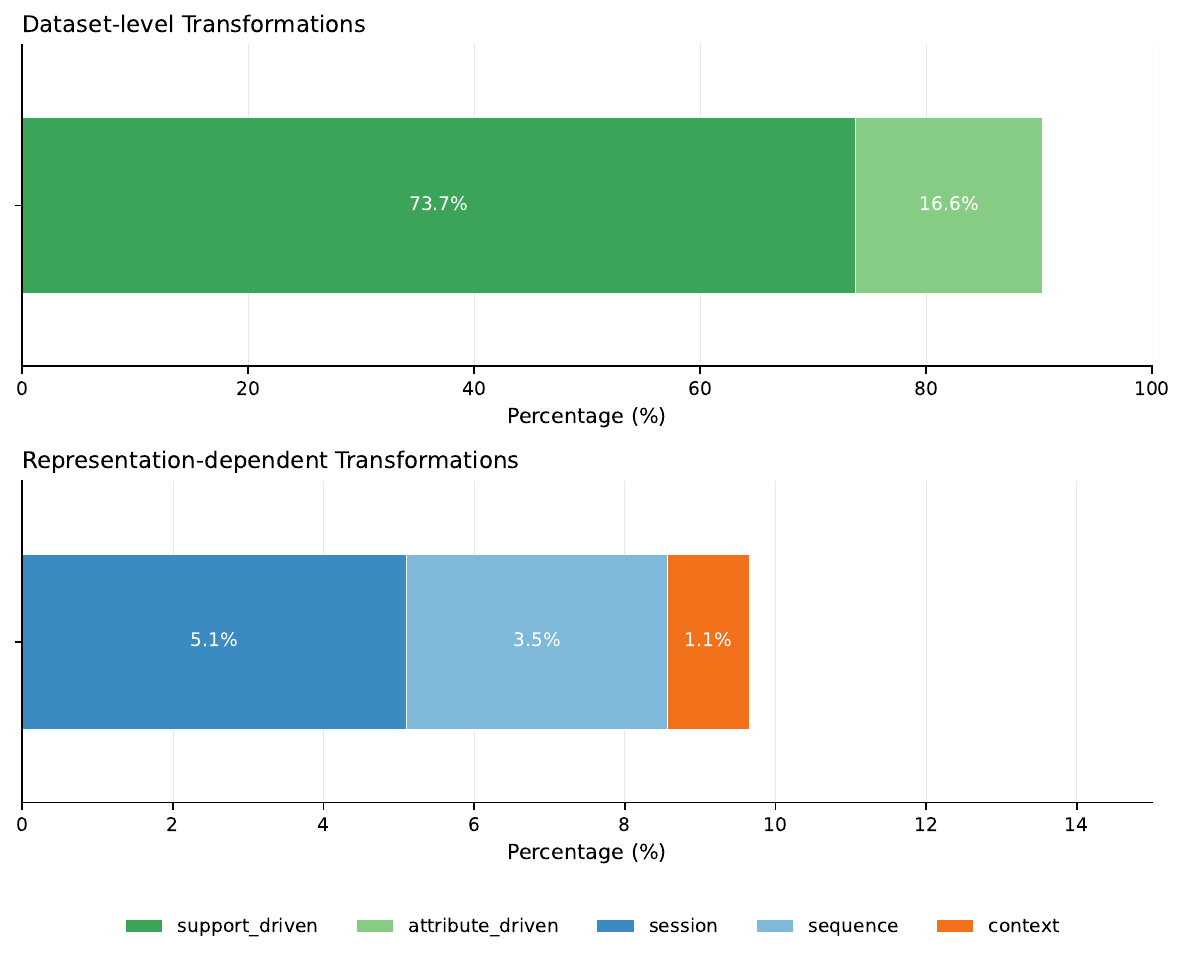}
    \caption{Structural distribution of transformation types using a dual-scale visualisation. The top panel shows the global prevalence of Dataset-level transformations (0-100\% scale), while the bottom panel provides a zoomed perspective (0-14\% scale) to illustrate the distribution of Representation-dependent transformations. All percentages represent the global frequency of each sub-category across the entire surveyed corpus (Section~\ref{sec:selected_papers}).}
    \label{fig:taxonomy_dist}
\end{figure}

As illustrated in Figure \ref{fig:taxonomy_dist}, the empirical landscape is overwhelmingly dominated by dataset-level operations, which are present in approximately 90.3\% of the analysed cases. To ensure a clear interpretation of these results despite the extreme numerical imbalance, we employ a dual-scale visualisation: the top panel displays the global dominance of dataset-level transformations, while the bottom panel provides a zoomed perspective on representation-dependent transformations.

It is important to note that all percentages shown are global, meaning they are calculated relative to the total number of observed transformation instances across the entire survey.

Within the dataset-level macro-category, support-driven transformations (e.g., $k$-core filtering) are the most prominent, accounting for 73.7\% of the total distribution. This confirms that the primary concern for researchers remains the mitigation of data sparsity. Attribute-driven operations follow at 16.6\%, representing cases where the log is pruned based on metadata availability.

In contrast, representation-dependent transformations account for only about 9.7\% of the practices. Among these, session-based adaptations (5.1\%) and sequential prunings (3.5\%) are the most common. Notably, transformations related to context are remarkably rare, appearing in only 1.1\% of the cases. This extreme scarcity highlights that, despite the theoretical importance of context-aware recommendation in the literature, the practical effort in data preparation is almost entirely consumed by basic interaction log pruning. Finally, graph-based representations, despite being a core category in our taxonomy, are virtually absent from the empirical distribution of the surveyed interaction logs.

This distribution emphasises that data preparation is largely viewed as a reduction process (cleaning the log) rather than a structural adaptation task. While advanced architectures are frequent in the literature, the fundamental bottleneck remains the management of the raw interaction log's density and support.

\subsubsection{Complexity of Data Preparation Pipelines}

\begin{figure}[h]
    \centering
    \includegraphics[width=\textwidth]{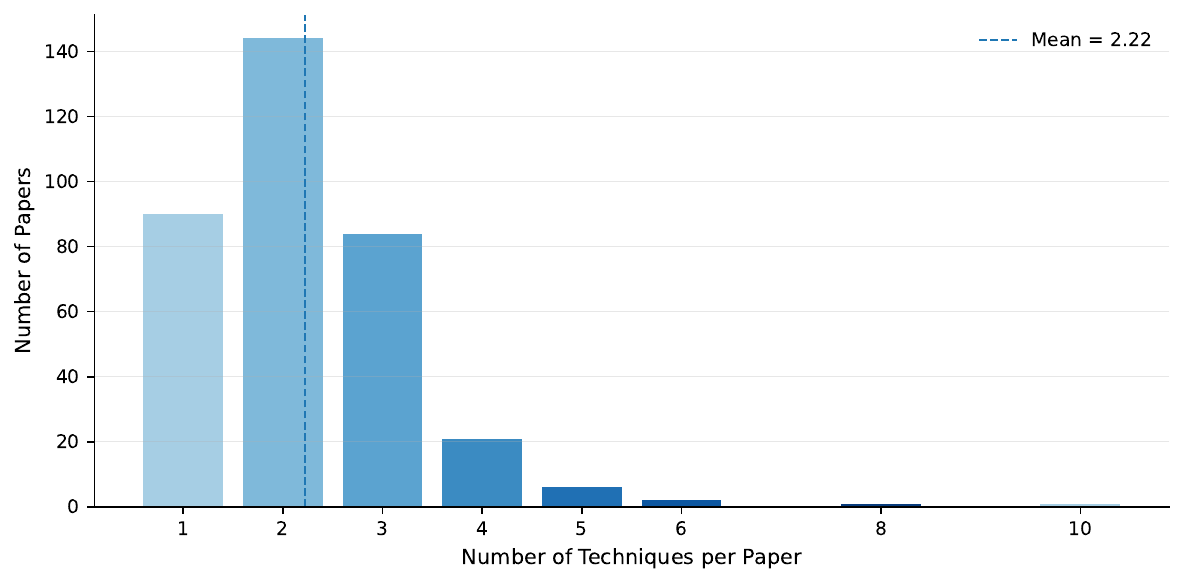}
    \caption{Distribution of the number of data preparation techniques applied per paper. The majority of studies employ a minimalist approach, with a mean of 2.22 operations, typically focusing on the standard $k$-core and binarisation pipeline.}
    \label{fig:pipeline_complexity}
\end{figure}

The final dimension of our analysis investigates the overall complexity of data preparation by measuring the number of distinct transformation techniques applied within each individual paper. As illustrated in Figure \ref{fig:pipeline_complexity}, the distribution reveals that most researchers employ relatively lean pipelines.  

The distribution is characterised by a mean of 2.22 techniques per paper, with a clear mode at two operations. This peak (exceeding 140 papers) corresponds to the standard pipeline identified in our co-occurrence analysis: the combination of a density filter ($k$-core) and a feedback transformation (binarisation). A significant number of studies (approx. 90) apply only a single technique, suggesting a minimalist approach where data preparation is seen as a secondary task rather than a core component of the experimental design.

As we move toward higher complexity, the frequency drops sharply. Papers utilising four or more techniques are in the minority, and instances of highly complex pipelines (8 to 10 operations) are extreme outliers. This trend suggests that while recommender system models are becoming increasingly sophisticated, the datasets on which they are evaluated undergo a relatively limited and standardised set of transformations.

This standardisation in pipeline construction, paired with the lack of explicit reporting, points to a potential risk of methodological stagnation in the field. The prevalence of 1 or 2 techniques does not suggest that simpler pipelines are inherently flawed; rather, it highlights a tendency toward ritualised data preparation. If the community continues to rely on a restricted and often opaque set of pruning operations without documenting their exact execution (e.g., one-pass vs iterative), it becomes difficult to discern whether model performance stems from genuine architectural innovation or from the specific characteristics of the dense sub-graphs produced by these standard filters. The goal for future research should not necessarily be to increase pipeline complexity, but to enhance the rationale and reproducibility of whatever data preparation steps are chosen.

\subsection{Empirical Trends in Side Information Transformations}
\label{sec:empirical_trends_mm}

This section empirically characterises the use of side information
transformations within the multimodal subset of the surveyed literature. It
operationalises the feature-extraction taxonomy introduced in
\Cref{subsec:feature_extraction} by examining the extractor category adopted
for each available modality, the specific encoders disclosed by the studies,
the provenance of the resulting feature representations, and the level of
detail with which these processing choices are reported.

The analysis focuses on the aspects of side information processing that can be
consistently recovered from published experimental descriptions. It therefore
does not attempt to infer unreported selection, standardisation, or structuring
decisions; whenever a modality is absent or the corresponding processing
pipeline is not disclosed, these cases are reported separately from declared
extractor choices.

\subsubsection{Corpus and method}
\label{subsec:empirical_corpus}

The analysis is based on a manually curated subset of the $517$ papers included
in the overall survey corpus. Specifically, it considers the $48$ studies that
evaluate recommendation methods on one or more datasets and use multimodal
side information. The selection procedure and inclusion criteria are described
in \Cref{sec:selected_papers}. Because a paper may evaluate its method on
multiple datasets, this subset comprises $130$ paper--dataset pairs.

For each pair, we manually coded the dataset, the availability of visual,
textual, and other side-information modalities, and the encoder used for each
available modality. The latter includes visual and textual encoders and, where
applicable, an additional encoder for audio or knowledge-graph information. We
also recorded whether the extractor was specified sufficiently to assign it to
one of the four categories and whether the representation was extracted by the
study or provided with the dataset.

Paper--dataset pairs are the unit of analysis for outcomes that may vary across
evaluation settings, such as dataset-specific encoder choices and the
provenance of the features used in an experiment. Whole papers are instead the
unit for paper-level reporting outcomes, such as whether the study fully
discloses its adopted encoders. Accordingly, pair-level counts describe
evaluation instances rather than statistically independent studies; each result
explicitly reports the denominator on which it is based.

\subsubsection{Encoder adoption by modality}
\label{subsec:empirical_families}

Each disclosed modality-specific extractor was assigned to one of the four
categories defined in \Cref{subsec:feature_extraction}. The unit of analysis is therefore a modality-specific paper--dataset observation: a single paper--dataset pair may contribute one observation for visual information, one for textual information, and, where applicable, one for audio or knowledge-graph information.

Although video is treated as a distinct side-information modality in the
conceptual taxonomy, it does not appear as a separate category in the empirical
analysis. Within the surveyed corpus, video-based inputs are consistently
decomposed into their visual component, their audio component, or both, rather
than processed through a dedicated temporally aware video encoder. Accordingly,
video-derived frames are included in the visual column, whereas extracted audio
streams are included in the audio component of the ``Other'' column.

\Cref{tab:empirical_family_distribution} reports the resulting distribution.
Visual extractors are declared for $110$ of the $129$ pairs in which visual
information is present, while textual extractors are declared for $110$ of the
$127$ pairs in which textual information is present. An audio or
knowledge-graph extractor is reported for only $19$ pairs.

\begin{table}[t]
\centering
\small
\setlength{\tabcolsep}{4pt}
\caption{Distribution of disclosed feature extractors by category. Counts refer
to modality-specific paper--dataset observations. For visual and textual
information, ``N/A'' denotes that the modality is absent. Video inputs are
included under visual information when processed as frames, and under
``Other'' when their audio stream is processed. In the ``Other'' column, pairs
without a reported audio or knowledge-graph extractor, combine cases in which the modality is absent and cases in which its extractor is not disclosed.}
\label{tab:empirical_family_distribution}
\begin{tabularx}{\linewidth}{@{}>{\raggedright\arraybackslash}Xrrr@{}}
\toprule
\textbf{Extractor category} & \textbf{Visual} & \textbf{Textual} &
\makecell{\textbf{Other}\\\textbf{(Audio/KG)}} \\
\midrule
(a) Hand-crafted                    &   0 &   3 &   0 \\
(b) Frozen pre-trained, uni-modal   &  71 &  82 &  12 \\
(b) Frozen pre-trained, cross-modal &   2 &   0 &   0 \\
(c) End-to-end trained              &   9 &  14 &   5 \\
(d) Dataset-shipped pre-computed    &  28 &  11 &   2 \\
\midrule
\textbf{Total declared}             & 110 & 110 &  19 \\
Extractor not declared              &  19 &  17 & --- \\
Modality not present                &   1 &   3 & --- \\
No other extractor reported         & --- & --- & 111 \\
\bottomrule
\end{tabularx}
\end{table}

The frozen pre-trained category, corresponding to family~(b) in the taxonomy,
is the most common declared option across all three source types. It accounts
for $73$ of the $110$ declared visual extractors, $82$ of the $110$ declared
textual extractors, and $12$ of the $19$ declared other-modality extractors.
Dataset-shipped representations are also substantial for visual information,
whereas end-to-end training is less frequent and hand-crafted extraction occurs
only for textual information.

\begin{figure}[t]
    \centering
    \includegraphics[width=\textwidth]{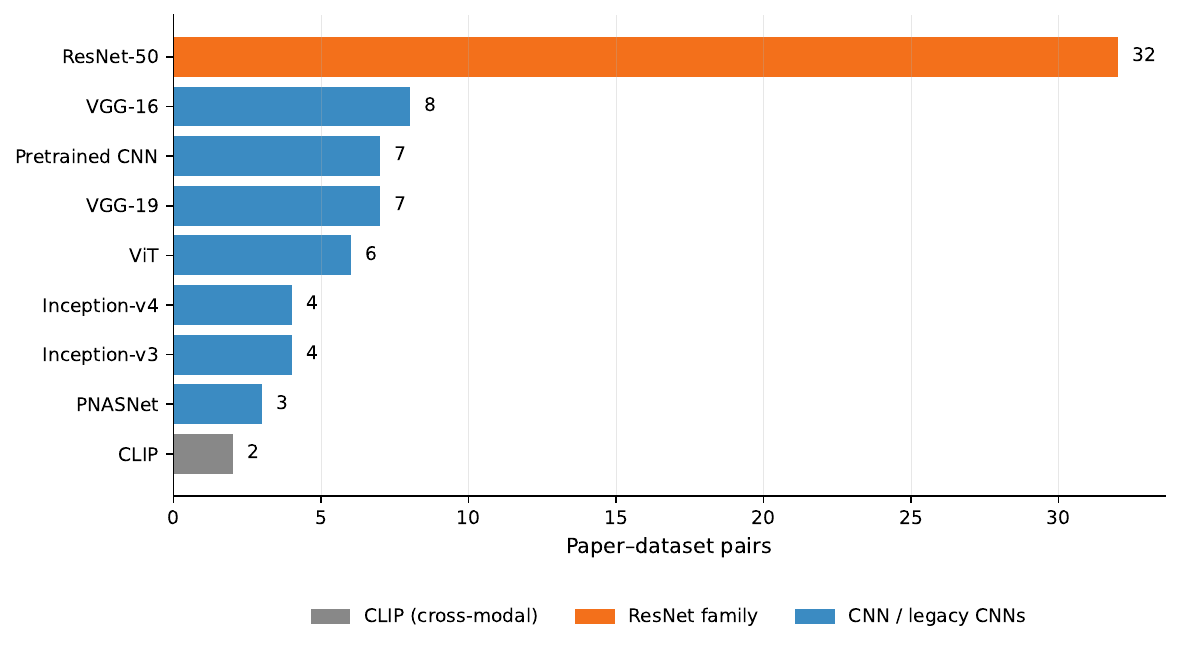}
    \caption{Distribution of reported visual extractors assigned to family~(b),
    over $73$ paper--dataset observations. CLIP (darker bar) is the only
    cross-modal entry; all other extractors are unimodal. ``Pretrained CNN''
    aggregates pairs in which the paper declares the use of a pre-trained CNN
    without specifying the architecture.}
    \label{fig:family_b_visual}
\end{figure}

\begin{figure}[t]
    \centering
    \includegraphics[width=\textwidth]{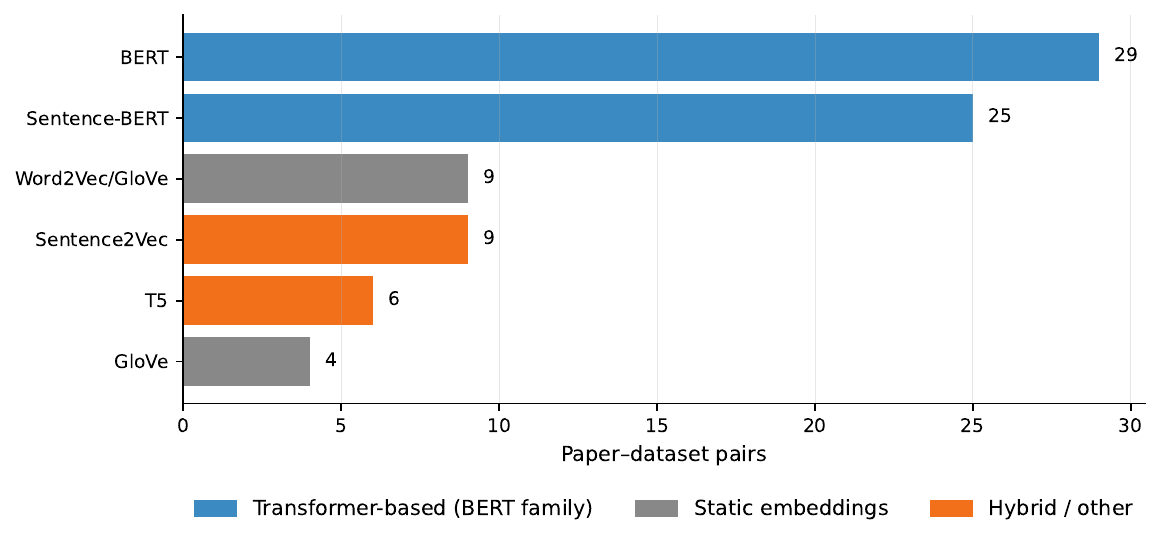}
    \caption{Distribution of reported textual extractors assigned to
    family~(b), over $82$ paper--dataset observations. ``Word2Vec/GloVe''
    aggregates pairs in which the paper declares a word-embedding pipeline
    (Skip-gram, gensim, or initialised GloVe) without further specification.}
    \label{fig:family_b_textual}
\end{figure}

Reported choices within family~(b) are concentrated on a small number of
models, as shown in \Cref{fig:family_b_visual,fig:family_b_textual}. For
visual information, ResNet-50~\citep{DBLP:conf/cvpr/HeZRS16} accounts for
$32$ of the $73$ family-(b) observations. The VGG family accounts for a
further $15$ observations, while ViT, Inception/PNASNet variants, and
generically reported pre-trained CNNs account for the remaining cases. For
textual information, BERT~\citep{DBLP:conf/naacl/DevlinCLT19} and
Sentence-BERT~\citep{DBLP:conf/emnlp/ReimersG19} jointly account for $54$ of
the $82$ family-(b) observations; Sentence2Vec, Word2Vec, and GloVe form a
long tail of pre-Transformer alternatives. For audio, VGGish
\citep{DBLP:conf/icassp/HersheyCEGJMPPS17} is used in $10$ of the $12$
declared family-(b) observations, all associated with TikTok or Kwai.

This concentration is consistent with a path-dependent pattern, in which
studies frequently reuse the extractors made available by established
benchmarks and training frameworks. Establishing this mechanism, however,
would require a direct analysis of framework reuse and encoder comparisons.

Dataset-shipped representations, corresponding to family~(d), account for
$28$ of the $110$ declared visual extractors ($25.5\%$), $11$ of the $110$
declared textual extractors ($10.0\%), and $2$ of the $19$ declared
other-modality extractors ($10.5\%). In these cases, the feature
representation is supplied with the dataset rather than extracted within the
reported experimental pipeline. This distinction is consequential: the study
does not control the encoder or preprocessing procedure unless it replaces the
supplied representation by extracting features anew from the raw side
information. Dataset-shipped representations should therefore be reported
separately from frozen pre-trained extraction.

End-to-end extractors, corresponding to family~(c), are comparatively
uncommon, accounting for $28$ modality-specific observations: $9$ visual,
$14$ textual, and $5$ other. These cases jointly optimise an encoder and the
recommendation objective, rather than treating the encoder as a fixed feature
source. A temporal analysis is required to determine whether the prevalence of
this category has changed over the surveyed period.

Hand-crafted extractors, corresponding to family~(a), are nearly absent from
the corpus. Their only occurrences are $3$ textual observations using SGRank,
a graph-based keyword-extraction method. No visual, audio, or knowledge graph
observation uses a hand-crafted extractor as its primary $\varphi_m$. Within
the surveyed corpus, hand-crafted representations therefore play only a
residual role relative to learned and dataset-provided features.

\chapter{Data Splitting Strategies}
\label{ch:chapter5}
Data splitting is the process of partitioning the available interaction data into distinct subsets, typically used for model training, hyperparameter selection, and final performance evaluation. It is a central component of offline evaluation in recommender systems. Although often treated as a technical detail, the chosen splitting protocol defines the relationship between training and evaluation data and can materially affect the observed performance of recommendation algorithms. More fundamentally, a splitting protocol specifies the experimental setting and the assumptions under which a recommender system is assessed. For example, a temporal split frames evaluation as the prediction of future user behaviour under an information-availability constraint, whereas a leave-one-out split evaluates performance on a single held-out interaction per user. Consequently, reported scores should be interpreted as properties of a model--protocol combination, rather than of the model in isolation.

\section{Taxonomy of Splitting Strategies}
The literature on recommender systems adopts a wide variety of data splitting protocols, often under partially overlapping names and definitions~\citep{DBLP:conf/recsys/MengMMO20, DBLP:conf/recsys/GusakVKVF25, DBLP:conf/sigir/MancinoBF0MPN25}. 
Because terminology is not fully standardised across the literature, the labels used in this taxonomy are descriptive categories adopted in this work. They synthesise recurrent splitting practices identified in the reviewed studies and are not intended as universally established or mutually exclusive protocol names.

Despite their apparent heterogeneity, these protocols can be grouped into a small number of recurring families. At a high level, the literature distinguishes between random (order-agnostic) strategies, temporal (time-aware) strategies, and a residual class of fixed, inherited, or hybrid protocols. This taxonomy aims to provide an initial map of the design space and a compact language for describing practices that are often presented inconsistently across papers.

\subsection{Random (order-agnostic) strategies}

Random splitting strategies assign interactions to training, validation, and test subsets without considering their chronological order. They are particularly common in classical collaborative filtering, rating prediction, and other settings in which the interaction matrix is treated as a static object rather than as the outcome of an evolving behavioural process.

The most direct instance is \emph{global random hold-out}, in which the entire set of observed interactions is randomly partitioned into training, validation, and test subsets according to predefined proportions, such as 80/10/10, 80/20, or 90/10. This strategy is easy to implement and preserves the target split proportions at the dataset level. However, because the split is performed globally, it does not guarantee that every user or item appears in all subsets, nor that their distributions are balanced across them.

A second common variant is \emph{user-wise random hold-out}, where interactions are first grouped by user and then randomly partitioned within each user's history. This strategy is often preferred when the evaluation protocol requires test users to be represented in the training data. Such representation, however, is guaranteed only when users have a sufficient number of interactions and when the protocol explicitly imposes eligibility constraints or filters users with shorter histories.

A more constrained variant is \emph{random leave-one-out}, in which exactly one interaction per user is randomly selected for testing, while the remaining interactions are used for training, possibly with an additional validation split carved out from the training portion. In this setting, the held-out item is the only positive target associated with each test user. The task is therefore to rank this target among a set of candidate items, effectively searching for a single relevant item in a much larger candidate space. In many top-$N$ recommendation benchmarks, this candidate space is constructed by combining the held-out positive item with sampled negative items. Negative sampling is thus a separate evaluation choice: it does not determine the split itself, but it substantially affects the difficulty and interpretation of the ranking task.

Two additional protocols appear less frequently but remain important. In \emph{repeated hold-out}, a random split is generated multiple times and performance is averaged across runs, reducing variance due to a single arbitrary partition. In \emph{$K$-fold cross-validation}, the dataset is partitioned into $K$ folds, and each fold is used as a test set in turn. These protocols differ from the previous variants in that they specify how the partitioning procedure is repeated or reused, rather than the unit at which interactions are assigned to a split. Cross-validation is more typical of earlier recommendation research, especially in matrix completion settings, and remains present in some collaborative, graph-based, and federated works.

Overall, random strategies are easy to implement and can facilitate comparisons across methods when the same protocol is shared. However, they disregard chronological order and may allow training data to include interactions that occur after those used for evaluation. For tasks intended to model future behaviour, this can make the evaluation setting less representative of deployment conditions.

\subsection{Temporal (time-aware) strategies}

Temporal strategies have become increasingly common in recommendation research, particularly with the rise of sequential and next-item prediction tasks. They are appropriate when the evaluation objective is to predict future user behaviour using only information that would have been available at the time of prediction. For this reason, time-aware protocols are often regarded as a closer approximation of online deployment, especially in sequential, session-based, and next-item recommendation settings.

The coarsest variant is \emph{global temporal hold-out}, where interactions are sorted chronologically over the entire dataset and split using fixed temporal thresholds. Typical examples include reserving the last day, week, or month of interactions for testing, while using earlier observations for training and validation. This strategy preserves global temporal dynamics, but may also produce substantial distributional shifts between train and test subsets. Among temporal protocols, global temporal hold-out most directly approximates a deployment setting in which a recommender system is trained using all information available up to a given point in time and is subsequently evaluated on interactions occurring after that point. This approximation is strongest when the same temporal constraint is also applied to feature construction, candidate availability, preprocessing procedures, and model updates.

A finer-grained variant is \emph{user-wise temporal hold-out}, in which interactions are first grouped by user and then ordered chronologically within each user's history. Earlier interactions are assigned to the training set, while more recent interactions are reserved for validation and testing. This strategy preserves within-user temporal order and is particularly common when the evaluation target is future user behaviour. Unlike global temporal hold-out, however, it does not necessarily enforce a common temporal boundary across users: training data may include interactions from other users that occur later than a given user's test interaction.

One of the most widely used protocols in sequential recommendation research is \emph{temporal leave-one-out}. In its simplest form, the last interaction of each user is used for testing, while all preceding interactions are used for training. A common extension introduces a validation set by assigning the second-to-last interaction to validation and retaining the remaining history for training. This protocol operationalises a next-item prediction setting at the user level. However, because users' last interactions may occur at different points in calendar time, it does not necessarily reproduce the common global cutoff used in a deployment-oriented temporal evaluation.

Another family of protocols uses sessions or temporal windows as the splitting unit. In \emph{session-wise temporal splitting}, complete sessions are assigned to training, validation, or test sets according to their temporal order. In \emph{time-window splitting}, interactions are instead partitioned using global calendar intervals, such as days, weeks, or months. Examples include reserving the sessions occurring on the last day as the test set or using interactions from the most recent calendar window for evaluation. These protocols are particularly common in session-based recommendation and in domains with strong short-term dynamics, such as e-commerce and news recommendation.

Compared with random protocols, time-aware strategies enforce temporal availability constraints and reduce the risk that future interactions directly influence model training. However, they presuppose the availability of reliable timestamp information, which is not uniformly satisfied across benchmark datasets. They may also accentuate sparsity, cold-start effects, and distribution shift, especially when recent users or items have limited overlap with the training data.

\subsection{Fixed, inherited, Hybrid, and undocumented protocols}

Random and temporal splitting describe the rule through which interactions are allocated to training, validation, and test data. However, the literature also differs with respect to the provenance of the adopted split, the combination of multiple allocation rules, and the degree to which the procedure is documented. These aspects are particularly relevant for reproducibility and comparability. For the empirical analysis presented later in this chapter, they are treated as distinct splitting families for classification purposes, even though they do not always constitute alternatives to random or temporal allocation rules.

A first family concerns \emph{fixed or pre-computed splits}, in which papers use train/validation/test partitions provided by the original dataset release, by a benchmark suite, or by a previous study. Closely related are \emph{inherited splits}, where authors explicitly adopt the same partition used by an earlier baseline, for example, by retrieving it from a public repository or directly from the original authors. In both cases, the partition may improve comparability across studies when it is publicly available and precisely identified. However, the mechanism originally used to construct the split is not always reported or recoverable. A fixed split may therefore have been generated through a random, temporal, or other procedure, while its underlying allocation rule remains unknown to subsequent users.

For this reason, fixed and inherited benchmark splits are treated as a separate family in our empirical classification. This category captures a practically important condition in which researchers evaluate methods on an externally defined partition, regardless of whether its original construction can be fully reconstructed. Such practices are common in benchmark-driven research communities, where influential baselines may establish widely reused datasets and evaluation pipelines. While this can support direct comparison with prior work, it can also propagate undocumented assumptions or methodological limitations embedded in the original split.

A separate family is represented by \emph{Hybrid or custom protocols}, which combine multiple splitting principles within the same evaluation setting or introduce task-specific allocation rules. For example, a study may use a temporal train/test boundary while constructing validation data through random hold-out. Other protocols may first divide users into disjoint groups and subsequently split the interaction histories of selected users, as in some cold-start evaluation settings. Cross-validation procedures may also be included in this family when they are combined with additional user-level, item-level, or temporal constraints. These protocols cannot always be described adequately through a single random or temporal label, and their interpretation depends on the full sequence of allocation decisions.

Finally, our literature review identifies studies in which the splitting protocol is only partially documented, delegated to previous papers without sufficient detail, or not declared at all. This is not a splitting strategy in itself, but a reporting condition that is sufficiently frequent to warrant explicit classification. When the construction of the split cannot be determined, it is impossible to assess whether the reported evaluation reflects a random, temporal, fixed, or hybrid setting. Incomplete reporting, therefore, limits reproducibility and prevents a well-founded assessment of the methodological comparability of empirical results.

\subsection{Discussion}
Two observations emerge from this classification. First, protocols presented under different names may rely on closely related allocation rules, while protocols assigned the same label may differ substantially in their unit of splitting, temporal constraints, eligibility criteria, or treatment of validation data. The terminology adopted in the literature is therefore not fully standardised, and protocol names alone are often insufficient to establish methodological equivalence.

Second, the choice of splitting protocol is rarely neutral. It defines the experimental setting in which a recommender system is assessed, including the information available during training, the type of user behaviour to be predicted, and the extent to which the evaluation approximates an intended deployment scenario. Reported performance should consequently be interpreted as a property of the model--protocol combination, rather than of the model in isolation. The provenance and documentation of a split are equally important: when a fixed or inherited partition cannot be traced back to its original construction, the assumptions underlying the evaluation remain partially unknown.

The remainder of this chapter develops this descriptive classification in three directions. We first examine the main dimensions along which splitting protocols vary across the literature. We then analyse the empirical Distribution of the identified splitting families and, finally, propose a unified formalisation that represents the most common procedures within a common framework.

\section{Dimensions of Variation Across Splitting Protocols}

While the taxonomy introduced above groups splitting protocols into broad methodological families, protocols belonging to the same family may still differ in ways that materially affect offline evaluation. Across the literature surveyed in this work, we identified several distinct, though not always independent, dimensions along which protocols vary. These dimensions help explain why methods described under the same label may still induce substantially different experimental settings.

\paragraph{Grouping unit.}
A first source of variation concerns the unit on which the split is applied. Some protocols operate globally on the entire interaction log, treating the dataset as a single pool of observations. Others apply splitting within smaller groups, most commonly users or sessions. For example, user-wise protocols split each user's interaction history independently, while session-based protocols operate on interaction sequences grouped into browsing sessions. The grouping unit strongly influences the structure of the resulting splits, the entities represented in each subset, and the behavioural patterns preserved during evaluation.

\paragraph{Ordering criterion.}
A second dimension concerns the ordering of interactions prior to their allocation. In random protocols, interactions are treated as exchangeable observations and are assigned to subsets through random sampling. In contrast, time-aware protocols explicitly preserve chronological order, assigning earlier interactions to training and reserving later interactions for validation or testing. This choice determines whether the evaluation enforces temporal availability constraints and prevents future interactions from directly influencing model training.

\paragraph{Selection rule and repetition scheme.}
Protocols also differ in the rules used to assign observations to training, validation, and test sets. Common allocation rules include proportion-based hold-out schemes, such as 80/10/10 or 70/10/20 splits, and leave-one-out procedures that reserve one or more interactions per user for evaluation. A further distinction concerns whether the partition is generated once or repeatedly reused. Repeated hold-out generates multiple random partitions and aggregates results across runs, whereas cross-validation rotates the test subset across multiple folds. Although these procedures are often grouped together in the literature, they differ in both the composition of the test data and the extent to which results depend on a particular partition.

\paragraph{Temporal cutoff and prediction horizon.}
Within time-aware protocols, the temporal boundary used to separate training from evaluation data constitutes an additional source of variation. A global temporal hold-out may reserve interactions from the last day, week, or month for testing, whereas a temporal leave-one-out may retain the final interaction of each user. These choices imply different prediction horizons, degrees of distributional shift, and approximations of deployment conditions. In particular, a global cutoff imposes a shared observation boundary across users, while user-wise temporal procedures define evaluation relative to each user's individual history.

\paragraph{Validation design.}
Another dimension concerns the role and construction of validation data in the evaluation pipeline. Some studies adopt simple train/test splits, while others explicitly reserve a validation subset for hyperparameter tuning and model selection. In sequential recommendation, a common design assigns the penultimate interaction of each user to validation and the most recent interaction to testing. Other works construct validation data by randomly sampling interactions from the training portion. Such choices affect both model selection and the effective information available at test time. In particular, when a temporal train/test split is combined with random validation sampling, the temporal consistency of the tuning procedure depends on how the validation interactions are selected and on the information made available during training.

\paragraph{Entity coverage and transductive constraints.}
A further distinction concerns whether users and items appearing in validation or test sets are guaranteed to be observed during training. Global protocols may produce evaluation instances involving users or items unseen during training, especially in sparse datasets. Many works, therefore, apply post-split filtering to enforce a transductive setting in which evaluation interactions involve entities already observed in the training data. Such cases do not necessarily represent a methodological problem, as unseen users or items may be intentionally retained to evaluate cold-start conditions. However, the treatment of unseen entities must be explicitly reported, since it determines whether the evaluation follows a transductive or an inductive setting.

\paragraph{Protocol provenance and reproducibility.}
Finally, protocols differ in how they are defined, documented, and reproduced. Some studies explicitly specify the full splitting procedure used in their experiments. Others reuse partitions from prior work, rely on dataset-provided splits, retrieve them from public repositories, or obtain them directly from the authors of previous studies. In such cases, the original allocation rule may not be fully documented or recoverable. In a non-negligible number of studies, the splitting procedure is only partially described or not reported at all. This dimension is methodological rather than algorithmic, but it has direct implications for reproducibility and for the comparability of empirical results.

Taken together, these dimensions reveal that protocols sharing the same label, such as \emph{random hold-out}, \emph{temporal hold-out}, or \emph{leave-one-out}, may correspond to substantially different evaluation settings. A meaningful comparison of experimental results, therefore, requires understanding not only the broad family of a split but also its grouping unit, ordering criterion, selection rule, temporal horizon, validation design, entity coverage, and provenance. Other evaluation choices, such as candidate-set construction and negative sampling, remain distinct from the splitting procedure itself, but may further affect the interpretation of the resulting performance estimates.
\section{A unified formalisation of splitting protocols}
\label{sec:unified_splitting}

The taxonomy introduced above shows that apparently different splitting protocols can often be understood as combinations of a limited number of design choices. Building on the notation introduced in the previous chapter, we formalise a splitting protocol as an assignment of observed interactions to training, validation, and test subsets. The rules used to construct this assignment may differ substantially across protocols, and may include grouping, ordering, temporal cutoffs, selection rules, filtering, and repeated resampling procedures.

\paragraph{Interaction log.}
Recall that an interaction is modeled as a tuple $r=(u,i,\tau)\in\Omega$, where $u\in\mathcal{U}$, $i\in\mathcal{I}$, and $\tau\in\Tau$ denotes the attributes associated with the event. The observed dataset is represented as a finite indexed collection of interactions
\[
\mathcal{D} = (r_1, r_2, \dots, r_n), \qquad r_k \in \Omega.
\]
We use an indexed collection rather than a plain set, since recommendation logs may contain repeated or otherwise indistinguishable interactions that should remain individually addressable.

Let
\[
\mathcal{N}=\{1,\dots,n\}
\]
denote the index set of the interaction log. We write
\[
\pi_u(r)=u, \qquad \pi_i(r)=i
\]
for the user and item projections. When the corresponding attributes are available, we also write.
\[
\pi_t(r), \qquad \pi_a(r), \qquad \pi_s(r)
\]
for the timestamp, action type, and session identifier associated with $r$.

\paragraph{Split assignment.}
Let
\[
\mathcal{L}=\{\mathrm{tr},\mathrm{val},\mathrm{te},\varnothing\}
\]
denote the set of split labels, where $\varnothing$ represents an interaction excluded from the final evaluation pipeline. A splitting protocol $\mathcal{P}$ induces an assignment function
\[
s_{\mathcal{P}}:\mathcal{N}^{0}\longrightarrow\mathcal{L},
\]
where $\mathcal{N}^{0}$ is the index set remaining after any initial data-cleaning operations. The training, validation, and test index sets are then defined as
\[
\mathcal{N}_{\ell}
=
\{k\in\mathcal{N}^{0}\mid s_{\mathcal{P}}(k)=\ell\},
\qquad
\ell\in\{\mathrm{tr},\mathrm{val},\mathrm{te}\}.
\]
The corresponding interaction sublogs are
\[
\mathcal{D}_{tr}=(r_k)_{k\in\mathcal{N}_{tr}}, \qquad
\mathcal{D}_{val}=(r_k)_{k\in\mathcal{N}_{val}}, \qquad
\mathcal{D}_{te}=(r_k)_{k\in\mathcal{N}_{te}}.
\]

By construction, these subsets are pairwise disjoint:
\[
\mathcal{N}_{tr}\cap\mathcal{N}_{val}
=
\mathcal{N}_{tr}\cap\mathcal{N}_{te}
=
\mathcal{N}_{val}\cap\mathcal{N}_{te}
=
\emptyset.
\]
When no interactions are excluded after splitting, they form a complete partition of the processed log:
\[
\mathcal{N}_{tr}\cup\mathcal{N}_{val}\cup\mathcal{N}_{te}
=
\mathcal{N}^{0}.
\]
Otherwise, the remaining indices are assigned the label $\varnothing$.

This assignment-based definition is intentionally general. It represents the final outcome of any split, independently of whether that outcome is generated through random sampling, chronological ordering, a fixed benchmark partition, or a composite protocol.

\paragraph{Constructing a split assignment.}
In the protocols surveyed in this work, the assignment function is commonly constructed through a combination of grouping, ordering, and selection decisions. These components should be understood as a useful procedural description rather than as mandatory stages of every protocol.

A grouping function may partition the interaction indices into groups:
\[
g:\mathcal{N}^{0}\longrightarrow\mathcal{G},
\]
where $\mathcal{G}$ is a set of group identifiers. Typical choices include
\[
g(k)=\star
\quad\text{(global protocol)},
\]
\[
g(k)=\pi_u(r_k)
\quad\text{(user-wise protocol)},
\]
and
\[
g(k)=\pi_s(r_k)
\quad\text{(session-wise protocol)}.
\]
For each group $h\in\mathcal{G}$, let
\[
\mathcal{N}^{(h)}
=
\{k\in\mathcal{N}^{0}\mid g(k)=h\}
\]
be the corresponding index subset.

Within each group, interactions may be treated as exchangeable and randomly permuted, or they may be ordered according to a criterion such as time. For a temporal ordering, we write
\[
r_{k_1}\preceq_h r_{k_2}
\quad\Longleftrightarrow\quad
\pi_t(r_{k_1})\leq\pi_t(r_{k_2}),
\]
whenever reliable timestamps are available. In practice, ties should be resolved through a deterministic rule, for example, using the original interaction index.

Given an ordered group
\[
(k_1,k_2,\dots,k_{n_h}),
\qquad
k_j\in\mathcal{N}^{(h)},
\]
a selection rule assigns indices to the three subsets. Under a proportion-based hold-out rule with
\[
\alpha_{tr}+\alpha_{val}+\alpha_{te}=1,
\]
a generic assignment is
\begin{equation}
\begin{aligned}
\mathcal{N}^{(h)}_{tr}
&=
\{k_1,\dots,k_{\lfloor\alpha_{tr}n_h\rfloor}\},\\
\mathcal{N}^{(h)}_{val}
&=
\{k_{\lfloor\alpha_{tr}n_h\rfloor+1},\dots,
k_{\lfloor(\alpha_{tr}+\alpha_{val})n_h\rfloor}\},\\
\mathcal{N}^{(h)}_{te}
&=
\{k_{\lfloor(\alpha_{tr}+\alpha_{val})n_h\rfloor+1},
\dots,k_{n_h}\}.
\end{aligned}
\label{eq:holdout_generic}
\end{equation}
When the ordering is random, this produces a random hold-out split; when it is chronological, it produces a temporal hold-out split.

A leave-one-out rule instead assigns the final element of an ordered group to test:
\[
\mathcal{N}^{(h)}_{te}=\{k_{n_h}\}.
\]
When validation data are used, the preceding interaction may be assigned to validation:
\[
\mathcal{N}^{(h)}_{val}=\{k_{n_h-1}\},
\]
with all earlier interactions assigned to training. Applied to user groups under chronological ordering, this construction yields temporal leave-one-out; under random ordering, it yields random leave-one-out.

The groupwise subsets are combined as
\[
\mathcal{N}_{tr}
=
\bigcup_{h\in\mathcal{G}}\mathcal{N}^{(h)}_{tr},
\qquad
\mathcal{N}_{val}
=
\bigcup_{h\in\mathcal{G}}\mathcal{N}^{(h)}_{val},
\qquad
\mathcal{N}_{te}
=
\bigcup_{h\in\mathcal{G}}\mathcal{N}^{(h)}_{te}.
\]

\paragraph{Global temporal cutoffs and block-based splits.}
Not all temporal protocols are naturally described as independent groupwise splits. In global temporal hold-out, a common cutoff is applied to the complete interaction log. Given two temporal thresholds $c_{tr}<c_{val}$, one may define
\[
\mathcal{N}_{tr}
=
\{k\in\mathcal{N}^{0}\mid \pi_t(r_k)\leq c_{tr}\},
\]
\[
\mathcal{N}_{val}
=
\{k\in\mathcal{N}^{0}\mid
c_{tr}<\pi_t(r_k)\leq c_{val}\},
\]
and
\[
\mathcal{N}_{te}
=
\{k\in\mathcal{N}^{0}\mid
\pi_t(r_k)>c_{val}\}.
\]
This construction imposes a common information boundary across all users and items.

Similarly, time-window protocols may first partition the log into ordered temporal blocks, such as days, weeks, or months, and then assign complete blocks to training, validation, or test. Unlike user-wise splitting, these protocols do not generally allocate interactions from every window to every subset. They therefore require an assignment at the level of temporal blocks rather than independent allocation within each block.

More complex protocols may apply multiple assignment stages. For example, a cold-start evaluation may first assign entire user groups to different roles and then split the histories of selected users further. Such hybrid procedures can still be represented by the final assignment function $s_{\mathcal{P}}$, even when their construction involves several sequential rules.

\paragraph{Repeated protocols.}
Repeated hold-out and cross-validation generate a family of assignments rather than a single split. Let $q\in\{1,\dots,Q\}$ index repetitions or folds. The protocol then yields
\[
\left\{
\left(
\mathcal{N}_{tr}^{(q)},
\mathcal{N}_{val}^{(q)},
\mathcal{N}_{te}^{(q)}
\right)
\right\}_{q=1}^{Q}.
\]
Repeated hold-out typically generates independent random assignments and aggregates the resulting performance estimates. In $K$-fold cross-validation, the folds are constructed so that each eligible observation, or each eligible group under grouped cross-validation, serves as test data in one fold.

\paragraph{Interaction-log preparation and split-dependent operations.}
As formalised in the previous chapter, the interaction log used for evaluation is generally the output of a data preparation pipeline rather than the raw dataset. The splitting protocol is therefore applied to the processed log $\tilde{\mathcal{D}}$, obtained by composing a sequence of parameterised preparation operators.

The placement of a preparation operator relative to the split is itself methodologically relevant. Operations based only on static or externally available information may be applied to the complete interaction log before splitting. In contrast, transformations whose parameters are estimated from interaction data, such as popularity statistics, learned representations, frequency-based filters, or normalisation statistics, must be fitted using training data only, or using the information available before the relevant temporal cutoff. Applying such operators globally before splitting may otherwise introduce information from validation or test interactions into the training pipeline.

Post-split filtering may also be applied to enforce transductive entity coverage. Let
\[
\mathcal{U}(\mathcal{D}')
=
\{\pi_u(r)\mid r\in\mathcal{D}'\},
\qquad
\mathcal{I}(\mathcal{D}')
=
\{\pi_i(r)\mid r\in\mathcal{D}'\}.
\]
A transductive filtering operator can be written as
\begin{multline}
\mathsf{F}(\mathcal{D}_{val},\mathcal{D}_{te};\mathcal{D}_{tr})\\
=
\big(
\{r\in\mathcal{D}_{val}\mid
\pi_u(r)\in\mathcal{U}(\mathcal{D}_{tr}),
\ \pi_i(r)\in\mathcal{I}(\mathcal{D}_{tr})\},\\
\{r\in\mathcal{D}_{te}\mid
\pi_u(r)\in\mathcal{U}(\mathcal{D}_{tr}),
\ \pi_i(r)\in\mathcal{I}(\mathcal{D}_{tr})\}
\big).
\end{multline}
This operation is not universally required: retaining unseen users or items may be an intentional choice in inductive or cold-start evaluation. The treatment of excluded interactions should therefore be reported as an explicit component of the protocol.

\paragraph{Auxiliary information under splitting.}
The split assignment acts directly on interaction events, but auxiliary information must also respect the information constraints of the evaluation setting. Static and exogenous attributes, such as a stable item category or a user-provided profile field, can be restricted to the entities appearing in each sublog. For a sublog $\mathcal{D}'$, define
\[
\mathcal{E}(\mathcal{D}')
=
\mathcal{U}(\mathcal{D}')
\cup
\mathcal{I}(\mathcal{D}').
\]
A static source $\phi_s$ can then be restricted to $\mathcal{E}(\mathcal{D}')$.

This inheritance principle does not apply automatically to time-varying or interaction-derived information. Item metadata that changes over time, popularity statistics, interaction graphs, learned embeddings, and representations built from textual or behavioural data must be restricted to information available at the relevant prediction time. In particular, such features should be fitted on training data or constructed using only observations preceding the applicable temporal cutoff.

\paragraph{Protocol provenance and external evaluation choices.}
A complete description of a splitting protocol also includes its provenance: whether the partition was generated by the study, provided with the dataset, inherited from a prior benchmark, recovered from a repository, or only partially documented. For fixed or inherited splits, the final assignment may be available even when the original random, temporal, or hybrid rule used to generate it cannot be reconstructed. Provenance is therefore a protocol-level metadata attribute rather than an allocation rule.

Finally, candidate-set construction and negative sampling are not components of the split assignment. They determine how held-out interactions are evaluated, particularly in top-$N$ ranking tasks, and should be reported alongside the splitting procedure as separate but complementary evaluation choices.

\section{Empirical Analysis of Splitting Practices}
\label{sec:empirical_analysis}

The dimensions described in the previous sections characterise the design space of splitting protocols in principle. Our literature review shows, however, that the protocols adopted in practice are far from uniformly distributed. Recurring patterns emerge across recommendation paradigms, reflecting differences in task formulation, benchmark reuse, and evaluation conventions.

To move beyond qualitative observations, we aggregate the annotated metadata extracted from the reviewed corpus. The analysis is conducted at the paper--dataset level and covers $1497$ paper--dataset pairs drawn from $517$ papers. This quantitative perspective allows us to examine how splitting families are distributed across paradigms, how their aggregate prevalence changes over time, and where reporting limitations persist.

\subsection{Overall prevalence of splitting families}
Table~\ref{tab:overall_split_family} summarises the distribution of splitting families across the $1497$ paper--dataset pairs in the surveyed corpus. At this aggregate level, random protocols account for the largest share of evaluation settings. Random global or user-wise hold-out represents $35.8\%$ of paper--dataset pairs, while random leave-one-out accounts for a further $21.2\%$. Explicitly time-aware protocols are also widely represented, with temporal leave-one-out and temporal hold-out accounting for $14.5\%$ and $12.3\%$, respectively.

This distribution indicates that a large share of offline evaluation continues to rely on protocols that do not impose a common temporal information boundary between training and testing. Such procedures may be appropriate when the task is framed as static matrix completion or reconstruction of unobserved interactions. Their interpretation, however, differs from that of protocols designed to evaluate the prediction of future user behaviour.

\begin{table}[htpb]
\centering
\small
\caption{Overall distribution of splitting families across the $1497$ surveyed paper--dataset pairs.}
\label{tab:overall_split_family}
\begin{tabular}{@{}lc@{}}
\toprule
\textbf{Splitting Family} & \textbf{Prevalence (\%)} \\
\midrule
Random (Global \& User-wise Hold-out) & $35.8\%$ \\
Random Leave-One-Out & $21.2\%$ \\
Temporal Leave-One-Out & $14.5\%$ \\
Temporal Hold-out (Global \& User-wise) & $12.3\%$ \\
Fixed / Inherited Benchmark Splits & $8.4\%$ \\
Hybrid / Custom / Cross-Validation & $4.1\%$ \\
Not Declared (ND) & $3.7\%$ \\
\bottomrule
\end{tabular}
\end{table}

Fixed or inherited benchmark splits account for $8.4\%$ of the corpus. This proportion is methodologically relevant because, for these evaluation settings, the final partition may be available while the random, temporal, or hybrid procedure originally used to generate it remains unknown. Hybrid, custom, and cross-validation procedures account for a further $4.1\%$, confirming that a non-negligible share of studies cannot be described adequately through a single hold-out or leave-one-out label.

\subsection{Paradigm-dependent conventions}

The Distribution of splitting families varies substantially across recommendation paradigms, as shown in Table~\ref{tab:category_split_normalised}. Rather than following one common evaluation convention, different research areas exhibit recurring protocol choices that align with their task definitions and benchmark ecosystems.

\paragraph{Random protocols in Collaborative Filtering and Cross-domain Recommendation.}
Collaborative filtering and cross-domain recommendation are predominantly evaluated using random protocols. In both paradigms, random global or user-wise hold-out and random leave-one-out account for the largest shares of paper--dataset pairs. This pattern is consistent with settings in which the user-item matrix is treated primarily as a static object, and the evaluation objective is the reconstruction or ranking of unobserved interactions, rather than prediction under a temporal availability constraint. Examples of this practice can be found in classical collaborative filtering \cite{DBLP:conf/uai/RendleFGS09,DBLP:conf/www/SedhainMSX15,DBLP:conf/www/HeLZNHC17} and cross-domain recommendation \cite{DBLP:conf/wsdm/ShuWTWL18}.

\paragraph{Temporal leave-one-out in Sequential Recommendation.}
Sequential recommendation shows the clearest concentration around a single family: temporal leave-one-out. In the reviewed corpus, this protocol accounts for the largest share of sequential paper--dataset pairs. The usual construction reserves the most recent interaction of each user for testing, often retains the penultimate interaction for validation, and uses the preceding history for training \cite{DBLP:conf/icdm/KangM18,DBLP:conf/cikm/SunLWPLOJ19,DBLP:conf/www/RendleFS10}. This design aligns the evaluation with a user-level next-item prediction task, although it does not necessarily impose a common global temporal cutoff across users.

\paragraph{Global temporal hold-out in Session-based Recommendation.}
Session-based recommendation is characterised primarily by temporal hold-out. Because sessions are bounded sequences and short-term dynamics are central to the task, many studies reserve the most recent temporal block---for example, the last day or week of sessions---for evaluation \cite{DBLP:journals/corr/HidasiKBT15,DBLP:conf/aaai/WuT0WXT19}. Compared with user-wise temporal leave-one-out, this approach more naturally supports a global deployment-oriented setting in which the model is trained on interactions available before a shared temporal boundary.

\paragraph{Fixed, inherited, and composite protocols in Graph-based Recommendation.}
Graph-based recommendation displays a comparatively large share of fixed, inherited, Hybrid, and custom procedures. In Table~\ref{tab:category_split_normalised}, these families are reported jointly for readability and account for $30\%$ of the graph-based paper--dataset pairs. This pattern is consistent with frequent reuse of benchmark pipelines introduced by influential baselines or distributed through public repositories. However, because the normalised table aggregates fixed/inherited and hybrid/custom procedures, it should not be interpreted as evidence that the entire share corresponds specifically to inherited benchmark splits.

\paragraph{Heterogeneity in LLM-based Recommendation.}
The LLM-based recommendation literature shows a heterogeneous distribution of splitting families, without a single family accounting for a majority of the reviewed paper--dataset pairs. Temporal leave-one-out, random hold-out, temporal hold-out, and the combined fixed/inherited or hybrid/custom family all occur in the reviewed paper--dataset pairs. This variability is consistent with the relative recency of the area and indicates that its evaluation conventions have not yet converged around a single dominant protocol.

\begin{table}[tbp]
\centering
\small
\setlength{\tabcolsep}{5pt}
\renewcommand{\arraystretch}{1.1}
\caption{Row-wise distribution of splitting families across recommendation paradigms. The combined column \emph{Fix./Inh. + Hyb./CV} groups fixed/inherited benchmark splits with hybrid/custom/cross-validation protocols; ND indicates protocols that are not declared or cannot be reliably reconstructed.}
\label{tab:category_split_normalised}
\begin{tabular}{@{}lcccccc@{}}
\toprule
\textbf{Category} &
\textbf{\shortstack{Rand.\\HO}} &
\textbf{\shortstack{Rand.\\LOO}} &
\textbf{\shortstack{Temp.\\LOO}} &
\textbf{\shortstack{Temp.\\HO}} &
\textbf{\shortstack{Fix./Inh.\\+ Hyb./CV}} &
\textbf{ND} \\
\midrule
Collaborative & 0.45 & 0.35 & 0.05 & 0.02 & 0.08 & 0.05 \\
Sequential    & 0.05 & 0.05 & \textbf{0.75} & 0.10 & 0.03 & 0.02 \\
Graph         & 0.28 & 0.08 & 0.02 & 0.25 & \textbf{0.30} & 0.07 \\
Session       & 0.05 & 0.00 & 0.05 & \textbf{0.65} & 0.15 & 0.10 \\
Cross-domain  & \textbf{0.55} & 0.20 & 0.00 & 0.15 & 0.08 & 0.02 \\
Federated     & 0.40 & 0.35 & 0.00 & 0.05 & 0.15 & 0.05 \\
Multimodal    & \textbf{0.60} & 0.20 & 0.05 & 0.00 & 0.05 & 0.10 \\
LLM           & 0.25 & 0.10 & 0.35 & 0.15 & 0.05 & 0.10 \\
\bottomrule
\end{tabular}
\end{table}

\subsection{Temporal trends: aggregate changes in protocol adoption}

\begin{figure}[tbp]
    \centering
    \includegraphics[width=\textwidth]{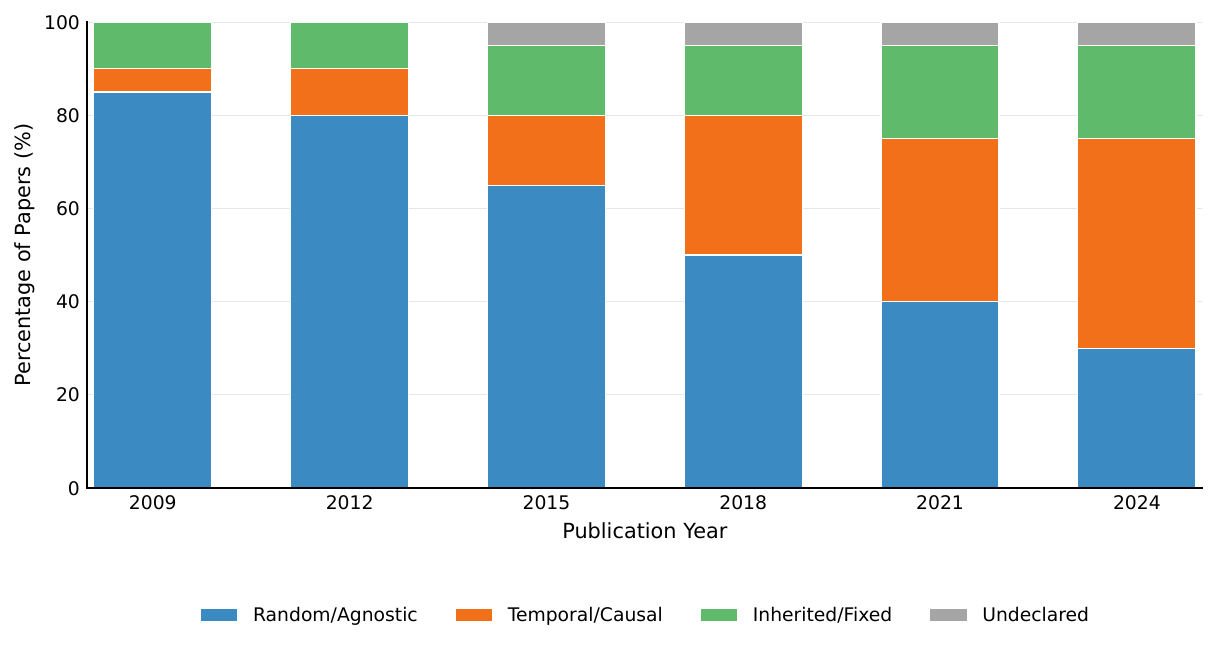}
    \caption{Aggregate evolution of splitting families over time, measured over paper--dataset pairs. Random protocols dominate the earlier years of the corpus, whereas temporal protocols become more prevalent with the growth of sequential and session-based recommendation.}
    \label{fig:temporal_trends}
\end{figure}

Figure~\ref{fig:temporal_trends} reports the aggregate Distribution of splitting families over time. Early portions of the reviewed literature are dominated by random hold-out and cross-validation procedures, consistent with the prominence of static matrix-completion perspectives in earlier recommendation research.

From approximately 2016 onward, temporal protocols have become more prevalent in the aggregate corpus, particularly temporal leave-one-out and temporal hold-out. This pattern coincides with the growing presence of sequential and session-based recommendation research. Fixed and inherited benchmark splits also become more visible in later years, alongside the expansion of benchmark ecosystems in graph-based recommendation.

These trends should be interpreted descriptively. Changes in the aggregate Distribution may reflect both evolving evaluation practices within individual paradigms and changes in the relative representation of different paradigms in the corpus. A rise in temporal protocols, for example, may partly result from the increased share of sequential and session-based studies rather than from a uniform field-wide shift toward temporal evaluation.

\subsection{Protocol reporting and reproducibility}

A final pattern concerns the completeness of reporting for splitting procedures. Some reviewed papers--dataset pairs provide only partial descriptions, refer vaguely to previous work, or do not declare the protocol used. These cases are annotated as \emph{Not Declared} (ND) in our analysis.

The prevalence of ND differs across paradigms. In the normalised Distribution, it is comparatively low in sequential recommendation and higher in session-based, multimodal, and LLM-based recommendation. This variation should be interpreted as an indicator of reporting completeness within the reviewed corpus, rather than as a direct measure of experimental validity.

Incomplete reporting limits reproducibility by preventing later researchers from reconstructing the allocation of interactions, the temporal constraints imposed by the evaluation, and the treatment of users or items absent from the training data. It also limits comparability: two reported scores cannot be interpreted as measuring the same empirical setting when the construction of the corresponding splits remains unknown. Clear disclosure of the allocation rule, grouping unit, ordering criterion, validation design, entity-coverage policy, and provenance of the split is therefore necessary for transparent offline evaluation.
\chapter{Conclusions: Towards Transparent and Task-Aware Offline Evaluation}
\label{sec:conclusion}

Offline evaluation is often presented as a comparison between recommender models. The evidence reviewed in this survey suggests that it is more accurately understood as a comparison between complete experimental pipelines. Dataset selection, interaction modelling, the construction and processing of side information, and data-splitting strategies jointly define the task that is eventually evaluated. They determine which signals are available to a model, which users and items are represented in the training data, what constitutes a relevant prediction, and whether a reported result reflects a static reconstruction problem or an attempt to anticipate future behaviour. Consequently, these choices are not merely implementation details: they are part of the experimental design and shape the interpretation of offline results.

The analysis of recommendation datasets highlights that a dataset should not be treated as a neutral user--item matrix. Its domain, feedback mechanism, temporal coverage, sparsity, entity population, and available side information all embed assumptions about the phenomenon under study. Even when two studies refer to the same named dataset, they may rely on different versions, subsets, or representations of it. Such differences may change the effective task as substantially as a change in model architecture. Dataset selection should therefore be motivated in relation to the intended recommendation scenario, rather than regarded solely as a conventional benchmark choice.

Data preparation further amplifies this point. Filtering interactions or entities, binarising or transforming feedback, constructing sequences or sessions, and extracting representations from textual, visual, or other side information all alter the evidence made available to the recommender. These operations may be necessary and well justified, but they also introduce analytical choices whose consequences cannot be separated from the observed performance of a model. The empirical patterns reviewed in this survey show that some of these choices are repeatedly adopted through established benchmark practice while their exact implementation is not always disclosed. This is particularly consequential for side information: the choice of modalities, feature extractors, pre-training sources, and representation provenance can materially affect the resulting recommendation problem. The objective should not be to prescribe a universally valid preprocessing pipeline, but to make the rationale and implementation of each pipeline explicit and reproducible.

The review of splitting strategies leads to a complementary conclusion. Splits do not simply partition a fixed problem into training, validation, and test data; they specify the prediction setting in which a method is assessed. Random and user-wise protocols are common in settings that frame recommendation as the ranking of unobserved interactions, whereas temporal protocols are more prevalent in sequential and session-based recommendation, where availability over time is central to the task. Fixed and inherited splits can facilitate comparison within an established benchmark ecosystem, but can also obscure the assumptions inherited from previous work. None of these choices is intrinsically preferable in all contexts. Their suitability depends on the deployment scenario and on the scientific question being asked. What matters is that the allocation rule, grouping unit, ordering criterion, validation design, treatment of previously unseen entities, and provenance of the split are stated clearly enough for the evaluated setting to be reconstructed.

Taken together, these observations have direct implications for how progress in recommender systems research should be interpreted. A gain in offline metrics cannot automatically be attributed to a modelling contribution when the datasets, transformations, or evaluation protocols differ across studies or remain only partially described. Conversely, a method that is less competitive under a widely used benchmark protocol may be better aligned with a different and more realistic operational setting. Comparability is thus not guaranteed by reporting scores on datasets with familiar names; it requires that studies make the pipeline that produced those scores visible. This is also essential for reproducibility, since later researchers cannot replicate or meaningfully extend an experiment without knowing how data were selected, prepared, and split.

Based on the surveyed literature, we encourage researchers to treat the following information as a core component of an offline evaluation report: the source and version of each dataset; the definition and semantics of interactions; all filtering and transformation steps; the construction and provenance of side-information features; and a complete specification of the training, validation, and test protocol. Whenever possible, these choices should be accompanied by released code, configuration files, and the actual data splits. More importantly, each choice should be justified with respect to the recommendation task and the intended use of the system. Standardised benchmarks and shared pipelines remain valuable, but their use should not replace an explicit account of the assumptions that they encode.

Several directions follow from this perspective. Future benchmark initiatives could provide versioned datasets together with declarative, reusable pipeline specifications, allowing researchers to distinguish clearly between model changes and data-processing changes. Reporting standards could make the description of preprocessing and splitting as routine as the description of model hyperparameters and evaluation metrics. Further work is also needed to assess how robust conclusions are to plausible variations in data preparation and splitting, especially in multimodal, LLM-based, and other emerging recommendation settings where conventions have not yet stabilised. Finally, offline protocols should be designed with greater attention to their relation to real deployment conditions, including temporal availability, cold-start conditions, and the gap between historical interaction logs and future user behaviour.

The central message of this survey is therefore not that recommender systems research should adopt one universal dataset, preprocessing pipeline, or splitting strategy. Such a prescription would be neither feasible nor desirable across the diversity of recommendation tasks. Rather, credible progress requires experiments whose data assumptions are explicit, whose protocols are task-aware, and whose results can be interpreted and reproduced by others. Improving the transparency of the full evaluation pipeline is a necessary step towards making offline evidence more comparable, more robust, and ultimately more informative for recommender systems research.

%BACKMATTER SEE DOCUMENTATION
\backmatter

\printbibliography

\end{document}